\documentclass[12pt]{article}\usepackage[]{graphicx}\usepackage[]{xcolor}
\makeatletter
\def\maxwidth{ %
  \ifdim\Gin@nat@width>\linewidth
    \linewidth
  \else
    \Gin@nat@width
  \fi
}
\makeatother

\definecolor{fgcolor}{rgb}{0.345, 0.345, 0.345}
\newcommand{\hlkwb}[1]{\textcolor[rgb]{0.69,0.353,0.396}{#1}}%

\usepackage{framed}
\makeatletter
 {\par\unskip\endMakeFramed%
 \at@end@of@kframe}
\makeatother

\definecolor{shadecolor}{rgb}{.97, .97, .97}
\definecolor{messagecolor}{rgb}{0, 0, 0}
\definecolor{warningcolor}{rgb}{1, 0, 1}
\definecolor{errorcolor}{rgb}{1, 0, 0}
\newenvironment{knitrout}{}{} % an empty environment to be redefined in TeX

\usepackage{alltt}
\usepackage[margin=1.2in]{geometry}
\usepackage{amsmath, amssymb, amsthm}
\usepackage[linesnumbered, ruled]{algorithm2e}
\usepackage{natbib}
\usepackage{enumitem}
\usepackage{bbold}
\usepackage{caption}
\usepackage{subcaption}
\usepackage{booktabs}
\usepackage{array}
\usepackage{placeins}
\usepackage{rotating}
\usepackage{xcolor}
\usepackage{xr}
\usepackage[hypertexnames=false]{hyperref}
\hypersetup{pdftex,colorlinks=true,allcolors=blue}

\newtheorem{prop}{Proposition}

\SetArgSty{textnormal} 

\newcommand\col{{\hspace{0.6mm}:\hspace{0.6mm}}}

\newcommand{\eps}{\epsilon}

\newcommand{\N}{\mathcal{N}}

\newcounter{example}
\title{Sampling from high dimensional, multimodal distributions using automatically tuned, tempered Hamiltonian Monte Carlo}
\author{Joonha Park\\
  Department of Mathematics, University of Kansas\\
  1460 Jayhawk Blvd. Lawrence, KS 66045, USA\\
  email: j.park@ku.edu\\
  ORCID id: 0000-0002-4493-7730}
\date{}
\IfFileExists{upquote.sty}{\usepackage{upquote}}{}
\begin{document}

\maketitle

\begin{abstract}
Hamiltonian Monte Carlo (HMC) is widely used for sampling from high dimensional target distributions with densities known up to proportionality.
While HMC exhibits favorable scaling properties in high dimensions, it struggles with strongly multimodal distributions.
Tempering methods are commonly used to address multimodality, but they can be difficult to tune, especially in high dimensional settings.
In this study, we propose a method that combines tempering with HMC to enable efficient sampling from high dimensional, strongly multimodal distributions.
Our approach simulates the dynamics of a time-varying Hamiltonian in which the temperature increases and then decreases over time.
In the first phase, the simulated trajectory gradually explores low-density regions farther from the mode; the second phase guides it back toward a local mode.
We develop efficient tuning strategies based on a time-scale transformation under which the Hamiltonian becomes approximately stationary.
This leads to a tempered Hamiltonian Monte Carlo (THMC) algorithm with automatic tuning.
We demonstrate numerically that our method scales more effectively with dimension than adaptive parallel tempering and tempered sequential Monte Carlo.
Finally, we apply our THMC to sample from strongly multimodal posterior distributions arising in Bayesian inference.

\end{abstract}

\renewcommand\floatpagefraction{0.6}

\section{Introduction} \label{sec:intro}

Hamiltonian Monte Carlo (HMC) is a class of Markov chain Monte Carlo (MCMC) algorithms that use Hamiltonian dynamics to construct efficient proposal mechanisms for sampling from unnormalized target densities \citep{duane1987hybrid}.
Compared to other commonly used MCMC methods, such as random-walk Metropolis or the Metropolis-adjusted Langevin algorithm (MALA), HMC exhibits superior scaling properties in high dimensional spaces.
This advantage arises from its use of local geometric information about the log target density to propose global moves \citep{gelman1997weak, roberts1998optimal, beskos2013optimal, neal2011mcmc}.
These favorable scaling properties have led to the widespread adoption of HMC in various domains, particularly Bayesian data analysis \citep{gelman2013bayesian,neal1996bayesian,brooks2009charmm,landau2021guide}.

However, in the case of strongly multimodal target distributions, HMC methods often encounter challenges in efficiently exploring multiple modes \citep{mangoubi2018does}.
These challenges manifest in constructed Markov chains that exhibit infrequent transitions between modes.
Furthermore, depending on the initial state, these chains may fail to visit globally dominant modes, potentially leading to a misrepresentation of the target distribution.
One potential strategy to address this issue involves running parallel chains with diverse initial states to enhance the likelihood of identifying dominant modes.
However, the proportion of chains settling into different local modes might not accurately reflect the relative probabilities associated with those modes.

Numerous strategies have been developed to enable efficient sampling from multimodal target distributions.
One class of methods uses optimization procedures to identify the locations and approximate shapes of the modes, and then constructs an MCMC kernel to facilitate transitions between them \citep{andricioaei2001smart, sminchisescu2011generalized, pompe2020framework}.
Darting Monte Carlo, for example, employs an independent Metropolis-Hastings (MH) sampler that proposes candidates near known mode locations \citep{andricioaei2001smart, sminchisescu2011generalized}.
A practical extension, proposed by \citet{ahn2013distributed}, adaptively tunes the independent MH sampler using parallel chains at regeneration times.
These approaches often approximate the target distribution using models such as mixtures of truncated normal distributions, with parameters estimated from the chain's history.
However, such approximations may become inaccurate and increasingly difficult to implement as the dimensionality of the space grows.

Tempering is a strategy that introduces a sequence of auxiliary distributions, typically constructed by raising the target density to a power known as the inverse temperature.
These intermediate distributions facilitate transitions between isolated modes by flattening the density landscape; points in low-density regions are more likely to be sampled when the inverse-temperature is low.
Simulated tempering, introduced by \citet{marinari1992simulated}, constructs a Markov chain targeting a mixture of tempered distributions at different temperature levels.
Effective sampling with simulated tempering requires careful selection of mixture weights for the tempered distributions, which may be achieved through adaptive tuning techniques, such as those proposed by \citet{wang2001efficient} and \citet{atchade2010wang}.
Parallel tempering, proposed by \citet{swendsen1986replica} and \citet{geyer1991markov}, involves constructing parallel chains, each targeting a different tempered distribution. 
Similarly, the equi-energy sampler, introduced by \citet{kou2006equi}, employs parallel chains targeting distributions at various temperatures.
However, unlike parallel tempering, state exchanges in the equi-energy sampler occur exclusively between points within the same potential energy band. 
The tempered transitions method, developed by \citet{neal1996sampling}, applies a series of transition kernels corresponding to a sequence of decreasing and increasing inverse temperature levels, facilitating exploration of the target distribution.
Tempered sequential Monte Carlo (TSMC) differs from the previously mentioned methods in that it incorporates tempering within the sequential Monte Carlo framework rather than within an MCMC framework \citep{neal2001annealed, delmoral2006sequential}.
However, MCMC kernels are still used to diversify the particles after each intermediate resampling step.

In this paper, we propose a method that incorporates tempering within Hamiltonian Monte Carlo to facilitate frequent mode transitions in high dimensional, multimodal target distributions.
Our approach simulates the dynamics of a time-varying Hamiltonian, in which the temperature increases and then decreases along each trajectory.
In the first half, the trajectory expands into broader, low-probability regions of the state space; in the second half, it contracts back toward a local mode.
This method is closely related to the velocity scaling approach proposed by \citet[Section~5.5.7]{neal2011mcmc}---under certain conditions, the method in \citet{neal2011mcmc} is equivalent to ours.
However, our method is more broadly applicable, as it facilitates adaptation to a wide range of target distributions.
Our contributions also include the development of adaptive tuning strategies for our tempered Hamiltonian Monte Carlo (THMC) method, based on an analysis of the time-varying Hamiltonian dynamics under a time scale transformation.
Incorporating the adaptive tuning algorithm yields an automatically tuned, tempered Hamiltonian Monte Carlo (ATHMC) method.

The remainder of the paper is organized as follows.
Section~\ref{sec:HMC_and_multimodality} provides a brief review of standard Hamiltonian Monte Carlo and discusses the challenges it faces when sampling from multimodal target distributions.
In Section~\ref{sec:THMC}, we introduce the tempered Hamiltonian Monte Carlo (THMC) algorithm and demonstrate its effectiveness using a mixture of log-polynomial distributions.
Section~\ref{sec:tuning} presents an adaptive tuning strategy for THMC.
In Section~\ref{sec:comparison}, we show that our automatically tuned, tempered HMC scales more effectively in high dimensions compared to parallel tempering and tempered sequential Monte Carlo methods.
Section~\ref{sec:applications} demonstrates the use of ATHMC in sampling from strongly multimodal distributions, including examples arising in Bayesian inference.
Section~\ref{sec:otherapproaches} reviews recent approaches for sampling from multimodal distributions.
Finally, Section~\ref{sec:discussion} concludes with a discussion of potential directions for further research.
An \textsf{R} package implementing our automatically tuned, tempered Hamiltonian Monte Carlo algorithm is available at \url{https://github.com/joonhap/athmc}.
All source codes used in the numerical experiments in provided as supplementary material.

\section{Hamiltonian Monte Carlo and multimodality}\label{sec:HMC_and_multimodality}
\subsection{Hamiltonian Monte Carlo}\label{sec:HMC}
Let $\pi(x)$ denote an unnormalized target density on $\mathsf X = \mathbb R^d$, with unknown normalizing constant $Z$.
A broad class of MCMC methods---including HMC and some variants of the bouncy particle sampler \citep{vanetti2017piecewise, bouchard2018bouncy, park2020markov}---employs an auxiliary \emph{momentum} variable $p \in \mathsf P = \mathbb R^{d}$ and targets the augmented density $\Pi(x,p) = \tfrac{1}{Z}\pi(x) \psi(p)$ defined on $\mathsf X \times \mathsf P = \mathbb R^{2d}$.
In HMC, $\psi(p)$ is typically chosen as the multivariate normal density with mean 0 and covariance matrix $M$.

A candidate for the next state of the Markov chain is obtained by simulating Hamiltonian dynamics governed by the Hamiltonian
\[
  H(x,p) = K(p) + U(x) = \frac{1}{2} p^\top M^{-1} p -\log \pi(x).
\]
Here $U(x) := -\log \pi(x)$ is referred to as the potential energy, and $K(p) = \frac{1}{2} p^\top M^{-1} p$ as the kinetic energy.
The matrix $M$ is conceptually interpreted as the generalized \emph{mass} of the particle and may be any symmetric, positive definite matrix.
The Hamiltonian $H(x,p)$ represents the total energy of a particle located at $x$ with momentum $p$.
The dynamics of a particle governed by this Hamiltonian follow the equations of motion 
\begin{equation}
  \frac{dx}{dt} = \frac{\partial H}{\partial p} = M^{-1} p,\qquad \frac{dp}{dt} = -\frac{\partial H}{\partial x} = -\frac{\partial U}{\partial x}.
\label{eqn:HEMxp}
\end{equation}

%$$
The exact solution of the HEM, denoted by $\Phi_t: (x(0), p(0)) \mapsto (x(t), p(t))$ and called the Hamiltonian flow, conserves the Hamiltonian \citep{leimkuhler2004simulating}:
\[
  H(x(0), p(0)) = H(x(t), p(t)), \qquad \forall t\geq 0.
\]
The Hamiltonian flow is symplectic, meaning that it satisfies
\[
  \left(\frac{\partial \Phi_t(x,p)}{\partial(x,p)}\right)^\top J^{-1} \left(\frac{\partial \Phi_t(x,p)}{\partial(x,p)}\right) = J^{-1},
  \quad \text{where} ~
  J = \begin{pmatrix} 0 & I_d \\ -I_d & 0 \end{pmatrix}.
\]
As a consequence of symplecticness, the volume element is also conserved by the Hamiltonian flow \citep{leimkuhler2004simulating, arnold1989mathematical}:
\[
  \left| \frac{\partial \Phi_t(x(0), p(0))}{\partial (x(0), p(0))} \right| = 1.
\]

Given the $i$-th state $X^{(i)}$ of the Markov chain, the Hamiltonian dynamics is numerically simulated starting from $x(0) = X^{(i)}$ with initial momentum $p(0)$ drawn from $\N(0, M)$.
Let $(x(T), p(T)) = \Psi_T(x(0), p(0))$ denote the end state of the simulated trajectory.
The proposed state $x(T)$ is accepted as $X^{(i+1)}$ if and only if
\begin{equation}
  \Lambda < \exp\big[ -H(x(T), p(T)) + H(x(0), p(0)) \big] \cdot \left|\frac{\partial (x(T),p(T))}{\partial (x(0), p(0))}\right|,
\label{eqn:acceptance_criterion}
\end{equation}
where $\Lambda$ is a $\text{Uniform}(0,1)$ random draw, independent of all other Monte Carlo variables.
If \eqref{eqn:acceptance_criterion} is not satisfied, the proposed move is rejected, and $X^{(i+1)}$ is set to $X^{(i)}$.

A commonly used numerical approximation method for solving the HEM is the leapfrog, or St\"ormer-Verlet, method \citep{hairer2003geometric, leimkuhler2004simulating}.
One leapfrog step approximately simulates the time evolution of the Hamiltonian dynamics for time duration $\epsilon$, referred to as the leapfrog step size.
It alternately updates the velocity and position $(x,v)$ in half steps as follows:
\begin{equation}
  \begin{split}
    p\left(t+\frac{\epsilon}{2}\right) &= p(t) - \frac{\epsilon}{2} \cdot \nabla U(x(t))\\
    x(t+\epsilon) &= x(t) + \epsilon \cdot M^{-1} p\left(t+\frac{\epsilon}{2}\right)\\
    p(t+\epsilon) &= p\left(t+\frac{\epsilon}{2}\right) - \frac{\epsilon}{2} \nabla U(x(t+\epsilon)). \label{eqn:leapfrog_M}
  \end{split}\end{equation}
Like the Hamiltonian flow $\Phi_t$, the numerical simulation map $\Psi_t$ is symplectic \citep{leimkuhler2004simulating}.
As a result, $\Psi_t$ preserves the volume element: $dx(t)dp(t) = dx(0)dp(0)$.
Moreover, the numerical simulation by the leapfrog method enjoys long-term stability---in particular, provided that $\epsilon$ is sufficiently small, we have
\[
H\{\Psi_t(x(0), p(0))\} = H(x(0), p(0)) + O(\epsilon^2)
\]
\citep{neal2011mcmc, leimkuhler2004simulating}.
Due to the long term stability and volume preservation, the probability of accepting the candidate $\Psi_t(x(0), p(0))$ according to the criterion \eqref{eqn:acceptance_criterion} can become arbitrarily close to one by employing a sufficiently small leapfrog step size.
%Since each line in \eqref{eqn:leapfrog_M} is a translation of either $v$ or $x$ by an amount determined by the other variable, the Jacobian determinant of the leapfrog update is equal to unity.
%Therefore, the numerical integrator constructed by the leapfrog method preserves the volume element:
%\[
%  \left|\frac{\partial \Psi_T(x,p)}{\partial(x,p)}\right| \equiv 1.
%\]
Finally, both $\Phi_t$ and $\Psi_t$ are time-reversible: writing $\mathcal T(x,p) := (x, -p)$, we have
\begin{equation}
  \mathcal T \circ \Psi_t \circ \mathcal T \circ \Psi_t (x,p) = (x,p), \quad \forall (x,p) \in \mathsf X\times \mathsf P.
  \label{eqn:TSTS}
\end{equation}
%The numerical stability of the leapfrog method depends on its step size $\epsilon$.
%For instance, if the target density is multivariate normal with covariance $\Sigma$, the numerical error diverges exponentially fast if and only if the step size $\epsilon$ is greater than twice the square root of the smallest eigenvalue of $\Sigma M$.
%From this observation, we can speculate that the numerical stability of the leapfrog method is critically dependent on the relative size of its step size to the inverse square root of the largest eigenvalue of the Hessian of the log target density.

\begin{figure}
  \centering
  \includegraphics[width=.89\linewidth]{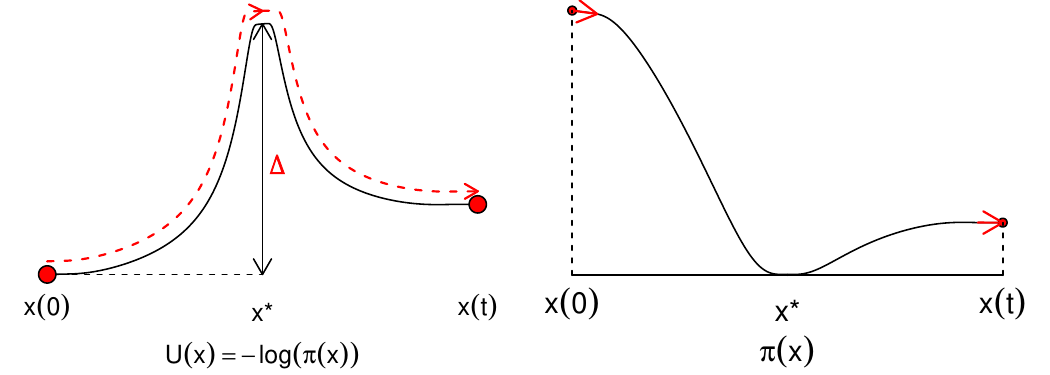}
  \caption{A: Illustrative plot of the potential energy $U(x)$ along a Hamiltonian trajectory in one dimension, showing traversal through a region of high potential energy.
  B: The corresponding target density $\pi(x)$, featuring two local modes. }
  \label{fig:potential_barrier}
\end{figure}
When the target distribution is multimodal, HMC typically fails to visit separated modes.
Denoting $(x(s), p(s)) = \Psi_s(x(0), p(0))$, we have
\[
  H\{x(0), p(0)\} = U\{x(0)\} + K\{p(0)\} \approx H\{x(s), p(s)\} = U\{x(s)\} + K\{p(s)\}.
\]
Consequently, the maximum potential energy increase along the trajectory,
\[
  \Delta := \max_{0\leq s \leq t} U\{x(s)\} - U\{x(0)\} = -\log\frac{\min_{0\leq s \leq t} \pi(x(s))}{\pi(x(0))},
\]
is approximately bounded above by the initial kinetic energy $K\{p(0)\}$.
Figure~\ref{fig:potential_barrier} schematically illustrates the potential energy $U(x)$ (panel A) and the target density $\pi(x)$ (panel B) corresponding to a trajectory that connects two isolated modes.
Since $p(0)$ is drawn from $\N(0,M)$, we have %$M^{1/2}v(0) \sim \N(0,I)$ where $M^{1/2}$ is the symmetric square-root matrix of $M$, and
\[
2K(p(0)) = p(0)^\top M^{-1} p(0) \sim \chi_d^2,
\]
where $\chi_d^2$ denotes the chi-squared distribution with $d$ degrees of freedom.
Therefore, if there are isolated modes in the target distribution $\pi$, the probability that a trajectory starting from one mode reaches another has a Chernoff bound
\begin{equation}
  \mathcal P (K(p(0)) > \Delta) = \mathcal P (\chi_d^2 > 2 \Delta) \leq \left(\frac{2\Delta}{d}\right)^{d/2} e^{\frac{d}{2}-\Delta}.
  \label{eqn:jumpprob}
\end{equation}
The probability \eqref{eqn:jumpprob} is independent of the choice of $M$ and decreases exponentially fast as $\Delta$ increases.
Due to this fact, standard HMC has a poor global mixing property for highly multimodal target distributions.

%%%%%%%%%%%%%%%%%%%%%%%%%%
%%% TEMPERED HMC %%%
%%%%%%%%%%%%%%%%%%%%%%%%%%
\section{Tempered Hamiltonian Monte Carlo}\label{sec:THMC}
\subsection{Incorporating tempering into HMC}\label{sec:motivation}
Various tempering techniques involve sampling from tempered distributions, whose densities are proportional to $\pi(x)^{1/\alpha}$ for some $\alpha \geq 1$.
MCMC targeting a tempered distribution with large $\alpha$ can more easily transition between isolated modes.
Since $\pi(x) \propto e^{-U(x)}$ in HMC, this corresponds to replacing the potential energy function $U(x)$ with $\alpha^{-1} U(x)$.
As a result, a tempered distribution with $\alpha > 1$ exhibits a relatively reduced degree of multimodality.

\addtocounter{algocf}{-1}
\begin{figure}[tp]
  \centering
  \begin{algorithm}[H]
    \SetKwInOut{Input}{Input}\SetKwInOut{Output}{Output}
    \Input{
      Current state of the Markov chain, $X^{(i)}$;~
      Simulation Temperature, $\alpha>1$;~
      Leapfrog step size, $\epsilon$;~
      Target number of acceptance, $N$;~
      Maximum number of proposals per iteration, $N_{\max}$
    }
    \vspace{1ex}
    
    Draw $\Lambda\sim \text{Uniform}(0,1)$\\
    Let $x(0) = X^{(i)}$ and draw $p(0) \sim \N(0, M)$ \hfill \texttt{// initial position and momentum}\\
    \For {$n \gets 1\col N_{\max}$} {
      Apply a leapfrog step for the modified Hamiltonian $H_\alpha$ given by \eqref{eqn:H_alpha}\\
      The obtained pair $(x(n\epsilon), p(n\epsilon))$ is acceptable if $\Lambda < \exp(-H\{x(n\epsilon), p(n\epsilon)\} + H\{x(0), p(0)\})$\\
      If $(x(n\epsilon), p(n\epsilon))$ is the $N$-th acceptable state, let $X^{(i+1)} \gets x(n\epsilon)$ and move on to the next (i.e., $i+1$-st) iteration
    }
    If fewer than $N$ states were acceptable, let $X^{(i+1)} \gets X^{(i)}$
    \caption{HMC with fixed temperature $\alpha > 1$ (toy algorithm, $i$-th iteration)}
    \label{alg:enhancedHMC}
  \end{algorithm}
\end{figure}

In HMC, tempering can be implemented via a modified Hamiltonian, defined as
\begin{equation}
  H_\alpha(x,p) = \frac{1}{2} p^\top M^{-1} p + \alpha^{-1} U(x),
  \label{eqn:H_alpha}
\end{equation}
where $\alpha \geq 1$.
A toy algorithm that simulates trajectories under $H_\alpha$ with a fixed $\alpha > 1$ is summarized in Algorithm~\ref{alg:enhancedHMC}.
Since the potential function is flattened for $\alpha > 1$, the simulated trajectories more easily traverse regions with high values of $U(x)$.
In this algorithm, we make use of time reversibility of the dynamics under $H_\alpha$ and the numerical stability afforded by the symplectic structure.
However, the acceptance probability is computed using the original Hamiltonian $H$ to ensure that the Markov chain leaves the target distribution $\pi(x) \propto e^{-U(x)}$ invariant.

%% Figure comparing HMC and THMC
\begin{figure}[tp]
  \centering
  \includegraphics[width=\textwidth]{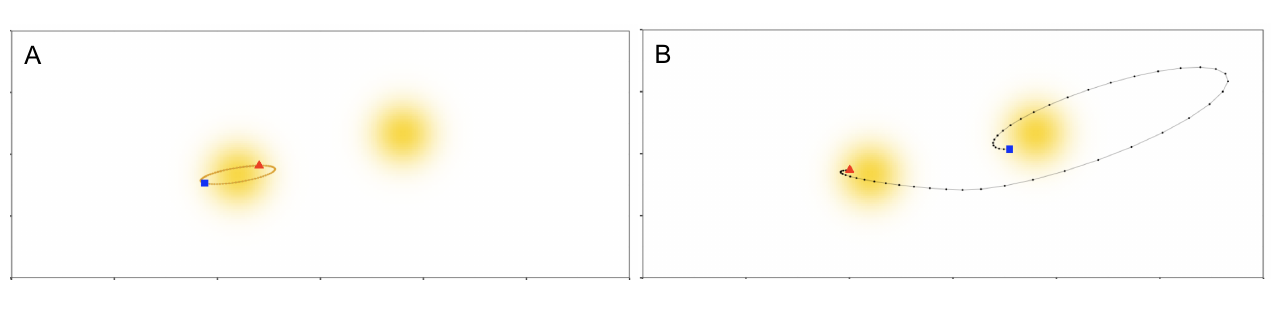}
  \caption{A: Simulated trajectory using standard HMC for a bimodal distribution.
    B: Simulated trajectory using tempered HMC (Algorithm~\ref{alg:THMC}).
    The two yellow clouds indicate regions of high target density.}
  \label{fig:HMC_THMC}
\end{figure}

\begin{figure}[t]
  \centering
  \begin{algorithm}[H]
    \SetKwInOut{Input}{Input}\SetKwInOut{Output}{Output}
    \Input{
      Current state of the Markov chain, $X^{(i)}$;~
      Numerical simulation length, $T$;~
      Temperature schedule, $\alpha(t)$, $0\leq t \leq T$
    }
    \vspace{1ex}
    
    Draw $\Lambda\sim \text{Uniform}(0,1)$\\
    Let $x(0) = X^{(i)}$ and draw $p(0) \sim \N(0, M)$ \hfill \texttt{// initial position and momentum}\\
    Numerically simulate the time-dependent Hamiltonian dynamics for $H_\alpha(x,p,t)$ for time duration $T$, starting from position $x(0)$ and momentum $p(0)$, using Algorithm~\ref{alg:THMC_sim}. Here, $H_\alpha$ is given by \eqref{eqn:H_alpha_t}.\\
    Let $(x(T), p(T))$ be the end state of this simulation\\
    \If {$\Lambda < \exp(-H\{x(T), p(T)\} + H\{x(0), p(0)\})$} {
      Accept $X^{(i+1)} \gets x(T)$\\
    } \Else {
      $X^{(i+1)} \gets X^{(i)}$
    }
    \caption{Tempered Hamiltonian Monte Carlo (THMC, $i$-th iteration)}
    \label{alg:THMC}
  \end{algorithm}
\end{figure}

%$$
A drawback of simulating trajectories for $H_\alpha$ with $\alpha>1$, however, is that the trajectories often terminate at points with high potential energy.
A \emph{sequential proposal strategy}, proposed by \citet{park2020markov}, can help mitigate this issue by continuing the trajectory until an acceptable state is found---specifically, one satisfying
\[
  \Lambda < \exp(-H\{x(nT), p(nT)\} + H\{x(0), p(0)\}).
\]
Using a common value of $\Lambda$ for all proposed candidates ensures that the Markov chain has $\pi$ as its invariant distribution \citep{park2020markov, campos2015extra}.
Algorithm~\ref{alg:enhancedHMC} summarizes this approach in a slightly more general setting, in which trajectories are continued until a specified number $N\geq 1$ of acceptable states are found.
This method can enable reasonably efficient sampling from multimodal distributions in low dimensions ($d\lesssim 5$); see Supplementary Section~\ref{sec:supp_enhancedHMC} for numerical demonstrations and further discussion of this algorithm.
However, transitions between isolated modes become increasingly rare in high dimensions, as it is unlikely for a trajectory simulated under $H_\alpha$ with $\alpha \gg 1$ to reach the small-volume region where $U(x)$ is sufficiently low.

This issue motivates us to consider a time-dependent Hamiltonian $H_\alpha(x,p,t)$, where the temperature $\alpha = \alpha(t)$ is a function of time that increases during the first half of the trajectory and decreases during the second. 
Increasing $\alpha$ during the first half enables the simulated particle to escape a local mode, while decreasing it during the second half encourages the particle to settle near a local minimum of $U(x)$.
Specifically, the time-dependent Hamiltonian is given by
\begin{equation}
  H_\alpha(x,p,t) = \frac{1}{2} p^\top M^{-1} p + \frac{1}{\alpha(t)} U(x),
  \label{eqn:H_alpha_t}
\end{equation}
where the temperature $\alpha(t)$ varies with time $t$.
The associated dynamics are described by
\begin{equation}
  \frac{dx}{dt} = \frac{\partial H}{\partial p} = M^{-1}p, \qquad \frac{dp}{dt} = -\frac{\partial H}{\partial x} = -\frac{1}{\alpha(t)} \frac{\partial U}{\partial x}.
  \label{eqn:HEM_alpha_t}
\end{equation}
We refer to this method as tempered Hamiltonian Monte Carlo (THMC), and it is summarized in Algorithm~\ref{alg:THMC}.
Symplectic numerical simulation of these time-dependent Hamiltonian dynamics, along with the construction of $\alpha(t)$, is discussed in Section~\ref{sec:tempering_HEM} and summarized in Algorithm~\ref{alg:THMC_sim}.
Figure~\ref{fig:HMC_THMC} illustrates the difference between standard HMC and our THMC approach: while the standard HMC trajectory remains confined to a single mode, THMC enables transitions between isolated modes by varying the temperature along the trajectory.

\subsection{Connection to the velocity scaling method of \citet{neal2011mcmc}}\label{sec:equivalence}

\citet[Section~5.5.7]{neal2011mcmc} proposed a method that incorporates tempering within a trajectory.
This method scales the momentum $p$, or equivalently the velocity $v = M^{-1}p$, by a certain factor $\xi$ or $\xi^{-1}$ after each leapfrog step.
This approach is equivalent to our tempered Hamiltonian Monte Carlo when the potential energy function is locally quadratic, but it is sub-optimal otherwise.

To see the equivalence between the two methods, we consider a new time scale $\breve t$ where $d\breve t = \alpha^{-1/2} dt$.
Throughout this paper, we will write
\[
  \eta = \tfrac{1}{2} \log \alpha.
\]
Let $\breve p = \alpha^{1/2} p = e^\eta p$.
Provided that $(x(t), p(t))$ obeys the time-dependent Hamiltonian dynamics \eqref{eqn:HEM_alpha_t}, $(x(\breve t), \breve p(\breve t))$ satisfy
\begin{equation}
  \begin{split}
    \frac{dx}{d\breve t} &= \frac{dx}{dt} \cdot \frac{dt}{d\breve t} = M^{-1} p \cdot e^\eta = M^{-1} \breve p,\\
    \frac{d\breve p}{d\breve t} &= \frac{d}{dt} \left( e^{\eta} p\right) \cdot \frac{dt}{d\breve t}
    = \left( -e^\eta \alpha^{-1} \frac{\partial U}{\partial x} + \frac{d \eta}{dt} \cdot e^\eta p \right) \cdot e^{\eta}
    = -\frac{\partial U}{\partial x} + \frac{d\eta}{d\breve t} \cdot \breve p.
  \end{split}
  \label{eqn:ODE_p_scaling}
\end{equation}
The equations in \eqref{eqn:ODE_p_scaling} match those in \eqref{eqn:HEMxp}, except for the additional term $(d\eta/d\breve t)\cdot \breve p$, which represents momentum scaling.
Numerically, each leapfrog step in \citet{neal2011mcmc}'s method corresponds to a fixed increment in $\breve t$.
The momentum scaling by $\xi^{\pm 1} = e^{\Delta \eta}$, where $\Delta \eta = \eta(\breve t + \Delta \breve t) - \eta(\breve t)$, accounts for the additional term in \eqref{eqn:ODE_p_scaling}.
Here, $d\eta/d\breve t$ is a positive constant during the first half of the trajectory and its negative during the second half.

In the supplementary text Section~\ref{sec:supp_neal2011}, we directly verify that the numerical simulation in \citet{neal2011mcmc}'s method with a constant step size $\bar \epsilon$ is equivalent to our tempered HMC method simulating \eqref{eqn:HEM_alpha_t} with a varying leapfrog step size $\epsilon = e^{\eta} \bar \epsilon$. 
While it may not be immediately clear whether \citet{neal2011mcmc}'s velocity scaling method preserves the long-term numerical stability of symplectic integrators, its equivalence to the time-dependent Hamiltonian dynamics under a reparameterized time scale ensures that such stability is attainable under certain conditions.

In particular, when the local growth rate of the potential $U(x)$ is quadratic---that is, when the polynomial degree $\gamma=2$---\citet{neal2011mcmc}'s method becomes identical to our method.
In Section~\ref{sec:tempering_HEM}, we will show that the optimal scaling of the leapfrog step size $\epsilon$ for the simulation of time-dependent Hamiltonian dynamics in \eqref{eqn:HEM_alpha_t} is given by
\[
  \epsilon = e^{2a\eta} \bar \epsilon,
\]
where $\bar \epsilon$ is a fixed reference step size and $a = \tfrac{2}{\gamma+2}$.
Since \citet{neal2011mcmc}'s velocity scaling method is equivalent to our method using $\epsilon = e^\eta \bar \epsilon$, it is optimal when $a=1/2$, or when $\gamma=2$.
For $\gamma\neq 2$, however, \citet{neal2011mcmc}'s method is suboptimal and may sometimes yield numerically unstable trajectories.

%$$
\subsection{Numerical simulation of the tempered Hamiltonian dynamics}\label{sec:tempering_HEM}
In this section, we develop a method for numerically simulating the time-dependent Hamiltonian dynamics described in \eqref{eqn:H_alpha_t} using a time-scale transformation. 
We consider the case where the potential function $U(x)$ grows locally like a polynomial of degree $\gamma$,
\[
  U(x) \propto \Vert x \Vert_B^\gamma := (x^\top B x)^{\gamma/2},
\]
where $B$ is a symmetric positive definite matrix.
Although our ultimate goal is to design an efficient MCMC algorithm for sampling from multimodal distributions, the analysis of this unimodal potential remains relevant.
This is because the net change in the Hamiltonian along a trajectory is largely determined by how the simulated particle moves away from and then contracts toward a local mode.
We found that the overall change in Hamiltonian---which governs the acceptance probability of a proposed state---is not highly sensitive to whether the trajectory crosses multiple modes during the middle portion, where the temperature $\alpha$ is high (see Figure~\ref{fig:path_1d_bimodal}).

Assuming $U(x) = \Vert x \Vert_B^\gamma$, we define transformed variables
\[
  \bar x = \alpha^{-\frac{1}{\gamma+2}} \cdot x = e^{-a\eta} \cdot x, \qquad
  \bar p = \alpha^{\frac{1}{\gamma+2}} \cdot p = e^{a\eta} \cdot p,
\]
where we denote
\[
  a = \frac{2}{\gamma+2} \quad \text{and} \quad \eta = \frac{1}{2} \log \alpha.
\]
The time-dependent Hamiltonian can then be expressed as
\begin{equation}
  \begin{split}
    H_\alpha(x,p,t) &= \frac{1}{2} p^\top M^{-1} p + \alpha(t)^{-1} U(x) \\
                    &= e^{-2a\eta} \cdot \frac{1}{2} \bar p^\top M^{-1} \bar p + e^{-2\eta} \cdot \Vert e^{a\eta} \bar x \Vert_B^\gamma\\
                    &=e^{-2a\eta} \cdot \frac{1}{2} \bar p^\top M^{-1} \bar p + e^{-\frac{4}{\eta+2}\eta} \Vert \bar x \Vert_B^\gamma \\
                    &= e^{-2a\eta} \left( \frac{1}{2} \bar p^\top M^{-1} \bar p + \bar U(\bar x) \right)\\
                    &=: \bar H_\alpha(\bar x, \bar p, t),
  \end{split}
  \label{eqn:H_barH}
\end{equation}
where $\bar U(\bar x) = \Vert \bar x \Vert_B^\gamma$.
The Hamiltonian dynamics corresponding to $\bar H_\alpha(\bar x, \bar p, t)$ is described by
\[
  \frac{d\bar x}{dt} = \frac{\partial \bar H_\alpha}{\partial \bar p} = e^{-2a\eta} \cdot M^{-1} \bar p,\qquad
  \frac{d\bar p}{dt} = -\frac{\partial \bar H_\alpha}{\partial \bar x} = -e^{-2a\eta} \cdot \frac{\partial \bar U}{\partial \bar x}.
\]
These equations are the same as the Hamiltonian equations of motion for $H$ in \eqref{eqn:HEMxp}, except for the presence of a scaling factor $e^{-2a\eta}$.
To reconcile this difference, we introduce a time rescaling:
\begin{equation}
  d\bar t = e^{-2a\eta} dt,
  \label{eqn:tbar}
\end{equation}
so that the dynamics of $(\bar x, \bar p)$ as a function of $\bar t$ becomes identical to the original Hamiltonian dynamics for $H$:
\begin{equation}
  \frac{d\bar x}{d\bar t} = \frac{d\bar x}{dt} \cdot \frac{dt}{d\bar t} = M^{-1} \bar p, \qquad
  \frac{d\bar p}{d\bar t} = \frac{d\bar p}{dt} \cdot \frac{dt}{d\bar t} = -\frac{\partial \bar U}{\partial \bar x}.
  \label{eqn:HEM_bar}
\end{equation}
Since we can simulate the dynamics in \eqref{eqn:HEM_bar} in a numerically stable manner, Equation~\ref{eqn:tbar} suggests that we simulate the Hamiltonian dynamics for $H_\alpha(x,p,t)$ using a leapfrog step size that scales as
\begin{equation}
  \epsilon = e^{2a\eta} \bar \epsilon,
  \label{eqn:bar_epsilon}
\end{equation}
where $\bar \epsilon$ is a fixed reference step size corresponding to a constant increment in the rescaled time $\bar t$.

\begin{figure}[t]
  \centering
  \begin{algorithm}[H]
    \SetKwInOut{Input}{Input}\SetKwInOut{Output}{Output}
    \Input{
      Initial position, $x(0)=X^{(i)}$;~
      Initial momentum, $p(0)\sim \N(0, M)$;~
      Temperature schedule, $\{\alpha_\kappa = e^{2\eta_\kappa}; \kappa = 0, \frac{1}{2}, 1, \dots, K-\frac{1}{2}, K\}$;~
      Reference leapfrog step size, $\bar \epsilon$;~
      Simulation time scale coefficient, $a$;~
    }
    \vspace{1ex}

    Let $t_0 = 0$\\
    \For {$k \gets 1\col K$} { 
      Let $\epsilon_{k-\frac{1}{2}} = e^{2a\eta_{k-\frac{1}{2}}}\bar \epsilon$\\
      Let $p(t_{k-1} + \tfrac{1}{2} \epsilon_{k-\frac{1}{2}}) = p(t_{k-1}) - \tfrac{1}{2} \epsilon_{k-\frac{1}{2}} \cdot \alpha_{k-\frac{1}{2}}^{-1} \frac{\partial U}{\partial x}(x(t_{k-1}))$\\\label{line:THMC_leapfrog1}
      Let $x(t_{k-1} + \epsilon_{k-\frac{1}{2}}) = x(t_{k-1}) + \epsilon_{k-\frac{1}{2}} \cdot M^{-1} p(t_{k-1}+\tfrac{1}{2}\epsilon_{k-\frac{1}{2}})$\\
      Let $p(t_{k-1} + \epsilon_{k-\frac{1}{2}}) = p(t_{k-1}+\tfrac{1}{2} \epsilon_{k-\frac{1}{2}}) - \tfrac{1}{2} \epsilon_{k-\frac{1}{2}} \cdot \alpha_{k-\frac{1}{2}}^{-1} \frac{\partial U}{\partial x}(x(t_{k-1} + \epsilon_{k-\frac{1}{2}}))$\\\label{line:THMC_leapfrog3}
      Let $t_k = t_{k-1} + \epsilon_{k-\frac{1}{2}}$\\
    }
    Let $T = t_K$ and consider $x(T) = x(t_K)$ as a candidate for the next state of the Markov chain
    \caption{Numerical simulation for tempered Hamiltonian Monte Carlo ($i$-th iteration)}
    \label{alg:THMC_sim}
  \end{algorithm}
\end{figure}

\begin{figure}[tp]
\begin{knitrout}
\definecolor{shadecolor}{rgb}{0.969, 0.969, 0.969}\color{fgcolor}

{\centering \includegraphics[width=0.8\linewidth]{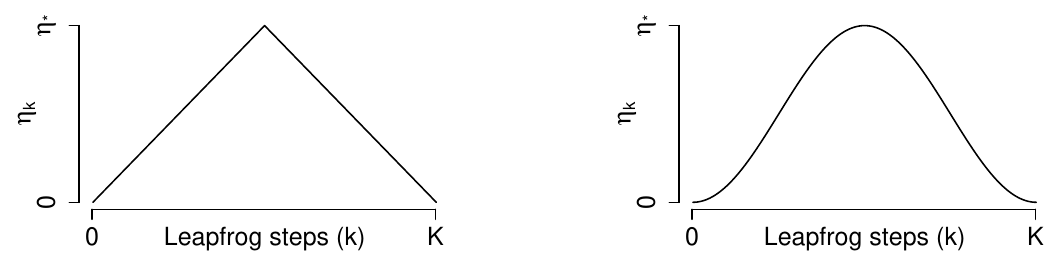} 

}

\end{knitrout}
\caption{Piecewise linear (left) and sinusoidal (right) temperature schedules, $\eta_k = \tfrac{1}{2} \log \alpha_k$.}
\label{fig:temperature_schedules}
\end{figure}

In the numerical simulation of the tempered Hamiltonian dynamics, each leapfrog step is understood as advancing the rescaled time $\bar t$ by a constant $\bar \epsilon$.
Thus, denoting the number of completed leapfrog steps $k$, we have $\bar t \propto k$.
We consider the following temperature schedules:
\begin{align}
  \text{piecewise linear }&\eta_k = \frac{2\eta_*}{K} \min(k, K-k),~~\text{or} \label{eqn:eta_piecewiselinear}\\
  \text{sinusoidal  }&\eta_k = \frac{\eta_*}{2} \left\{ 1- \cos\left(\frac{2\pi k}{K}\right)\right\}, \label{eqn:eta_sinusoidal}
\end{align}
for $0 \leq k \leq K$, where $K$ is the total number of leapfrog steps in a simulated trajectory.
See Figure~\ref{fig:temperature_schedules} for a graphical illustration of these schedules.
In both equations, $\eta_k$ represents the value of $\eta$ at $\bar t = k \bar \epsilon$, and $\eta_* >0$ is the maximum value of $\eta$.
We let $t_0 = 0$ and
\[
  t_k = t_{k-1} + e^{2a\eta_{k-\frac{1}{2}}}\bar \epsilon, \qquad k=1, \dots, K.
\]
Here we use half-integer index $\eta_{k-\frac{1}{2}}$ to ensure that the simulate trajectory possesses the time reversibility given in \eqref{eqn:TSTS}.
Moreover, the temperature schedule should satisfy $\eta_\kappa = \eta_{K-\kappa}$ for any integer and half-integer $\kappa \in \{0, \tfrac{1}{2}, 1, \dots, K-\tfrac{1}{2}, K\}$ and $\eta_0 = \eta_K = 0$.
Algorithm~\ref{alg:THMC_sim} summarizes the numerical simulation method for THMC.

We provide the proof of the following result in the appendix.
\begin{prop}\label{prop:THMC_reversibleMC}
  %The trajectory constructed by Algorithm~\ref{alg:THMC_sim} is symplectic. Moreover,
  If the temperature schedule is symmetric---that is, if $\eta_\kappa = \eta_{K-\kappa}$ for every $0\leq \kappa\leq K$---then the tempered Hamiltonian Monte Carlo algorithm (Algorithm~\ref{alg:THMC}) constructs a reversible Markov chain that leaves the target density $\pi/Z$ invariant.
\end{prop}

While half-integer values of $k$ are used to scale the leapfrog step size by $\eps = e^{2a\eta_{k-\frac{1}{2}}} \bar \epsilon$, integer values for $k$ are used for defining the Hamiltonian after each leapfrog step:
\[
  H_\alpha(x(t_k), p(t_k), t_k) = \frac{1}{2} p(t_k)^\top M^{-1} p(t_k) + e^{-2\eta_k} U(x(t_k)).
\]
Since \eqref{eqn:HEM_bar} implies that $\bar H_\alpha(\bar x, \bar p, \bar t)$ is approximately conserved, Equations~\ref{eqn:H_barH} suggest that the Hamiltonian $H_\alpha$ approximately scales as $e^{-2a\eta}$:
\begin{equation}
  \begin{split}
    H_\alpha(x(t_k), p(t_k), t_k) &= \bar H_\alpha(\bar x(k\bar\epsilon), \bar p(k\bar\epsilon), k\bar\epsilon)\\
    &= e^{-2a\eta_k} \left\{ \frac{1}{2} \bar p(k\bar\epsilon)^\top M^{-1} \bar p(k\bar\epsilon) + \bar U(\bar x(k\bar\epsilon))\right\}\\
    &\approx e^{-2a\eta_k} \left\{ \frac{1}{2} \bar p(0)^\top M^{-1} \bar p(0) + \bar U(\bar x(0)) \right\}\\
    &= e^{-2a\eta_k} H_\alpha(x(0), p(0), 0).
  \end{split}
  \label{eqn:Halpha_scaling}
\end{equation}
As a result, the original Hamiltonian $H(x,p)$ is approximately conserved after the simulation of a full temperature cycle, since $\eta_K = 0$:
\[
  H(x(t_K), p(t_K)) %&= \frac{1}{2} p(t_K)^\top M^{-1} p(t_K) + U(x(t_K)) 
  = H_\alpha(x(T_K), p(t_K), t_K)
  \approx e^{-2a\eta_K} H_\alpha(x(0), p(0), 0)
  = H(x(0), p(0)).
\]
This implies that the acceptance probability of the terminal state of a tempered trajectory can be close to 1, provided that $\bar \epsilon$ is sufficiently small to maintain numerical accuracy.

\begin{figure}[t]
\begin{knitrout}
\definecolor{shadecolor}{rgb}{0.969, 0.969, 0.969}\color{fgcolor}

{\centering \includegraphics[width=1\linewidth]{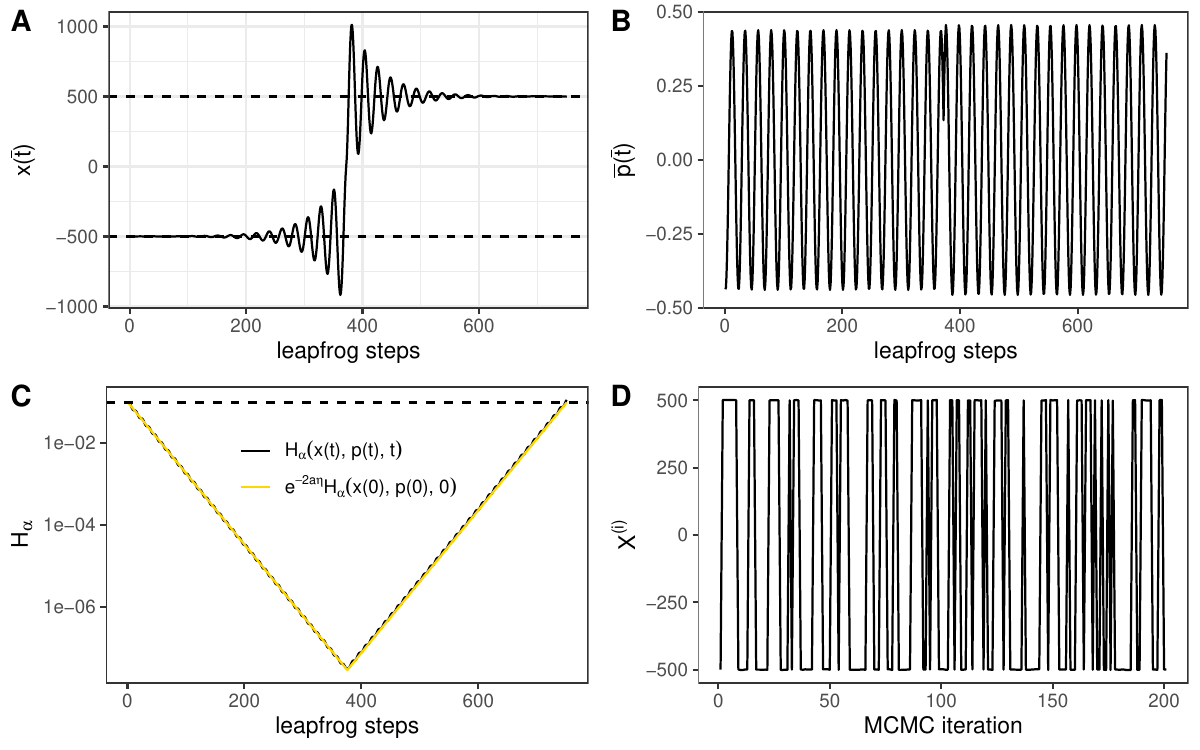} 

}

\end{knitrout}
\caption{A: Simulated trajectory for Example~\ref{ex:modulated_bimodal_1d} using Algorithm~\ref{alg:THMC_sim}.
  The centers of the two density components are indicated by horizontal dashed lines.
  B: Transformed momentum, $\bar p(\bar t)$, as a function of the number of leapfrog steps.
  C: Hamiltonian $H_\alpha(x(t), p(t), t)$ over the number of leapfrog steps.
  For comparison, the graph of $e^{-2a\eta} H_\alpha(x(0), p(0), 0)$ is shown in orange, with the initial Hamiltonian level indicated by a horizontal blue dashed line.
  D: Markov chain $X^{(i)}$ constructed over 200 MCMC iterations.
}
\label{fig:path_1d_bimodal}
\end{figure}

\subsection{Demonstrations of tempered HMC on toy examples}\label{sec:THMC_demo}
%% Example 1D bimodal

\setcounter{example}{0}
\refstepcounter{example}
\paragraph{Example~\theexample: Mixture of two one-dimensional Gaussian components}\label{ex:modulated_bimodal_1d}
We demonstrate tempered HMC using a mixture of two Gaussian components
\[
\frac{1}{2} \N(-500, 1^2) + \frac{1}{2} \N(500, 1^2).
\]
Panel~A of Figure~\ref{fig:path_1d_bimodal} shows $x(\bar t)$ for a simulated trajectory generated by Algorithm~\ref{alg:THMC_sim}.
A piecewise linear temperature schedule was used, given by $\eta_k = (2\eta_*/K)\cdot \min(k, K-k)$, with $\eta_* = 15$ and $K=750$.
The leapfrog step size varied as $\epsilon = e^{2a\eta} \cdot \bar\epsilon$, with $a = 2/(\gamma+2) = \tfrac{1}{2}$ and $\bar\epsilon = 0.2$.
The trajectory originates from a point near $-500$, and as the temperature increases, it oscillates with increasing amplitude.
During the second half, as the temperature decreases, the trajectory settles near a different mode at $500$.
Panel~B shows the transformed momentum $\bar p(\bar t)$, which exhibits stationary oscillatory behavior except during transitions between the two modes.
Panel~C displays the Hamiltonian $H_\alpha(x(t), p(t), t)$, which is approximately proportional to $e^{-2a\eta}$, as predicted by Equation~\ref{eqn:Halpha_scaling}.
The crossing between the two modes introduces only a minor perturbation in $H_\alpha$ near the midpoint of the trajectory; as a result, the final value is nearly equal to the initial Hamiltonian.
Panel~D shows a trace plot of a Markov chain constructed using Algorithm~\ref{alg:THMC}.
Out of 200 MCMC iterations, 68 transitions occurred between the two modes.

%% Example Nd bimodal
\refstepcounter{example}
\paragraph{Example~\theexample: Mixture of two log-polynomial densities in high dimension} \label{ex:modulated_bimodal_d10000}
We consider target densities given by
\begin{equation}
  \pi(x) \propto e^{-\Vert x-\mu_1\Vert^\gamma} + e^{-\Vert x - \mu_2\Vert^\gamma}, \quad x\in \mathbb R^{10000},
  \label{eqn:bimodal_d10000}
\end{equation}
where $\Vert \mu_1 - \mu_2 \Vert = 400$ and $\gamma$ is varied.

\begin{figure}
\begin{knitrout}
\definecolor{shadecolor}{rgb}{0.969, 0.969, 0.969}\color{fgcolor}

{\centering \includegraphics[width=\maxwidth]{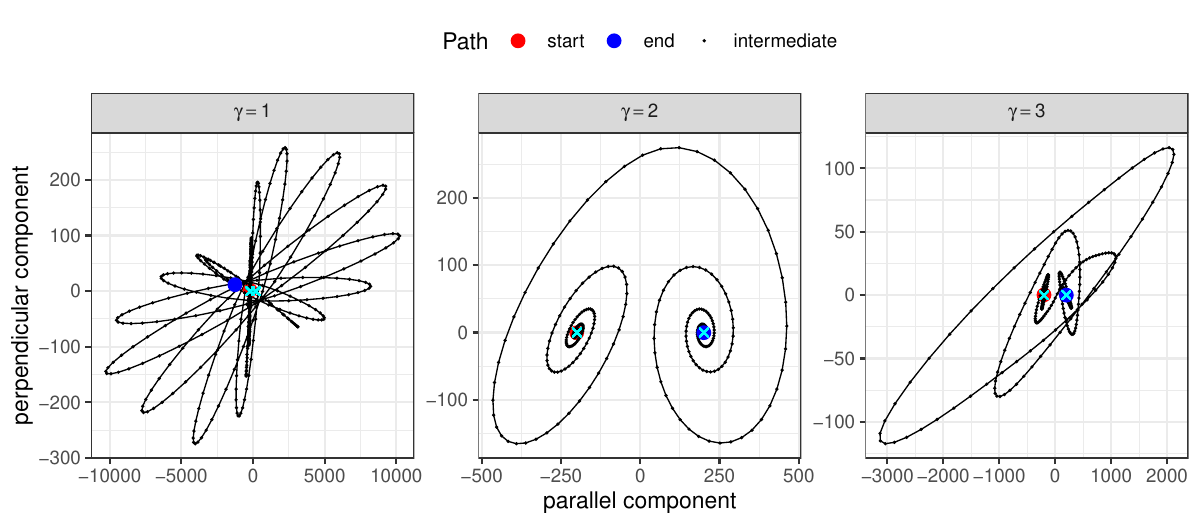} 

}

\end{knitrout}
\caption{
  Example trajectories simulated for $d=10{,}000$ dimensional target density \eqref{eqn:bimodal_d10000} for $\gamma=1$, 2, and 3.
  The $x$-axis shows the coordinate in the direction parallel to the vector $\mu_2-\mu_1$, and the $y$-axis displays the coordinate in a random direction perpendicular to $\mu_2-\mu_1$.
  The initial and the terminal points are marked in red and blue, respectively. 
  The centers of the two mixture components, $\mu_1$ and $\mu_2$, are indicated by cyan `$\times$'.
}
\label{fig:modulated_bimodal}
\end{figure}

Figure~\ref{fig:modulated_bimodal} shows example trajectories simulated for the target density \eqref{eqn:bimodal_d10000} for $\gamma=1$, 2, and 3 using piecewise linear $\{\eta_k\}$ described in \eqref{eqn:eta_piecewiselinear}.
Each trajectory was initialized from the first mode, centered at $\mu_1$.
As the total energy increases due to the rising mass, the particle moves away from the initial mode and searches for isolated modes in the 10{,}000 dimensional space, guided by the gradient of the potential energy.
During the second half of the trajectory, the temperature decreases, causing the particle to settle near a mode.
We note that for $\gamma=1$, the endpoint of the trajectory was not as close to a mode as in the cases $\gamma=2$ or 3, due to the slower growth rate of the potential energy function.
%, where $\eta_*$, $a$, $K$, and $\bar \epsilon$ were automatically tuned using Algorithm~\ref{alg:autotuning}.

\begin{figure}
\begin{knitrout}
\definecolor{shadecolor}{rgb}{0.969, 0.969, 0.969}\color{fgcolor}

{\centering \includegraphics[width=\maxwidth]{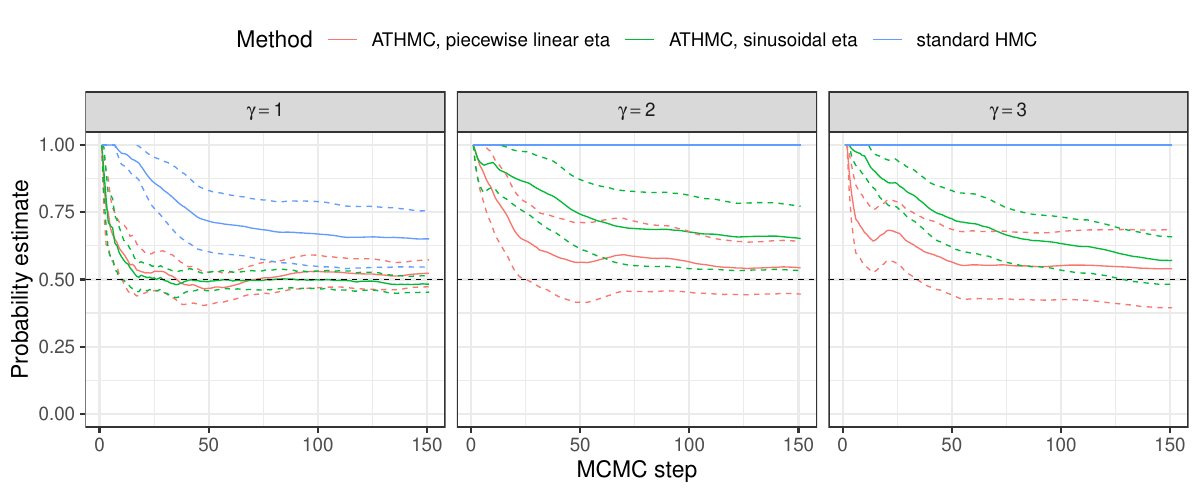} 

}

\end{knitrout}
\caption{Running estimates of the probability $P[\Vert X-\mu_1\Vert < \Vert X-\mu_2\Vert]$ for the target density \eqref{eqn:bimodal_d10000} where tempered HMC (Algorithm~\ref{alg:THMC}) is used to construct chains.
  ATHMC with a piecewise linear $\{\eta_k\}$ (Equation~\ref{eqn:eta_piecewiselinear}) and a sinusoidal $\{\eta_k\}$ (Equation~\ref{eqn:eta_sinusoidal}) are compared with standard HMC, in which $\eta_k = 0$ for all $k$.
  The solid curves show the averages over 20 independently constructed chains, and the dashed curves mark $\pm 2$ standard errors.
}
\label{fig:closest_mode}
\end{figure}

%$$
Figure~\ref{fig:closest_mode} shows the running estimates of the probability $P[\Vert X-\mu_1\Vert < \Vert X-\mu_2\Vert]$, whose exact value is $\frac{1}{2}$ due to the symmetry of the two modes.
The plots in Figure~\ref{fig:closest_mode} show the average estimates over 20 independently constructed chains, along with error bands corresponding to two standard errors.
Rapid convergence to the ergodic mean $\frac{1}{2}$ indicates that the Markov chains frequently transition between the two isolated modes.
For $\gamma=1$, standard HMC exhibits slower convergence compared to tempered HMC.
For $\gamma=2$ and 3, the target distribution is strongly multimodal, since $\Delta - \frac{d}{2} \approx \Vert\frac{\mu_1-\mu_2}{2}\Vert^\gamma - \frac{d}{2}$ in Equation~\ref{eqn:jumpprob} is on the order of $10^4$ or $10^6$.
In these cases, standard HMC produced no transitions between the modes across all replications.
In contrast, tempered HMC enabled reasonably frequent transitions.
The numerical results were comparable between the piecewise linear and the sinusoidal $\{\eta_k\}$ sequences, while the former led to slightly faster convergence.
%These results suggest that the smoothness of $\eta$ does not have significant impact on the numerical accuracy of simulations.

\refstepcounter{example}
\paragraph{Example~\theexample: Mixture of many anisotropic Gaussian densities}\label{ex:mixtureGaussian_many}
\begin{figure}[t]
\begin{knitrout}
\definecolor{shadecolor}{rgb}{0.969, 0.969, 0.969}\color{fgcolor}

{\centering \includegraphics[width=0.6\linewidth]{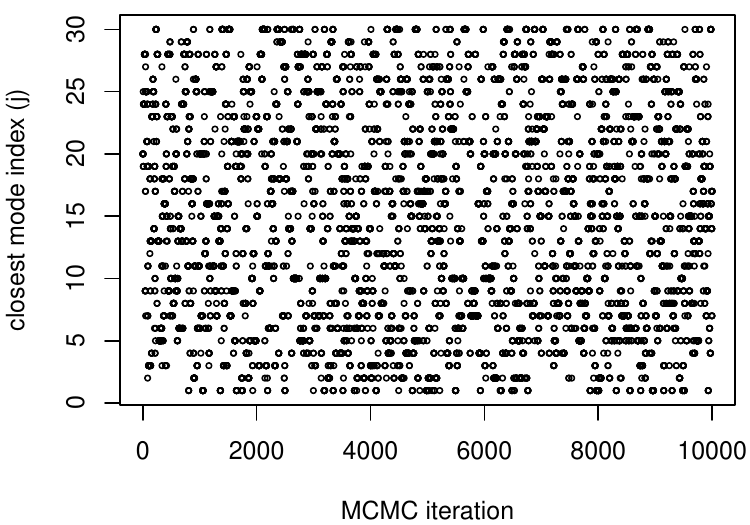} 

}

\end{knitrout}
\caption{Trace plot of the closest modes for a Markov chain constructed using tempered HMC for Example~\ref{ex:mixtureGaussian_many}.}
\label{fig:mixtureGaussian_many}
\end{figure}

We consider a mixture of $J=30$ Gaussian components,
\[
  \frac{1}{J} \sum_{j=1}^J \N(\mu_j, \Sigma_j),
\]
in a $d=50$-dimensional space.
The means $\mu_j$ are selected in three different ways:
\begin{enumerate}[nosep, label=(\alph*)]
\item For $j \in 1\col 10$, each $\mu_j$ is $s=1000$ times the coordinate unit vector $e_j$.
\item For $j \in 11\col 20$, each $\mu_j$ has zero entries except along ten randomly selected coordinate axes; the projection of $\mu_j$ onto those ten axes is drawn from the ten-dimensional multivariate normal distribution with mean zero and variance $(s/\sqrt{10})^2 I_{10}$.
\item For $j \in 21\col 30$, each $\mu_j$ is drawn from the multivariate normal distribution $\N(0, (s/\sqrt{d})^2 I_d)$.
\end{enumerate}
The norms $\Vert \mu_j \Vert$ for $j\in 1\col 30$, as well as the pairwise distances $\Vert \mu_j - \mu_{j'} \Vert$ for $j \neq j'$, are all on the order of $s=1000$.
The covariance matrices, $\Sigma_j$, for $j\in 1\col J$, are anisotropic and drawn from the inverse Wishart distribution $\text{InvWishart}(I_d/(2d), 2d)$.
The condition numbers---defined as the ratio of the largest to the smallest singular value---of the $\Sigma_j$ have a mean of 27.6 and a standard deviation of 4.0.

A Markov chain of length $10{,}000$ targeting the mixture distribution was constructed using tempered Hamiltonian Monte Carlo (Algorithm~\ref{alg:THMC}).
We used a piecewise linear log-temperature schedule $\eta_k$ with a maximum value of $\eta_* = 13$ and a rate of change $|d\eta/d\bar t| = 2\eta_*/(K\bar\epsilon) = 0.075$, resulting in a trajectory length of $K=1733$.
The reference leapfrog step size $\bar\epsilon$ was set to 0.2.
Figure~\ref{fig:mixtureGaussian_many} shows the index $j$ of the mode closest to each state of the Markov chain.
This plot indicates that transitions between modes occur frequently: out of 10{,}000 MCMC iterations, 1591 resulted in jumps between distinct modes.
In contrast, when we ran standard HMC with the same leapfrog step size $\epsilon=0.2$ and a similar number of leapfrog steps per trajectory $K=1700$, no mode transitions were observed over 10{,}000 iterations (results not shown).

% 
%%%% TUNING %%%%
\section{Tuning tempered Hamiltonian Monte Carlo}\label{sec:tuning}
We develop guidelines for tuning tempered Hamiltonian Monte Carlo (Algorithms~\ref{alg:THMC} and \ref{alg:THMC_sim}).

\newcommand\opt{\text{opt}}
\textbf{Tuning $\bar \epsilon$.}
Tempered HMC simulates trajectories using leapfrog steps sizes $\epsilon = e^{2a\eta} \bar\epsilon$ as summarized in Algorithm~\ref{alg:THMC_sim}.
The choice of the reference step size $\bar \epsilon$ involves a trade-off between computational speed and numerical accuracy, as in standard HMC.
Increasing $\bar\epsilon$ reduces the number of leapfrog steps needed to construct a trajectory of fixed length but tends to result in a larger net increase in the Hamiltonian, thereby lowering the acceptance probability.

We propose an adaptive approach to tuning $\bar \epsilon$.
Starting from an arbitrary initial point, we run standard HMC while adjusting the step size to achieve an average acceptance rate close to a target value $p^*_{\text{pilot},\text{acc}}$.
The tuning process is continued until both $U(x)$ and the leapfrog step size stabilize; the stabilization of $U(x)$ suggests that the Markov chain has reached a local mode.
Tuning the step size by targeting the average acceptance rate is a standard practice, supported both empirically and theoretically in the literature \citep{creutz1988global, neal2011mcmc, beskos2013optimal}.
For example, the step size can be tuned via the update rule
\[
  \log \bar\epsilon^{(i+1)} \gets \log \bar \epsilon^{(i)} + \frac{1}{(i+1)^{0.6}}(p^{(i)}_{\text{pilot}, \text{acc}} - p^*_{\text{pilot},\text{acc}}),
\]
where $p^{(i)}_{\text{pilot},\text{acc}}$ denotes the probability of accepting the proposed candidate at the $i$-th iteration of the pilot run.
Across the examples we considered, target acceptance rates $p^*_{\text{pilot},\text{acc}}$ in the range $[0.5, 0.9]$ yielded reference step sizes $\bar\epsilon$ that led to satisfactory performance of tempered HMC.
The resulting step size is then used as the reference step size in tempered HMC.

% $$
This approach is justified because the transformed variables $(\bar x(\bar t), \bar p(\bar t))$ approximately satisfy the same equations of motion as the original variables $(x(t), p(t))$, as shown in Equation~\ref{eqn:HEM_bar}.
Therefore, the leapfrog step size that produces small net change in Hamiltonian under standard HMC is also expected to yield a small net change under tempered HMC.
Tuning based on local behavior near a mode is appropriate because the primary source of numerical error---and hence the main contributor to changes in the Hamiltonian---occurs during the temperature ramp-up and ramp-down phases, rather than during transitions between modes.
Our empirical results support these tuning guidelines.
However, to account for potential scale differences across distinct modes, $\bar\epsilon$ may be reduced from the tuned value by a fixed multiplicative factor, since a newly discovered mode with a smaller scale could cause a previously acceptable step size to produce unstable trajectories.

\begin{figure}
\begin{knitrout}
\definecolor{shadecolor}{rgb}{0.969, 0.969, 0.969}\color{fgcolor}

{\centering \includegraphics[width=\maxwidth]{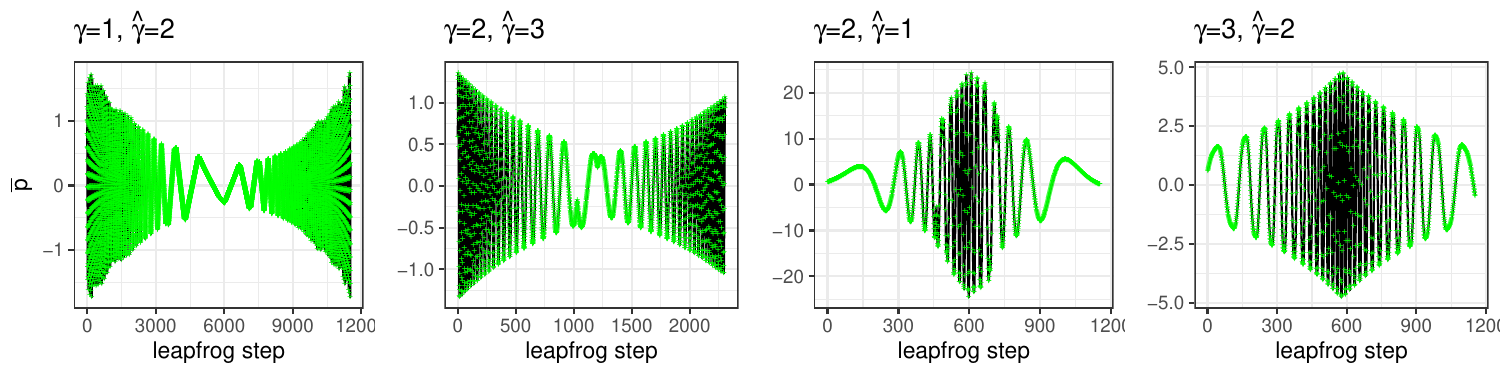} 

}

\end{knitrout}
\caption{
  Transformed momenta $\bar p = \exp\{\frac{2}{\hat\gamma+2} \eta\} \cdot p$ for simulated trajectories where the polynomial degree $\gamma$ of $U(x)$ was incorrectly estimated (i.e., $\hat\gamma \neq \gamma$). Green dots indicate the values of $\bar p$ after each leapfrog step. Piecewise linear sequences for $\eta_k$ were used, with $\eta_* = 12$.
}
\label{fig:pbar_gamma_misspecified}
\end{figure}

\textbf{Tuning $a = \frac{2}{\gamma+2}$.}
The scaling of the leapfrog step size $\epsilon = e^{2a\eta} \bar\epsilon$ depends on the parameter $a$, whose optimal value is given by $\frac{2}{\gamma+2}$.
In some situations, the degree $\gamma$ of the local polynomial growth of $U(x)$ can be identified by inspecting the closed-form expression of the posterior density.
However, when this is not possible, $\gamma$ can be estimated by using the fact that the transformed momentum $\bar p = e^{a\eta} p = e^{\frac{2}{\gamma+2}\eta} p$ exhibits stationary oscillatory behavior as a function of the rescaled time $\bar t$.
If the estimate $\hat\gamma$ exceeds the true $\gamma$---so that the corresponding $\hat a = \tfrac{2}{\hat\gamma+2}$ underestimates $a$---then the amplitude of oscillations in $\bar p$ decreases as $\eta$ increases.
Furthermore, in this case, the oscillation frequency with respect to the number of leapfrog steps also decreases with increasing $\eta$.
This behavior arises because the leapfrog step size scales as $\epsilon = \exp(2\hat a \eta) \cdot \bar \epsilon$, while the optimal time scale transformation corresponds to $dt = \exp(2a\eta) \cdot d\bar t$.
Figure~\ref{fig:pbar_gamma_misspecified} illustrates these effects using a mixture of log-polynomial densities, $\pi(x) \propto e^{-\Vert x-\mu_1\Vert^\gamma} + e^{-\Vert x - \mu_2\Vert^\gamma}$. 
The first two plots exhibit decreasing oscillation amplitude and frequency of $\bar p$ as $\eta$ increases when $\gamma$ is overestimated.
In contrast, the next two plots show that both the amplitude and frequency of $\bar p$ increase with $\eta$ when $\gamma$ is underestimated.

Using this fact, the value of $\hat\gamma$ can be tuned adaptively as follows.
Given an initial state $x(0) = X^{(i)}$ in the neighborhood of a local basin of $U(x)$, we simulate a trajectory with step size scaling as $\epsilon = \exp(2\hat a\eta)\cdot \bar \epsilon$ using an initial guess $\hat a$.
If $\bar p = \exp(\hat a\eta) \cdot p$ exhibits decreasing amplitude and frequency with increasing $\eta$, the value of $a$ should be increased, and in the opposite case, it should be decreased.
%The amount of change in $a$ can be determined as follows.
Specifically, as the oscillation amplitude of $\bar p = e^{(\hat a-a)\eta} \cdot e^{a\eta}p $ scales as $e^{(\hat a-a)\eta}$, we can update our estimate of $a$ as follows.
Let
\begin{equation}
  r_j := \frac{\max_{\frac{3K}{8} \leq k < \frac{K}{2}} |\bar p_j(t_k)|} {\max_{0\leq k < \frac{K}{8}} |\bar p_j(t_k)|}, \quad j\in 1\col d, \label{eqn:amp_ratio}\\
\end{equation}
where $|\bar p_j(t_k)|$ denotes the absolute value of the $j$-th coordinate value of the scaled momentum $\bar p(t_k)$ at the end of the $k$-th leapfrog step.
We then have approximately
\[
  r_j \approx e^{(\hat a-a)\Delta \eta},
\]
where $\Delta \eta = \eta_{[\frac{7K}{16}]} - \eta_{[\frac{K}{16}]}$ is the increase in $\eta_k$ between the two time intervals.
The optimal value of $a$ can thus be approximated by 
\[
  a \approx \hat a - \text{median} \{\log r_j; j\in 1\col d\} / \Delta \eta.
\]
We use two well-separated intervals, $[0,\tfrac{K}{8})$ and $[\tfrac{3K}{8}, \tfrac{K}{2})$, to reduce variability in the estimate of $a$.
Numerical results suggest, however, that the stability of the tuning cycles can be improved by decreasing the size of innovation, for instance by setting
\begin{equation}
  \hat a \gets a - 0.3 \cdot \text{median} \{\log r_j; j\in 1\col d\} / \Delta \eta.
  \label{eqn:a_opt}
\end{equation}

\begin{figure}
\begin{knitrout}
\definecolor{shadecolor}{rgb}{0.969, 0.969, 0.969}\color{fgcolor}

{\centering \includegraphics[width=0.7\linewidth]{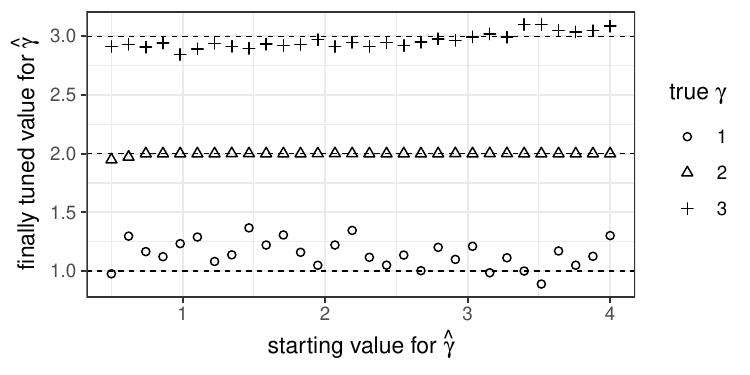} 

}

\end{knitrout}
\caption{Automatically tuned $\hat\gamma$ values using the automatic tuning algorithm (Algorithm~\ref{alg:tuning_THMC}) for varied starting values.
  The bimodal target density was given by \eqref{eqn:bimodal_d10000} with $d=10{,}000$, and $\gamma$ was varied over $\{1,2,3\}$.
}
\label{fig:estimated_powers}
\end{figure}

Figure~\ref{fig:estimated_powers} shows the tuned values of $\hat\gamma = (2/\hat a)-2$ for the target density given by \eqref{eqn:bimodal_d10000} with $\gamma \in \{1,2,3\}$.
Tuning started with $\hat\gamma$ in the range $[0.5, 4]$ and stopped when $|\text{median}\{\log r_j\}|$ was less than 0.2.
The plot indicates that our method can estimate $\gamma$ with reasonable accuracy.
%Over iterated MCMC steps, the tuned values of $\hat\gamma$ generally approach even closer the true value $\gamma$ (see Figure~\ref{fig:tuned_values}).

\textbf{Tuning $\eta_*$.}
The maximum value $\eta_*$ of the sequence $\{\eta_k; 0\leq k \leq K\}$ determines how far the simulated Hamiltonian trajectories can reach.
To escape a mode with a log-polynomial potential function $U(x) = \Vert x \Vert^\gamma$, we require
\[
  \eta_* \gtrsim O\{\log(\text{the depth of the mode of }U(x))\},
\]
since the rescaled potential $\bar U(\bar x) = \Vert \bar x \Vert^\gamma = e^{-\gamma a \eta} U(x)$ remains approximately constant in magnitude.
Using a large $\eta_*$ increases the likelihood of discovering isolated modes that are far away.
However, larger values of $\eta_*$ also tend to cause a greater net increase in the Hamiltonian, thereby lowering the acceptance probability of proposals and reducing computational efficiency.

A reasonable value for $\eta_*$ can be found by adaptively tuning it such that the simulated trajectory meets a predefined search scope criterion in a specified proportion of iterations (e.g., $\tfrac{2}{3}$).
For instance, given a suitably chosen reference point $x^0 \in \mathbb R^d$ and coordinate-wise desired search scales $\{s_j; j\in 1\col d\}$, a rectangular search scope can be characterized by:
\begin{equation}
  \max_{k\in 0\col K} (x_j(t_k) - x_j^0)_+ \geq s_j \text{ and } 
  \max_{k\in 0\col K} (x_j(t_k) - x_j^0)_- \geq s_j
  \text{ for at least $\frac{d}{2}$ components $j$},
  \label{eqn:search_criterion}
\end{equation}
where $x_j(t_k)$ denotes the $j$-th coordinate of the position $x(t_k)$ at the end of the $k$-th leapfrog step.
The value of $\eta_*$ determined through tuning tends to increase as $s_j$ increases---choosing $s_j$ too small introduces a risk of missing a mode, while setting it too large $s_j$ can reduce computational efficiency.
Another way to define a search scope is to require that the simulated trajectory reaches a point where the potential energy $U(x)$ exceeds a prescribed threshold.
%In order to facilitate the tuning of other parameters ($a$, $\bar \epsilon$, and $K$), we suggest to let the starting value of $\eta_*$ be somewhat lower than the anticipated optimum for $\eta_*$, so that numerically stable paths can be constructed.

\begin{figure}[tp]
\begin{knitrout}
\definecolor{shadecolor}{rgb}{0.969, 0.969, 0.969}\color{fgcolor}

{\centering \includegraphics[width=\maxwidth]{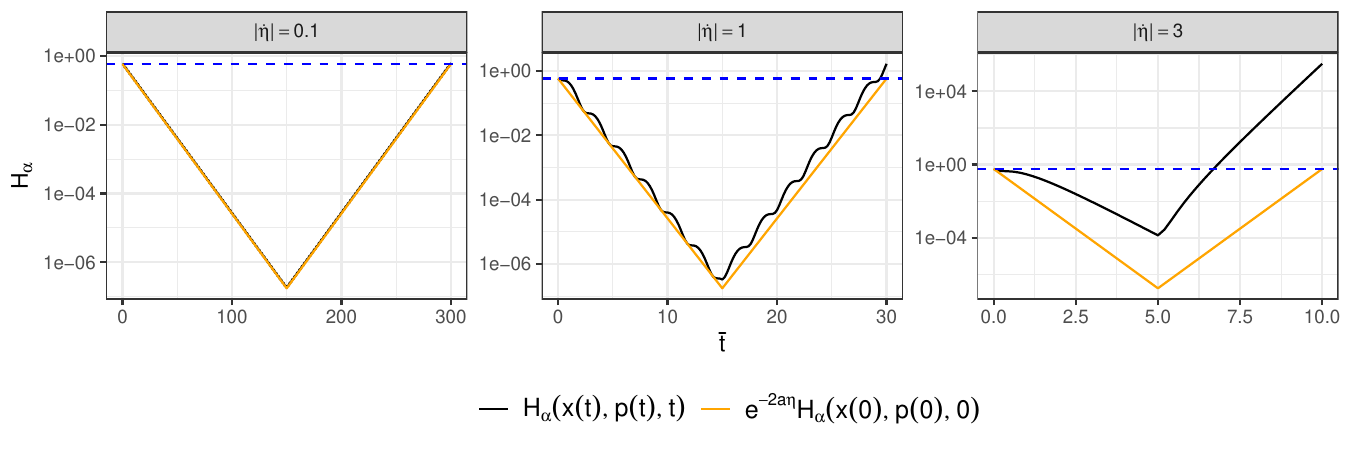} 

}

\end{knitrout}
\caption{The value of the Hamiltonian $H_\alpha(x(t), p(t), t)$ along tempered trajectories constructed using Algorithm~\ref{alg:THMC_sim} with varying tempering rates $|\dot\eta|$. The $x$-axis shows the rescaled time $\bar t$. The initial value of the Hamiltonian, multiplied by $e^{-2a\eta}$, is shown in orange.}
\label{fig:etaRate}
\end{figure}

\textbf{Tuning the length of the temperature schedule ($K$).}
For a piecewise linear schedule defined by $\eta_k = (2\eta_*/K) \min(k,K-k)$, the absolute rate of change in $\eta$ with respect to rescaled time $\bar t$ can be expressed as
\[
  \left| \dot \eta \right| := \left| \frac{d\eta}{d\bar t} \right| = \frac{2\eta_*}{K\bar\epsilon},
\]
since each leapfrog step advances time $\bar\epsilon$ in $\bar t$.
We select the length of the temperature schedule, $K$, by tuning the tempering rate $|\dot\eta|$.
Figure~\ref{fig:etaRate} shows that the net increase in Hamiltonian generally grows with increasing $|\dot\eta|$.
Thus if $|\dot\eta|$ is too large, the acceptance probability can become extremely low.
Conversely, if $|\dot\eta|$ is too small, reaching the same $\eta_*$ with fixed $\bar\epsilon$ requires a large $K$, increasing the computational cost per MCMC iteration.

We propose to adaptively tune $|\dot\eta|$ so that the acceptance rate of tempered HMC approaches a target value, denoted $p^*_{\text{acc}}$.
The recursive update can be performed using, for example,
\[
  \log|\dot\eta|^{(i+1)} \gets \log|\dot\eta|^{(i)} + \frac{1}{(i+1)^{0.6}} (p_{\text{acc}}^{(i)} - p_{\text{acc}}^*),
\]
where $p_{\text{acc}}^{(i)} = \min(1, \exp\{-H(x(t_K), p(t_K)) + H(x(0), p(0))\})$ is the acceptance probability at the $i$-th iteration.
%We found that, when tuning $\eta_*$ and $|\dot\eta|$ simultaneously, updating $|\dot\eta|$ only when the specified search criterion was met led to a more stable tuning process. (Not true.)
The optimal target acceptance rate $p^*_{\text{acc}}$ may depend on the global geometry of the potential energy function $U(x)$.
In general, increasing $p^*_{\text{acc}}$ can improve the rate of transitions per MCMC iteration, but this comes at the cost of requiring more leapfrog steps per trajectory. 
For the examples considered in this paper, $p^*_{\text{acc}}$ values in the range $[0.05, 0.2]$ achieved an approximately optimal trade-off.

%$$
\begin{figure}[t]
  \centering
  \begin{algorithm}[H]
    %\SetAlgoRefName{S1}
    \SetKwInOut{Input}{Input}\SetKwInOut{Output}{Output}
    \Input{
      Target acceptance ratio $p_{\text{pilot},\text{acc}}^*$ for pilot HMC (for tuning $\bar \epsilon$);~
      Search criterion (for tuning $\eta_*$) and an initial guess $\eta_*^{(1)}$;~
      Target acceptance ratio $p_{\text{acc}}^*$ for tempered HMC (for tuning $|\dot\eta|$ and $K$) and an initial guess $|\dot\eta|^{(1)}$;~
      The local polynomial degree $\gamma$ of $U(x)$ (if $\gamma$ is known), or an initial guess $\hat\gamma^{(1)}$ (otherwise);~
    }
    \vspace{1ex}

    Pilot run standard HMC to locate a mode of the target distribution and tune the reference leapfrog step size $\bar\epsilon$ using a target acceptance rate $p_{\text{pilot},\text{acc}}^*$\\
    (Optional) Reduce the tuned step size: $\bar\epsilon \gets \tau \bar\epsilon$ where $0< \tau < 1$\\
    Let $X^{(1)}$ be the current state of the Markov chain from the pilot HMC run\\ 
    \For {$i \geq 1$} {
      Let $x(0) = X^{(i)}$ and draw $p(0) \sim \N(0, M)$\\
      Simulate a tempered Hamiltonian trajectory using Algorithm~\ref{alg:THMC_sim} with $\eta_* = \eta_*^{(i)}$, $K = K^{(i)} = \lceil 2\eta_*^{(i)} / (|\dot\eta|^{(i)}\bar\epsilon) \rceil$, and $a = \frac{2}{\gamma+2}$ (or $\hat a^{(i)} = \frac{2}{\hat\gamma^{(i)}+2}$ if $\gamma$ is unknown)\\
      Compute $p_{\text{acc}}^{(i)}$, the acceptance probability for the final state $(x(t_{K^{(i)}}), p(t_{K^{(i)}}))$\\
      \textbf{if} $\gamma$ is unknown \textbf{then} update $\hat a$ using Equation~\ref{eqn:a_opt} and let $\hat\gamma^{(i+1)} \gets (2/\hat a^{(i+1)})-2$\\
      Let $\log |\dot\eta|^{(i+1)} \gets \log |\dot\eta|^{(i)} + \frac{1}{(i+1)^{0.6}} (p_{\text{acc}}^{(i)} - p_{\text{acc}}^*)$\\
      Let $\eta_*^{(i+1)} \gets \eta_*^{(i)} + \frac{1}{(i+1)^{0.6}} \left(2 - 3 \cdot \mathbf 1[\text{the trajectory meets the search criterion}]\right)$\\
      Let $X^{(i+1)} \gets X^{(i)}$ with probability $p_{\text{acc}}^{(i)}$
    }
    \caption{Automatically tuned Tempered Hamiltonian Monte Carlo (ATHMC)}
    \label{alg:tuning_THMC}
  \end{algorithm}
\end{figure}

\textbf{Automatically tuned Tempered Hamiltonian Monte Carlo (ATHMC).}
Algorithm~\ref{alg:tuning_THMC} gives a summary of the tuning procedure for tempered HMC. 
By incorporating the adaptive tuning strategies into tempered HMC, we obtain an automatically tuned tempered Hamiltonian Monte Carlo (ATHMC) algorithm.
Once all parameters have been tuned, the plot of the rescaled momentum $\bar p$ should exhibit approximately steady oscillations, with each oscillation cycle comprising a sufficient number of leapfrog steps (typically $\gtrsim 10$).
Although the adaptive tuning procedure breaks the Markovian property of the resulting chain, the ergodicity of adaptive MCMC algorithms is guaranteed under the \emph{simultaneous uniform ergodicity} and \emph{diminishing adaptation} conditions \citep{andrieu2008tutorial, roberts2007coupling}.
As an alternative to the adaptive MCMC, one can freeze the tuned parameters once reasonable values have been found.
The code used for the numerical experiments in this paper is available at: \url{https://github.com/joonhap/athmc}.

\section{Comparison with parallel tempering (PT) and tempered sequential Monte Carlo (TSMC)}\label{sec:comparison}
In this section, we numerically compare automatically tuned, tempered HMC (Algorithm~\ref{alg:THMC}) with parallel tempering (PT) and tempered sequential Monte Carlo (TSMC).
Specifically, we evaluate the variances of the Monte Carlo estimates produced by each method under equal overall computational cost, measured by the total number of leapfrog steps performed. 

We consider a strongly multimodal target distribution given by a mixture of two Gaussian components,
\[
  w \N\left(\frac{\mu}{2}, \Sigma_1\right) + (1-w) \N\left(\frac{\mu}{2}, \Sigma_2\right),
\]
where $\Vert \mu \Vert = 10{,}000$.
As demonstrated in Example~\ref{ex:mixtureGaussian_many} (Section~\label{sec:THMC_demo}), transitions between the modes of a bimodal distribution in Markov chains generated by ATHMC imply the possibility of inter-mode transitions in more complex, multimodal settings.
This is because the feasibility of such transitions depends primarily on whether a tempered trajectory can be constructed that starts in one mode and reaches another, rather than on the number of modes present.

For parallel tempering, we employ $K$ chains each targeting a tempered density proportional to $\pi^{1/\alpha_k}$, $k\in 1\col K$ where $1 = \alpha_1 < \alpha_2 < \cdots < \alpha_{K}$.
Adopting the strategy proposed by \citet{syed2022nonreversible}, swaps between adjacent pairs of parallel chains were carried out in an alternating manner, such that in even numbered iterations the pairs $(k, k+1)$ were swapped for even $k$'s, and in odd numbered iterations the pairs $(k,k+1)$ were swapped for odd $k$'s.
The temperature levels were adaptively tuned using the method proposed by \citet{miasojedow2013adaptive}.
Specifically, let $p_k^{(i)}$ the acceptance probability for the swap between the $k$-th and the $k+1$-st chains at the $i$-th MCMC iteration.
We set
\[
  \alpha_1^{(i+1)} = 1, \quad \alpha_{k+1}^{(i+1)} = \alpha_k^{(i+1)} + e^{\rho_k^{(i+1)}}, ~~k\in 1\col K-1,
\]
where
\[
  \rho_k^{(i+1)} = \rho_k^{(i)} + \frac{1}{(i+1)^{0.6}}(p_k^{(i)} - 0.234).
\]
The target acceptance probability of 0.234 follows the recommendation of \citet{kone2005selection} and \citet{atchade2011towards}.
In addition, we adaptively increased the number of parallel chains so that the highest-temperature chain satisfied a specified search criterion.
Specifically, every 50 MCMC iterations, a new chain was added above the current highest temperature level unless the highest chain had visited both the intervals $(-\infty, -10000)$ and $(10000, \infty)$ in at least half of the $d$ coordinates over the past 100 iterations.
The addition of chains stopped after the search criterion had been satisfied in 100 iterations overall.
The leapfrog step size for each chain was adaptively tuned to target an acceptance rate of 0.9, using a diminishing adaptation rate proportional to $(i+1)^{-0.6}$.
Each HMC kernel used 50 leapfrog steps per iteration.

%$$
For tempered HMC, we adaptively tuned both the maximum log-temperature level $\eta_*$ and the rate of change $|\dot\eta|$, as described in Algorithm~\ref{alg:tuning_THMC}.
The reference step size was tuned in a pilot run of standard HMC by targeting an acceptance rate of 0.9.
The resulting step size was then scaled by a factor of 0.5 to set the reference step size $\bar\epsilon$.
We tuned $\eta_*$ using a rectangular search criterion, aiming for the constructed trajectories to visit both $(-\infty, -10000)$ and $(10000, \infty)$ in at least $d/2$ coordinates in approximately two thirds of the iterations.
To tune $|\dot\eta|$, we used a target acceptance rate of 0.2.

Tempered sequential Monte Carlo (TSMC) is a recursive algorithm that evolves an ensemble of particles toward the target distribution through alternating importance sampling and MCMC-based particle diversification steps \citep{neal2001annealed}.
TSMC applies the general SMC sampler framework developed by \citet{delmoral2006sequential} to sample from multimodal distributions by introducing intermediate distributions that form a bridge between a base distribution and the target distribution.
In our implementation, the intermediate tempered distributions are defined as
\[
  \pi_k(x) \propto g_0(x)^{1-\beta_k} \pi(x)^{\beta_k},
\]
where $g_0(x)$ is the density of a base distribution $\N(0,(20000^2/d) I)$ and $\beta_k$ is the $k$-th inverse temperature.
The base distribution is broad in scale and designed to encompass potential modes of the target distribution.
We adopted the strategy proposed by \citet{buchholz2021adaptive} to recursively tune the inverse temperatures $\beta_k$, starting from $\beta_1=0$, so that the effective sample size at each importance sampling step is approximately half the ensemble size.
To replenish particle diversity, we applied an HMC kernel five times in succession after each resampling step.
Each HMC kernel used ten leapfrog steps, with the step size tuned during a pilot run to achieve an acceptance rate of approximately 0.9.
Unlike \citet{buchholz2021adaptive}, we did not tune the number of HMC kernel applications based on the empirical correlation coefficient, as we found this measure unreliable at high inverse temperature levels where the target distribution is strongly multimodal.

A fundamental challenge in sampling from high dimensional, multimodal target distributions arises when isolated modes have significantly different scales.
This issue limits the global mixing of tempering based methods, such as parallel tempering and simulated tempering \citep{woodard2009sufficient, woodard2009conditions, bhatnagar2016simulated}.
The difficulty stems from the fact that, in high dimensions, a mode with a relatively small scale occupies an extremely small volume.
Consequently, its probability under the tempered distribution $\pi^\beta(x)$ becomes vanishingly small for low inverse temperatures ($\beta \ll 1$).
As a result, such modes are rarely, if ever, visited at high temperature levels, and remain unvisited as the temperture cools.

This limitation also affects tempered Hamiltonian Monte Carlo (THMC), as trajectories rarely visit modes of small scale when the temperature is high.
Consider two modes with comparable probability mass, but one with a much smaller scale than the other.
A trajectory starting in the narrower mode has a low probability of being accepted if it ends in the broader mode, because the target density at the endpoint is much lower. 
Conversely, a trajectory starting in the broader mode has low probability of ending in the narrower mode due to the small volume that must be hit precisely.
\citet[Section~5]{neal1996sampling} discusses this issue in more detail.
\citet{tawn2020weight} propose a method for stabilizing the mixture weights of tempered distributions by incorporating the Hessian of the log target density.
However, this approach requires prior knowledge of the locations and the local geometry of the modes.

For parallel tempering and tempered HMC, the variance of Monte Carlo estimates hinges on the rate of global mixing, which is determined by how frequently the sampler transitions between modes---assuming local mixing within each mode is fast.
For tempered sequential Monte Carlo (TSMC), \citet{paulin2019error} established bounds on the asymptotic variance in the central limit theorem for SMC estimates \citep{beskos2014stability}.
Notably, in the supplementary material \citep{paulin2017supplement}, they derived a bound that applies even when the MCMC kernel used for particle diversification exhibits no mixing between the modes.
This bound, however, involves a \emph{growth-within-mode} constant that depends on the maximum ratio of the probabilities of local modes across different temperatures.
As a result, when the target distribution has modes with significantly different scales in high dimensions, TSMC suffers from the same challenge of high Monte Carlo variability as PT and THMC.
To the best of our knowledge, theoretical results on the finite sample bias and variance of TSMC remain unavailable.

\begin{figure}[tp]
\begin{knitrout}
\definecolor{shadecolor}{rgb}{0.969, 0.969, 0.969}\color{fgcolor}

{\centering \includegraphics[width=\maxwidth]{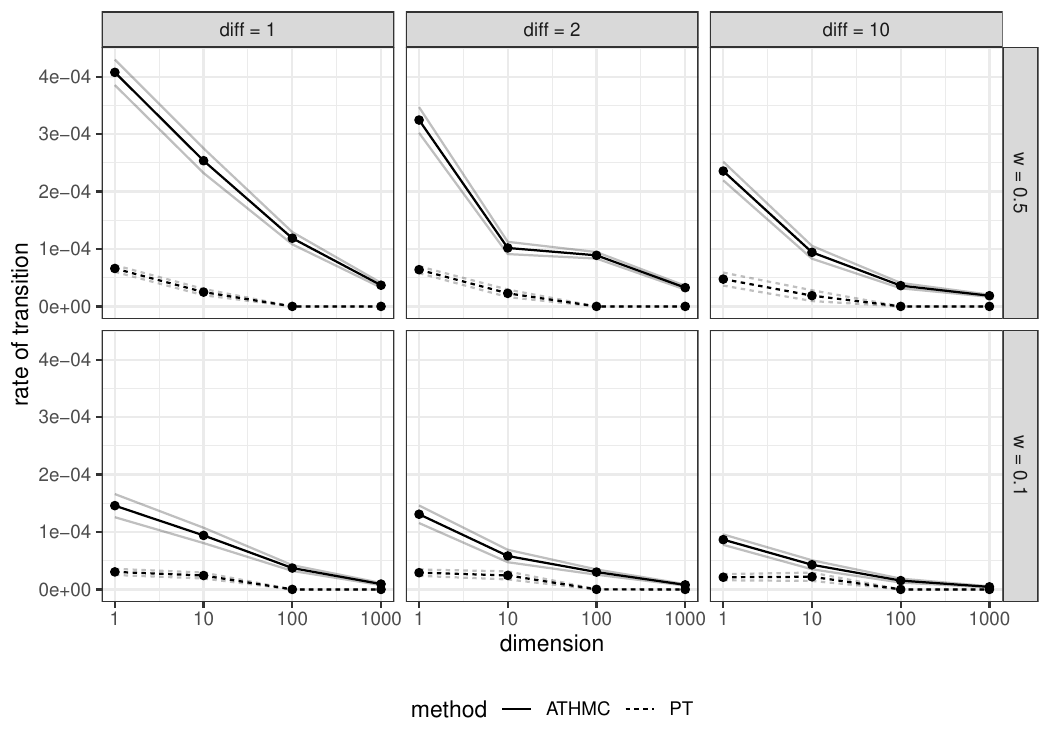} 

}

\end{knitrout}
\caption{Number of inter-mode transitions divided by the total number of leapfrog steps, for the bimodal target density~\eqref{eqn:bimodal_d10000}.
  ATHMC (solid line) and parallel tempering with adaptive tuning (dashed line) are compared under various settings for the dimension $d$, mixture weight for the first component $w$, and the scale difference between the two covariances ($c$, labelled as \emph{diff}).
  The average transition rate over 20 replications is shown, with $\pm 1$ standard deviation indicated by upper and lower bounding lines in light gray.
}
\label{fig:athmc_pt_transition_rate_iso}
\end{figure}

For numerical comparison, we first considered isotropic covariance matrices, where $\Sigma_1 = I$ and $\Sigma_2 = c^{-1/d}I$, where the scale difference factor $c$ varied across values $1$, $2$, and $10$.
We varied the dimension $d$ over $1$, $10$, $100$, and $1000$, and the mixture weight for the first component over $w=0.5$ and $0.1$.
Figure~\ref{fig:athmc_pt_transition_rate_iso} shows the inter-mode transition rate, defined as the number of transitions divided by the overall computational cost, measured by the total number of leapfrog steps.
The average transition rates over 20 replicated experiments are shown, each of which ran for 5000 MCMC iterations.
The transition rates for parallel tempering were substantially lower than those of ATHMC under all test settings.
Diagnostic plots for parallel tempering, provided in the supplementary text (Figures~\ref{fig:pt_iso_tempseq} and \ref{fig:pt_iso_cumtr}), show that in high dimensions, a large number of parallel chains are needed and the corresponding temperature sequence requires a long tuning process.
For $d=1$ and $10$, transitions between modes occurred at the lowest temperature level chain only after the temperature sequence stabilized.
In higher dimensions, tuning had not stabilized by the 5000-th MCMC iteration, and no inter-mode transitions occurred.

For ATHMC, transitions between modes occurred reasonably frequently even as $d$ increased.
Moreover, ATHMC exhibited transition rates that were relatively insensitive to differences in scale between the modes.
In our setup, the second mixture component had a volume $c$ times smaller than the first, with $c$ varying over $1$, $2$, and $10$.
However, if the ratio of scale widths between the modes was fixed (and not equal to 1) while the dimension $d$ increased, the relative volume would scale exponentially with $d$, and, consequently, all the Monte Carlo methods we considered would exhibit an exponentially decreasing rate of mixing.

\begin{figure}[tp]
\begin{knitrout}
\definecolor{shadecolor}{rgb}{0.969, 0.969, 0.969}\color{fgcolor}

{\centering \includegraphics[width=\maxwidth]{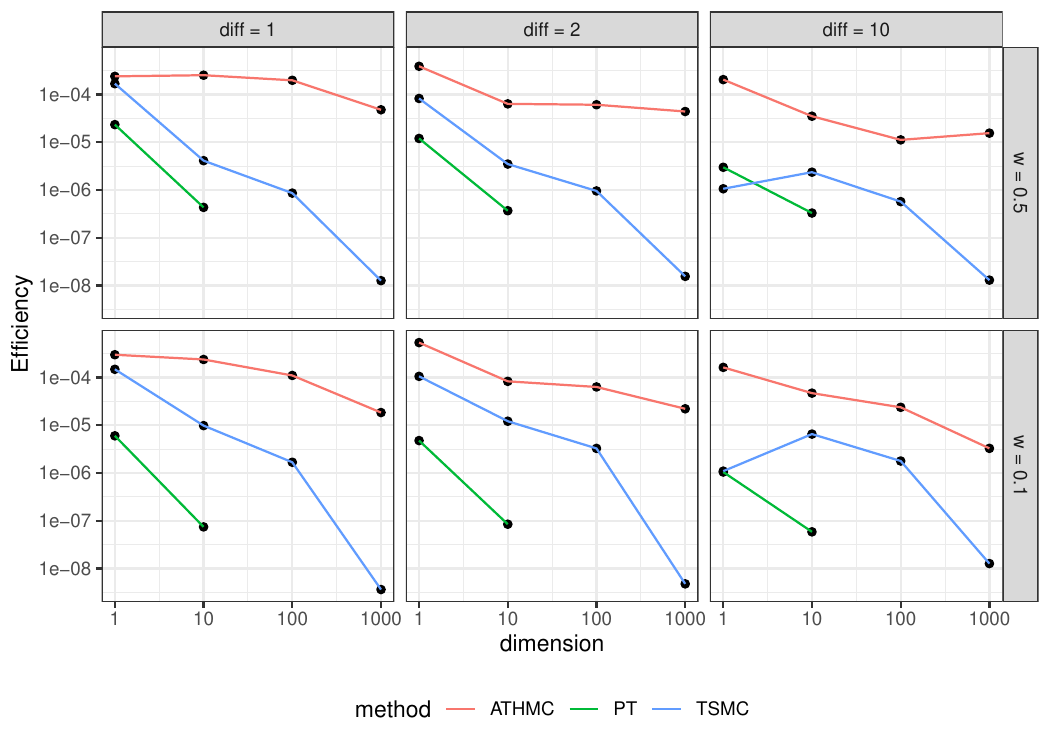} 

}

\end{knitrout}
\caption{Monte Carlo efficiency, evaluated as the effective number of draws divided by the total number of leapfrog steps, for ATHMC, PT, and TSMC.
  Efficiency is shown on a logarithmic scale.
  For PT, the efficiency for $d=100$ and $1000$ was not computed, as there occurred no inter-mode transitions. }
\label{fig:comparison_eff}
\end{figure}

To compare the efficiency of ATHMC, PT, and TSMC, we applied each method in 20 replications and estimated the mixture weight for the first mode, $w \approx P[\Vert X - \mu_1 \Vert < \Vert X - \mu_2\Vert]$.
For ATHMC and PT, we used Markov chains of length 5000, and for TSMC, we used an ensemble of 5000 particles.
For each method under each setting, we computed the empirical mean squared error (MSE) of $w$ as $\tfrac{1}{20}\sum_{b=1}^{20} (\hat w_b - w)^2$ where $\hat w_b$ is the estimate of $w$ from the $b$-th replication.
The effective number of Monte Carlo draws was then calculated as $w(1-w)$ divided by the MSE.
Overall Monte Carlo efficiency was defined as the effective number of draws divided by the total number of leapfrog steps.

%$$
Figure~\ref{fig:comparison_eff} presents the efficiency computed for ATHMC, PT, and TSMC.
Efficiency was not computed for PT in dimensions $d=100$ and $1000$, as there were no transitions between modes.
The plots in Figure~\ref{fig:comparison_eff} show that the efficiency of ATHMC was substantially higher than that of PT or TSMC in nearly all scenarios.
Crucially, ATHMC maintained meaningful efficiency in high dimensions, where the efficiency of the other methods dropped considerably.
These numerical results indicate that our automatically tuned, tempered HMC enables efficient global sampling from strongly multimodal, high dimensional distributions.
Numerical results for a mixture of two Gaussian densities with anisotropic covariances showed similar patterns, as illustrated in Figures~\ref{fig:supp_athmc_pt_transition_rate_aniso} and \ref{fig:supp_comparison_eff_aniso} of the supplementary text.

\section{Applications}\label{sec:applications}

\subsection{Bayesian mixture models}\label{sec:Bayesian_mixture}
Bayesian mixture models naturally induce multimodal posterior distributions due to label-switching symmetry.
When the prior distributions on model parameters are exchangeable, the multiple modes arising from label permutations can, in principle, be recovered by permuting the variables in the MCMC samples corresponding to a single labelling \citep{stephens1997bayesian}.
However, if the Markov chain cannot transition between modes, we have limited confidence that all plausible mixture configurations (beyond label switching) have been explored, or that the identified mode is one of the globally dominant posterior modes.
In this section, we present a demonstrative example showing that ATHMC (Algorithm~\ref{alg:tuning_THMC}) can explore all isolated posterior modes in a Bayesian mixture model.

\begin{figure}[tp]
\begin{knitrout}
\definecolor{shadecolor}{rgb}{0.969, 0.969, 0.969}\color{fgcolor}

{\centering \includegraphics[width=\maxwidth]{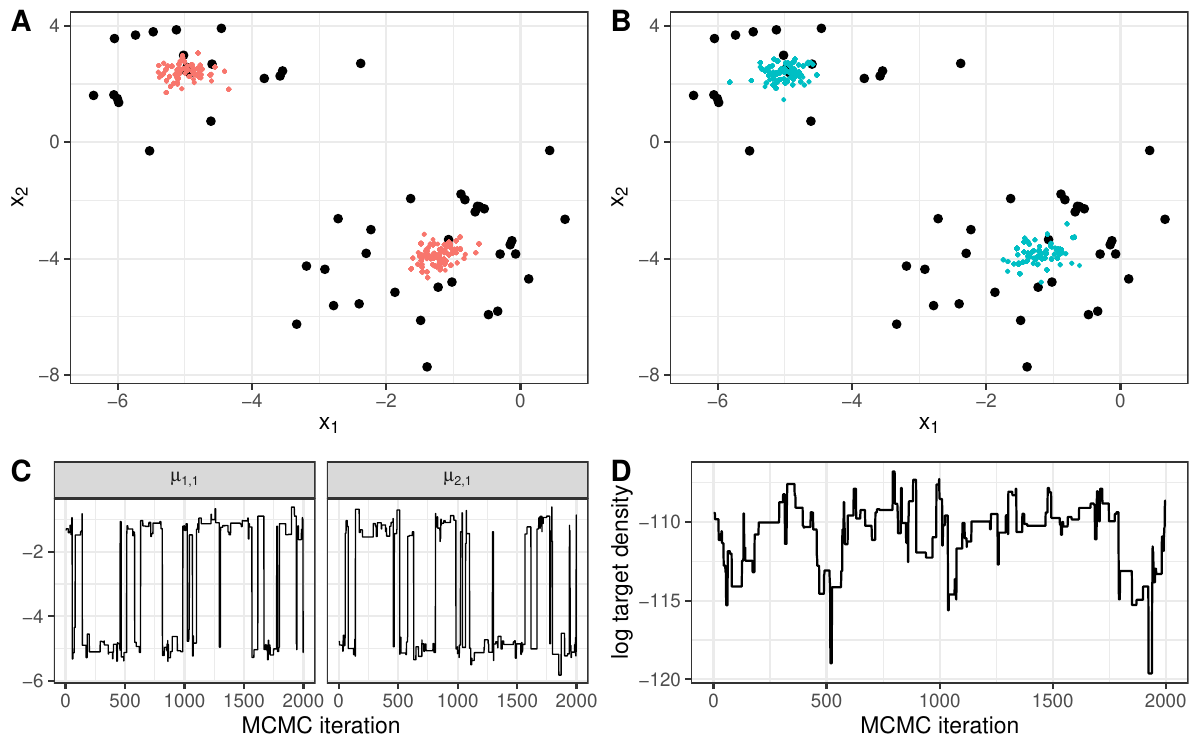} 

}

\end{knitrout}
\caption{A: Means of the first mixture component sampled using ATHMC (Algorithm~\ref{alg:tuning_THMC}).
  Black dots indicate the observed data points.
  B: Means of the second mixture component sampled.
  C: Trace plots of the first coordinates of the mixture component means.
  D: Trace plot of the log target densities evaluated at the sampled model parameters.}
\label{fig:bayesmix_athmc}
\end{figure}

We consider iid observations $X_1,\dots X_n \in \mathbb R^d$ drawn from a Gaussian mixture model
\[
  \sum_{j=1}^{n_{mix}} w_j \N(\mu_j, \Sigma_j).
\]
The normalized weights $w_j$ are expressed in terms of unnormalized weights $W_j$ via
\[
  w_j = \frac{W_j}{\sum_{j'} W_{j'}}.
\]
We parameterize the precision matrices $\Omega_j := \Sigma_j^{-1}$ using their unique Cholesky decompositions, $\Omega_j = L_j L_j^\top$, where $L_j$ is lower triangular with positive diagonal entries.
Up to an additive constant, the log-likelihood is given by
\[
  \ell(W_j, \mu_j, L_j; X_{1:n}) = \sum_{i=1}^n \left[ \log\left( \sum_{j=1}^{n_{mix}} W_j \cdot \det L_j \cdot \exp\left\{ -\frac{1}{2} \Vert L_j^\top (X_i - \mu_j) \Vert^2 \right\} \right) - \log \left( \sum_{j=1}^{n_{mix}} W_j \right) \right].
\]
To facilitate sampling with THMC for the corresponding posterior distribution, we employ the following parameterizations.
First, since the log-likelihood $\ell$ decreases approximately like $\log W_j$ as $W_j\to 0$ and like $-\log W_j$ as $W_j \to \infty$, we define
\[
  \log W_j = \operatorname{sign}(v_j-1) \cdot (v_j-1)^2,
\]
which induces a one-to-one correspondence between $W_j \in (0,\infty)$ and $v_j \in (-\infty, \infty)$.
Second, consider the diagonal entries $L_{j,rr}$, $1\leq r \leq d$, of the Cholesky factor of $\Omega_j$.
As $L_{j,rr} \to 0$, the log-likelihood $\ell$ decreases like $\log L_{j,rr}$; as $L_{j,rr}\to\infty$, it decreases approximately quadratically in $L_{j,rr}$ with a negative coefficient.
Based on this behavior, we define
\[
  L_{j,rr} = \begin{cases} e^{-(\lambda_{j,r} -1)^2} & \text{ if } \lambda_{j,r}<1,\\ \lambda_{j,r}  & \text{ if } \lambda_{j,r} \geq 1, \end{cases}
\]
which defines a one-to-one correspondence between $L_{j,rr} \in (0,\infty)$ and $\lambda_{j,r} \in (-\infty, \infty)$.
The off-diagonal entries $L_{j,rs}$ for $r\neq s$, as well as the mean parameters $\mu_j$, are estimated without transformation.

%$$
We place independent priors on the parameters as follows.
The transformed weights $v_j$ are assigned iid priors $v_j \sim \N(1, (\tfrac{1}{\sqrt 2})^2)$.
The component means $\mu_j$ are given iid priors $\N(0, 10^2\cdot I)$.
For the precision matrices, we use a Wishart prior: $\Omega_j \sim \text{Wishart}(\nu, I/\nu)$.

%$$
We generate $n=50$ observations $X_i \in \mathbb R^2$ from a mixture of $n_{mix} = 2$ Gaussian components.
The true parameters are drawn as follows: $v_{j,\text{true}} \sim \N(1, (\tfrac{1}{\sqrt 2})^2), \mu_{j,\text{true}} \sim \N(0, 2^2\cdot I)$, $\Omega_{j,\text{true}} \sim \text{Wishart}(\nu, I/\nu)$ where we use $\nu=3$.
In this model, the partial derivatives of the log-posterior density with respect to some variables approach zero as the parameter values tend to $-\infty$, $\infty$, or both.
In Hamiltonian Monte Carlo, when the simulated Hamiltonian trajectory enters a region where the potential energy $U$ has a flat landscape with respect to a particular variable, the trajectory can continue indefinitely in that direction, since the momentum undergoes minimal change.
To address this issue, we apply momentum reflection when the value of a variable moves outside a specified interval.
Specifically, we use bounds of $(-1,3)$ for $v_j$, $(-10,10)$ for each component $\mu_{j,r}$, and $(-1,\infty)$ for each $\lambda_{j,r}$, for $j=1,2$ and $r=1,2$.
The off-diagonal entries $L_{j,rs}$, with $r\neq s$, are left unbounded.
The target posterior distribution remains invariant under both standard and tempered HMC when the momentum is reflected according to a specific rule, as explained in the supplementary text (Section~\ref{sec:supp_bounce}).

%$$
We first located a posterior mode and simultaneously tuned the reference leapfrog step size by running 20 iterations of standard HMC.
Subsequently, we ran ATHMC, during which both the maximum log-temperature $\eta_*$ and the rate of change $|\dot\eta|$ were tuned adaptively.
To tune $\eta_*$, we used a search criterion requiring that the potential energy $U$ reach at least 300 at some point along the trajectory.
For tuning $|\dot\eta|$, we targeted an acceptance rate of 0.05.
Panels~A and B of Figure~\ref{fig:bayesmix_athmc} display the mixture component means $\mu_1$ and $\mu_2$ sampled using ATHMC.
Panel~C shows trace plots of the first coordinates of the two means over two thousand MCMC iterations.
Panel~D confirms that the two posterior modes approximately have the same log target density.
Since tempered HMC does not include any ad hoc mechanisms specifically designed to encourage label switching, the fact that the Markov chain transitioned between the two modes corresponding to both label assignments indicates that no other mixture configurations have posterior densities comparable to those of the two dominant modes.
Trace plots of all model parameters are provided in the supplementary material (Section~\ref{sec:supp_bayesmixfig}). 

\begin{figure}[tp]
\begin{knitrout}
\definecolor{shadecolor}{rgb}{0.969, 0.969, 0.969}\color{fgcolor}

{\centering \includegraphics[width=\maxwidth]{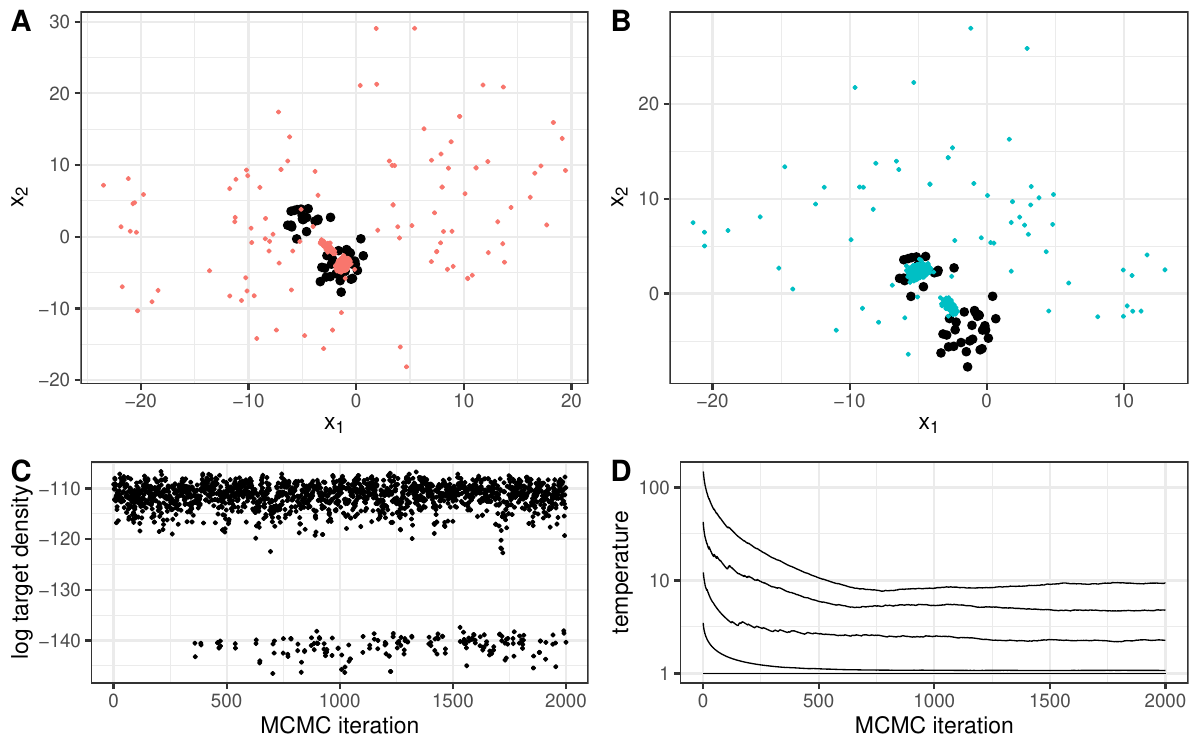} 

}

\end{knitrout}
\caption{A: Means of the first mixture component sampled using parallel tempering with adaptive temperature tuning.
  Black dots indicate the observed data points.
  B: Means of the second mixture component sampled.
  C: Trace plot of the log target densities evaluated at the sampled model parameters. 
  D: Trace plot of the tuned temperature levels for the five parallel chains. }
\label{fig:bayesmix_pt}
\end{figure}

For comparison, we employed parallel tempering to sample from the same posterior distribution using $K=5$ chains.
The temperature levels were adaptively tuned as described in Section~\ref{sec:comparison}.
Panels~A and B of Figure~\ref{fig:bayesmix_pt} show the sampled means of the two mixture components.
These sample draws can be grouped into two distinct classes.
In the first class, each mixture mean is located near the center of one of the two observed point clouds.
In the second class, one component captures both point clouds, while the other is placed at seemingly arbitrary positions.
The mean of the mixture component that includes all data points is located near the weighted average of the two clouds.

Interestingly, one of the global posterior modes---corresponding to a label permutation of the first class of MCMC draws---was not sampled.
Panel~C of Figure~\ref{fig:bayesmix_pt} shows the log posterior densities of the obtained samples.
The first class of draws, in which each point cloud is assigned to a separate mixture component, has log target densities around -110.
In contrast, the second class, where both point clouds are assigned to a single component, has log densities around -140.

The fact that the second class of draws---despite its substantially lower posterior density---appears with noticeable frequency suggests poor global mixing of the collection of parallel chains.
During adaptive tuning, the temperature levels gradually decreased, as shown in Panel~D of Figure~\ref{fig:bayesmix_pt}.
The less likely configuration that places both clouds in a single component is frequently sampled by the chain at the second-lowest temperature level during the early phase of tuning.
This occurs because that configuration occupies a relatively larger volume of the parameter space, even though it has lower posterior probability than the dominant modes.
As the temperature decreases, the second-lowest temperature chain does not mix fast enough to adapt to the changing target distribution.
As a result, samples from the non-dominant mode---oversampled early on---percolate into the lowest temperature chain due to the reduced temperature gap between adjacent chains.
This behavior reflects insufficient joint mixing across the ensemble of parallel chains.
This phenomenon poses a practical challenge, as it is difficult to determine whether the overall sampling scheme has reached stationarity.
In contrast, ATHMC samples from both dominant modes through direct inter-mode transitions.

\subsection{Bayesian sparse regression using a spike-and-slab prior}\label{sec:sparse_regression}

\begin{figure}[tp]
\begin{knitrout}
\definecolor{shadecolor}{rgb}{0.969, 0.969, 0.969}\color{fgcolor}

{\centering \includegraphics[width=\maxwidth]{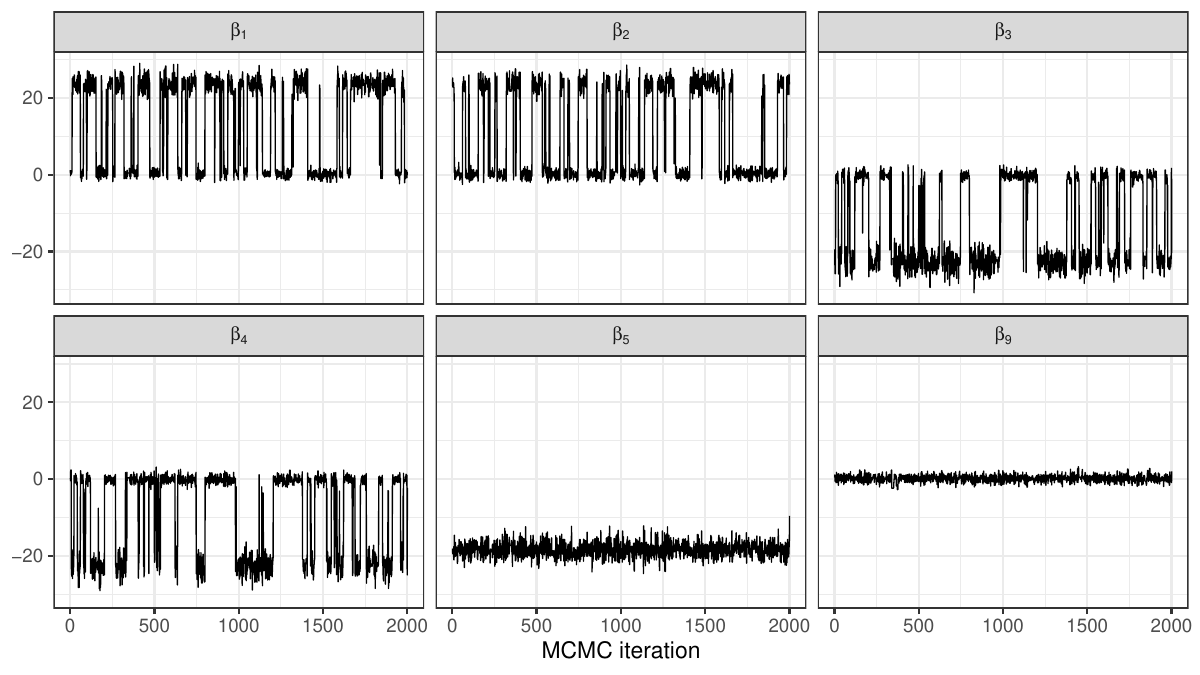} 

}

\end{knitrout}
\caption{Trace plots of the coefficients $\beta_1$, $\beta_2$, $\beta_3$, $\beta_4$, $\beta_5$, and $\beta_9$ in the Bayesian sparse linear regression model.}
\label{fig:sparse_bayes_traceplots}
\end{figure}

We applied ATHMC to sample from the posterior distribution of a Bayesian sparse regression model with a spike-and-slab prior.
We considered the linear model
\begin{equation}
  Y_i = X_i^\top \beta + \epsilon_i, \quad \epsilon_i \overset{iid}\sim \N(0,1), \quad X_i \in \mathbb R^{100}, \quad i =1,\dots,30.
  \label{eqn:linearmodel}
\end{equation}
The sparsity-inducing spike-and-slab prior was specified as
\[
  \beta_j \overset{iid}\sim 10^{-4} \cdot \N(0, 10^2) + (1-10^{-4}) \cdot \N(0, 1^2). \quad j=1, \dots, 100.
\]
The observed responses $Y_i$, $i=1,\dots, 30$, were generated according to the linear model~\eqref{eqn:linearmodel}, with only the first eight coefficients being nonzero: $\beta = (10, 15, -10, -15, -15, -10, 10, 20, 0, 0, \dots)$.
Each covariate $X_{i,j}$ was independently drawn from $N(0,1)$.

To induce multimodality in the posterior distribution, we introduced exact collinearity by setting $X_{\cdot, 1} = X_{\cdot, 2}$ and $X_{\cdot, 3} = X_{\cdot, 4}$.
Because of this collinearity, the likelihood is invariant under changes to $\beta_1$ and $\beta_2$ (or $\beta_3$ and $\beta_4$) that preserve the sums $\beta_1+\beta_2$ and $\beta_3+\beta_4$.
However, the spike-and-slab prior induces a posterior distribution in which only one of $\beta_1$, $\beta_2$ and one of $\beta_3$, $\beta_4$ is nonzero with high probability.

%$$
We used tempered HMC with a reference step size of $\bar \epsilon = 0.02$ and $K = 2000$ leapfrog steps per MCMC iteration.
Figure~\ref{fig:sparse_bayes_traceplots} shows trace plots of the first five coefficients and $\beta_9$.
The two pairs, $(\beta_1, \beta_2)$ and $(\beta_3, \beta_4)$, were almost perfectly anti-correlated, with exactly one coefficient in each pair being active---except for a single instance out of 2000 iterations. % the only exception being beta3=beta4=1 in MCMC iter 167.
A coefficient was considered active if its absolute value exceeded 5.
The active parameter between $\beta_1$ and $\beta_2$ were approximately equal to the sum of the their true values, $\beta_1 = 10$ and $\beta_2 = 15$.
The same pattern held for $\beta_3$ and $\beta_4$, whose true values were $-10$ and $-15$, respectively.
The fifth coefficient, $\beta_5$, with a true value of -15, was always estimated to be active, whereas $\beta_9 = 0$ was consistently estimated to be inactive.
In summary, ATHMC successfully identified all four dominant modes of the posterior distribution through reasonably frequent inter-mode transitions.

\subsection{Ising model}\label{sec:ising}

\begin{figure}[tp]
\begin{knitrout}
\definecolor{shadecolor}{rgb}{0.969, 0.969, 0.969}\color{fgcolor}
\includegraphics[width=\maxwidth]{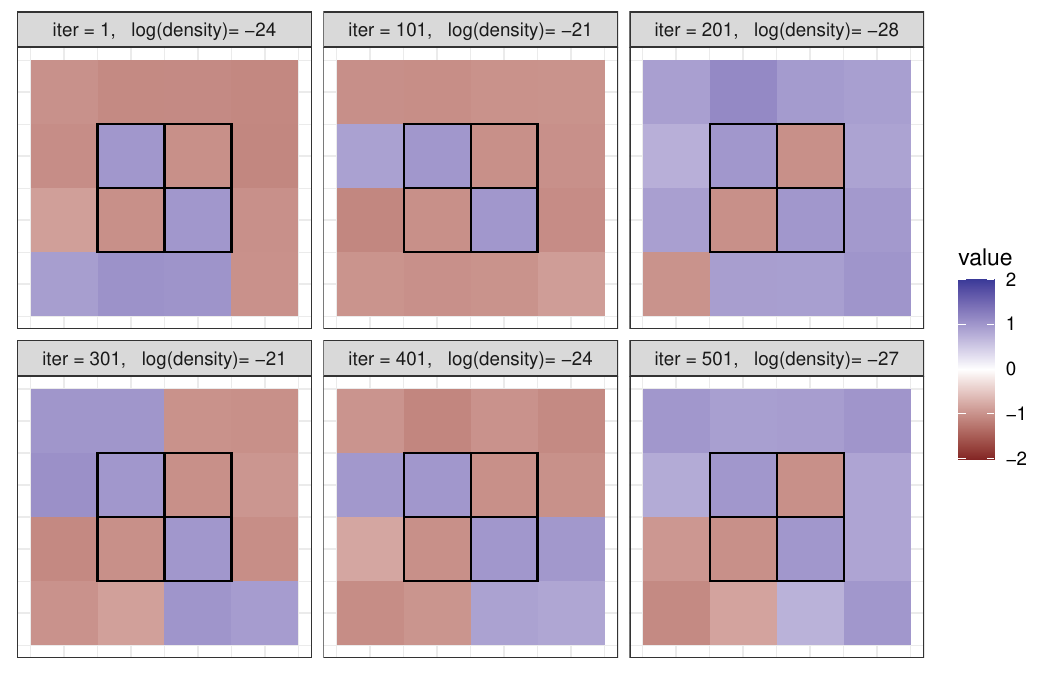} 
\end{knitrout}
\caption{Spin configurations sampled using ATHMC for the Ising model.}
\label{fig:ising}
\end{figure}

We considered an Ising model on a square lattice with four rows and four columns, where the spin at each site is a continuous-valued variable.
Sites $a=(i,j)$ and $b=(i',j')$ are considered adjacent if $i=i'$ and $|j-j'|=1$ or $j=j'$ and $|i-i'|=1$.
If $a$ and $b$ are adjacent, we write $a\sim b$.
Denoting the spin at site $a$ as $x_a$, the potential function is defined as
\begin{equation}
  U(x) = \rho_1 \sum_{a\sim b} (x_a - x_b)^2 + \rho_2 \sum_a (x_a^2-1)^2.
  \label{eqn:isingU}
\end{equation}
The first term encourages adjacent sites to have similar spin values, while the second term promotes spins near $\pm 1$.
We used $\rho_1 = 0.5$ and $\rho_2 = 20$.
We fixed the spins at the four center sites to $(-1, 1, 1, -1)$ to create interesting spatial patterns, as shown in Figure~\ref{fig:ising}.

We applied ATHMC to sample from the Boltzmann distribution with density $\pi(x) \propto e^{-U(x)}$.
After identifying a local mode found using standard HMC, we ran tempered HMC with a reference step size of $\bar\epsilon = 0.02$.
A rectangular search criterion was used, with coordinate-wise center $x_a^0 = 0$ and scale $s_a = 1.5$ for every peripheral site $a$.
The leapfrog step size was varied as $\epsilon = e^{2a\eta} \bar \epsilon$, where $a = \tfrac{2}{\gamma+2}$ with the log-polynomial degree $\gamma = 4$ chosen based on the second term in the potential function $U(x)$ defined in \eqref{eqn:isingU}.

Figure~\ref{fig:ising} shows six sample spin configurations from 500 MCMC iterations, illustrating frequent transitions between isolated modes of the target distribution.
Indeed, the signs of the spins changed in 48 out of 500 iterations.
In contrast, no sign changes were observed when standard HMC was used, as shown in Figure~\ref{fig:ising_hmc} in the supplementary text.

\subsection{Self-localization of a sensor network}\label{sec:sensor_localization}

\begin{figure}
\begin{knitrout}
\definecolor{shadecolor}{rgb}{0.969, 0.969, 0.969}\color{fgcolor}

{\centering \includegraphics[width=\maxwidth]{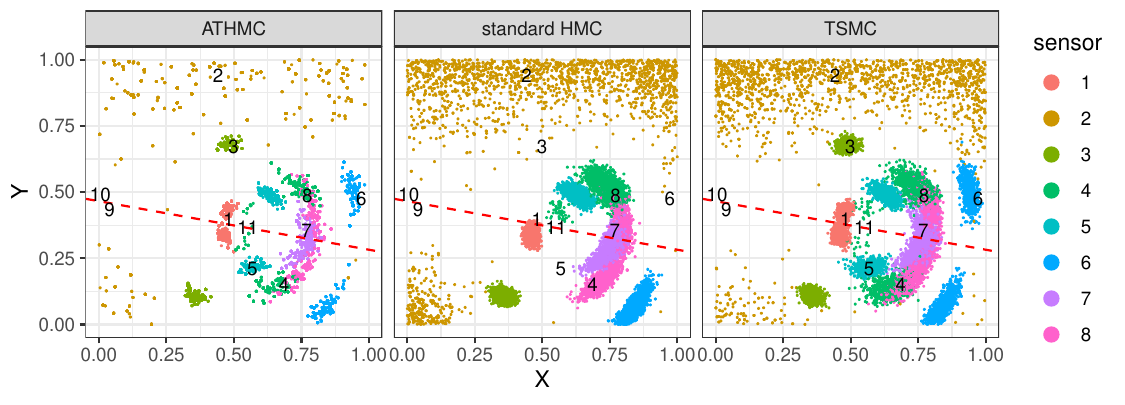} 

}

\end{knitrout}
\caption{
  The sample points for the eight sensor locations obtained by ATHMC, standard HMC, and TSMC for the sensor self-localization example with the posterior density given by \eqref{eqn:sensor_posterior}.
  The true location of each sensor is marked by its number ID.
  A line approximately connecting the three sensors of known locations (9, 10, 11) is marked by a red dashed line.
  The true posterior density is bimodal, with the locations of each sensor corresponding to the two modes being approximately mirror images of each other with respect to the red dashed line.
  %The parameters $R=0.3$ and $\sigma_e=0.02$ were assumed to be known.
}
\label{fig:sensor_marginal_ATHMC_HMC}
\end{figure}

We apply ATHMC to a sensor network self-localization problem previously considered by \citet{ihler2005nonparametric}.
Noisy pairwise distance measurements are available, and the goal is to localize the positions of eight sensors (labelled 1 through 8) within a two dimensional square, $[0,1]^2$.
Additionally, there are three sensors (labelled 9, 10, and 11) at known locations---these sensors would uniquely determine the locations of the others if the distances were measured without noise.
However, we consider a scenario in which these three sensors are positioned approximately collinearly, so that the locations of the other eight sensors are approximately identifiable only up to reflection about the line connecting the three reference sensors.
The true locations of all eleven sensors are marked in the plots in Figure~\ref{fig:sensor_marginal_ATHMC_HMC}, where the reflection symmetry is indicated by a red dashed line.

The distance between sensors labelled $t$ and $u$ is measured with noise following a normal distribution, $\N( \Vert \mathbf{x}_t - \mathbf{x}_u\Vert, \sigma_e^2)$, where $\sigma_e = 0.02$.
However, not all pairwise distances are measured.
The probability that two sensors $t$ and $u$, located at $\mathbf{x}_t=(x_t, y_t)$ and $\mathbf{x}_u=(x_u, y_u)$, have a distance measurement is given by $e^{-\Vert \mathbf{x}_t - \mathbf{x}_u\Vert^2 / (2R^2)}$, where $R=0.3$.
Distance measurements were generated according to this model.

For Bayesian inference, we assume independent uniform prior distributions on the square $[0,1]^2$ for the locations of sensors labelled 1-8.
Denoting by $d_{t,u}$ the distance measurements between sensors $t$ and $u$ and by $\iota_{t,u} \in \{0,1\}$ the binary variable indicating whether the distance is measured, $1\leq t \leq 8$ and $t<u\leq 11$, the posterior density for $\mathbf{x}_{1:8}$ given the measurement data is given by
\begin{multline}
  \pi(\mathbf{x}_{1:8}|\{(\iota_{t,u}, d_{t,u}); 1\leq t \leq 8, t < u \leq 11\})\\
  \propto \prod_{\substack{1\leq t \leq 8,\\ t < u \leq 11}} \left[ (e^{-\Vert \mathbf{x}_t - \mathbf{x}_u\Vert^2/2R^2})^{\iota_{t,u}} (1- e^{-\Vert \mathbf{x}_t - \mathbf{x}_u \Vert^2/2R^2})^{1-\iota_{t,u}} \right]\\
  \cdot \prod_{\substack{1\leq t \leq 8,\\ t < u \leq 11,\\ \iota_{t,u} = 1}} \frac{1}{2\pi \sigma_e^2} e^{-(\Vert \mathbf{x}_t - \mathbf{x}_u\Vert - d_{t,u})^2/2\sigma_e^2}
  \cdot \prod_{t=1}^8 \mathbf{1}\left[\mathbf{x}_t \in [0,1]^2 \right].
  \label{eqn:sensor_posterior}
\end{multline}
Due to the reflection symmetry about the line connecting the three sensors of known locations ($\mathbf{x}_{9:11}$), the posterior distribution of the unknown sensor locations is bimodal.

We applied ATHMC, standard HMC, and TSMC across 20 replications.
Whenever a sensor position reached the boundary of the square, the simulated trajectory rebounded (see the supplementary section~\ref{sec:supp_bounce} for the construction of this rebound step, which ensures the reversibility of the resulting chains).
The HMC kernel within TSMC also incorporated this rebounding mechanism.

The reference leapfrog step size for ATHMC was tuned via a pilot run of standard HMC and then scaled by a factor of 0.5.
ATHMC used a piecewise linear $\{\eta_k\}$ sequence \eqref{eqn:eta_piecewiselinear} and a rectangular search scope centered at $(\tfrac{1}{2}, \tfrac{1}{2})$, with coordinate-wise scale $\frac{1}{6}$ in both $x$ and $y$ directions for each sensor.
The target acceptance probability was set to 0.05.
For standard HMC, we used the same step size and 300 leapfrog steps per trajectory.
Both ATHMC and standard HMC were run for 2000 iterations.

For TSMC, we used 2000 particles, with temperature level and step size adaptively tuned at each stage.
The base distribution for TSMC consisted of independent copies of the uniform distribution over $[0,1]^2$ for each sensor.
Particles were diversified using a standard HMC kernel with 10 leapfrog steps, repeated 5 times.

\begin{figure}
\begin{knitrout}
\definecolor{shadecolor}{rgb}{0.969, 0.969, 0.969}\color{fgcolor}

{\centering \includegraphics[width=0.8\linewidth]{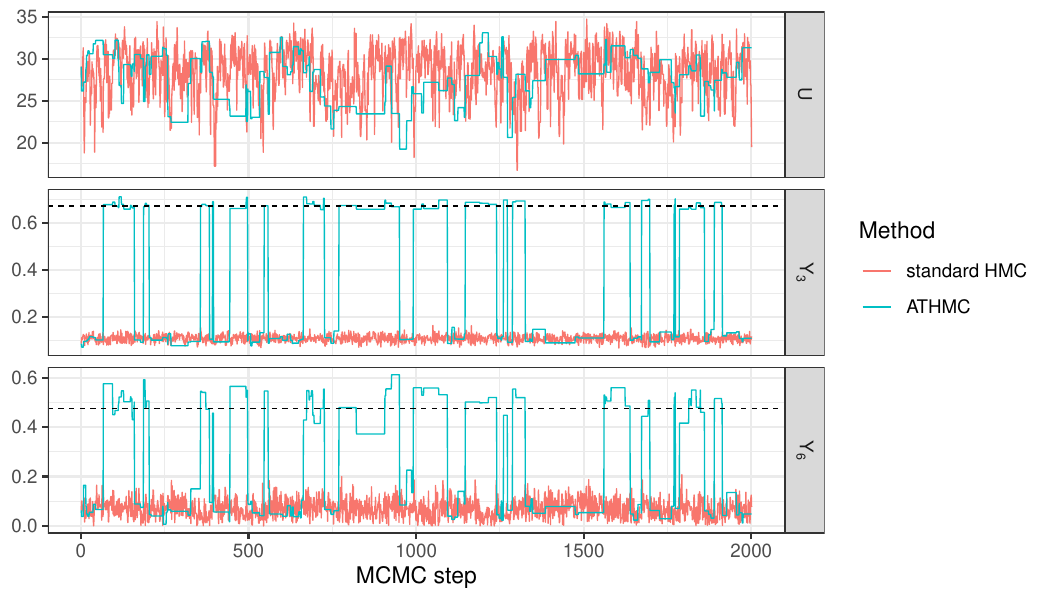} 

}

\end{knitrout}
\caption{
  Trace plots of the potential energy $U(\mathbf{x}_{1:8}) = -\log\pi(\mathbf{x}_{1:8})$ and the $y$-coordinates of the third and the sixth sensors for one of the 20 chains constructed by ATHMC and standard HMC.
  %The parameters $R=0.3$ and $\sigma_e=0.02$ were assumed to be known.
  The true $y$-coordinates for the third and the sixth sensors are indicated by horizontal dashed lines.
}
\label{fig:sensor_traceplots}
\end{figure}

Figure~\ref{fig:sensor_marginal_ATHMC_HMC} shows the sampled sensor locations, colored by sensor ID, from one of the 20 replications of each method.
The true sensor locations are indicated by their corresponding number labels.
Sample points obtained by ATHMC and TSMC exhibit both modes of the posterior density, as seen clearly in the marginal draws for sensor \#3 (dark green) and sensor \#6 (light blue).
In contrast, all sample points obtained by standard HMC remain within a single posterior mode, with the estimated locations differing from the true locations.
The same pattern was observed in all 20 replications of each method (see Figures~\ref{fig:sensor_marginal_ATHMC_only} and \ref{fig:sensor_marginal_HMC_only} in the supplementary text).
We note that, although ATHMC produced fewer unique sample points compared to other methods, the number of unique samples can be readily increased by intermittently incorporating standard HMC kernels for local exploration within a mode.
Our primary interest is whether both posterior modes can be sampled.
%Sensor \#2 has no distance measurements with any other sensors, so its marginal posterior draws are scattered where there are no sensors nearby.

Figure~\ref{fig:sensor_traceplots} shows trace plots of the potential function $U$ and the $y$-coordinates of sensors \#3 and \#6.
Both ATHMC and standard HMC chains exhibit similar levels of $U$, suggesting that the chains remain near one of the two posterior modes.
However, the $y$-coordinates for sensors \#3 and \#6 reveal frequent transitions between modes in ATHMC, while no such transitions occur in HMC.
Supplementary Section~\ref{sec:supp_sensorfig} presents results in which ATHMC is used within a Gibbs sampler to jointly estimate $R$ and $\sigma_e$.

Table~\ref{tab:sensor_athmc_tsmc} compares ATHMC and TSMC based on the estimates of the $y$-coordinate of sensor \#3 ($Y_3$), which has a strongly bimodal marginal posterior distribution.
The average of the estimates $\hat Y_3$ across 20 replications are similar for both methods, suggesting that each correctly samples from the bimodal target distribution.
The standard deviations of the estimates are comparable.
Efficiency---defined as the reciprocal of the product of the Monte Carlo variance of $\hat Y_3$ and the number of leapfrog steps---is similarly comparable.
TSMC performs reasonably well in this example, likely due to the restricted support and moderate dimensionality of the target distribution.

\begin{table}

\centering
\begin{tabular}{lllll}\hline
  method & mean($\hat Y_3$) & s.e.($\hat Y_3$) & Tot. LF steps & Efficiency \\\hline
  ATHMC & 0.283 & 0.048 & \ensuremath{4.29\times 10^{5}} & 0.0010 \\
  TSMC & 0.300 & 0.037 & \ensuremath{1.8\times 10^{6}} & 0.0004 \\\hline
\end{tabular}
\caption{ 
  Bayesian estimates of the $y$-coordinate of sensor \#3 obtained using ATHMC and TSMC.
  The mean and standard deviation of the estimates across 20 replications are reported.
  Also shown are the total number of leapfrog steps and the efficiency, defined as the reciprocal of the product of the Monte Carlo variance of the estimate and the number of leapfrog steps.
}
\label{tab:sensor_athmc_tsmc}
\end{table}

\newcommand\xt[1][]{x_T^\text{#1}}
\newcommand\vt[1][]{v_T^\text{#1}}
\section{Other recent approaches to sampling from multimodal distributions}\label{sec:otherapproaches}
In this section, we briefly review some of the recent approaches to sampling from multimodal distributions.
Continuous tempering is a strategy originally developed for, akin to various other tempering methods, simulating molecular dynamics where the free energy function has multiple isolated modes \citep{gobbo2015extended, lenner2016continuous}.
It extends the Hamiltonian system by including a variable $\xt \in \mathbb R$ linked to the temperature level and the associated velocity variable $\vt$.
The extended Hamiltonian can be written in the form of
\[
\hat H(x, p, x_T^{}, p_T^{}) = H(x,p) - f(x_T^{}) G(x,p) + w(x_T^{}) + \frac{p_T^2}{2 m_T^{}} 
\]
where $f$, $G$ and $w$ are some functions and $m_T^{} \in \mathbb R$ represents the mass associated with the added variable $x_T^{}$.
The link function $f$ is chosen such that $f(x_T^{}) = 0$ for a certain interval, say for $x_T^{} \in (-c_T^{}, c_T^{})$.
The states in the constructed Markov chain for which $x_T^{} \in (-c_T^{}, c_T^{})$ may then be considered as draws from the original target density $\Pi(x,v) \propto e^{-H(x,v)}$.
\citet{graham2017continuously} considered the case
\[
f(x_T^{}) = 1- \beta(x_T^{}), \quad G(x,p) = U(x) + \log g(x)
\]
where $\beta(x_T^{})$ represents the inverse temperature and $g(x)$ is the normalized density function of a certain base distribution.
In this case, the extended Hamiltonian defines a smooth transition between the target density $e^{-U(x)}$ and $g(x)$ such that the distribution of $X$ given $\beta(x_T^{}) = \beta^*$ has density
\[
p(x | \beta(x_T^{}) = \beta^*) \propto e^{-U(x)\beta^*} g(x)^{1-\beta^*}.
\]
This density function has the same form as the bridging density commonly used by annealed importance sampling \citep{neal2001annealed}.
Like simulated tempering, the continuous tempering strategy becomes efficient when the temperature variable is marginally evenly distributed across its range.
If the marginal distribution is highly concentrated on low temperature values, transitions between isolated modes may occur with extremely low probability.
On the contrary, if the distribution is concentrated on high temperature values, the samples from the original target distribution may be obtained rarely.
In order to achieve an even distribution of temperature, techniques such as adaptive biasing force have been used \citep{darve2001calculating}.
\citet{luo2018thermostat} used adaptive biasing force to continuously tempered Hamiltonian Monte Carlo, and further extended the method to settings with mini-batches by introducing Nos\'e-Hoover thermostats \citep{nose1984unified, hoover1985canonical}.
However, adaptive biasing techniques may exhibit slow adaptation.

Darting Monte Carlo uses independence Metropolis-Hastings proposals to facilitate transitions between isolated modes \citep{andricioaei2001smart, sminchisescu2011generalized}.
The modes of the target density are often found by a deterministic gradient ascent method started at different initial conditions to discover as many local maxima as possible.
A mixture of density components centered at the discovered modes is often used by the independence Metropolis-Hastings (MH) sampler as a proposal distribution.
The proposal distribution in this independence MH sampler can be adaptively tuned at regeneration times \citep{ahn2013distributed}.
Darting Monte Carlo methods can be efficient for finding the relative probability masses of the discovered density components, but one of its drawbacks is that an external procedure for finding the modes and approximating the shapes of the modes needs to be employed.
Another issue is the unfavorable scaling properties with increasing dimensions, as noted by \citet{ahn2013distributed}.

Wormhole Hamiltonian Monte Carlo \citep{lan2014wormhole} connects the known locations of the modes by modifying the metric so that the modes are close to each other under the modified metric.
The method then runs Riemannian manifold Hamiltonian Monte Carlo, which takes into account the given metric while simulating the Hamiltonian trajectories \citep{girolami2011riemann}.

\citet{tak2018repelling} developed a novel strategy that attempts a MH move that favors \emph{low} target probability density points before attempting another MH move that favors high density points.
In their algorithm, the proposal is first repelled from the the current state and then attracted by a local mode, which may be a different mode than that it started from.
The authors, however, could only develop their method for symmetric proposal kernels, such as zero-mean random walk perturbations.
Due to the use of random walk kernels, the scaling rate with respect to the space dimension is not likely to be more favorable than methods based on HMC.
Moreover, the random walk variance greatly affects the probability of transitions from one mode to another, but the tuning may not be straightforward in practice.

There are interesting recent developments in sampling techniques that offer alternatives to the MCMC framework.
For instance, \citet{qiu2024efficient} developed a method for sampling from multimodal distributions by applying a series of invertible maps constructed by neural networks to draws from a simple base distribution \citep{hoffman2019neutra, kingma2016improved}.
These invertible maps bridge a sequence of intermediate tempered distributions, trained in a way that approximates the Wasserstein gradient flow.
One potential drawback of this method is that the normalizing constants of the tempered distributions need to be estimated by importance sampling, which often does not scale to high dimensions.
In general, the construction of a manageable pullback distribution for a complex target distribution can complement MCMC sampling, and vice versa.
%%[[ TODO: Discuss Ge, Lee, Risteski (2018) Tempering Langevin Monte Carlo]]

%and Tawn, Moores, Roberts (2021) Annealed leap point sampler for multimodal target distributions (not completed, the main idea seems to be given by Tawn, Roberts, Rosenthal (2019) Weight preserving simulated tempering, StatComp
%and Huang, Jiao, Kang, Liao, Liu (2021) Schrodinger Follmer sampler Sampling without ergodicity, (not specific to multimodal distribution, and not applicable unless the target density is analytically simple.)

%
%
\section{Discussion}\label{sec:discussion}
We developed a tempered Hamiltonian Monte Carlo method for sampling from high dimensional, strongly multimodal distributions by simulating Hamiltonian dynamics with a time-varying temperature.
In applications to mixtures of log-polynomial distributions, our method enabled frequent transitions between modes even in extremely high dimensions ($d=10{,}000$) and with large mode separation ($\Vert \mu_1 - \mu_2 \Vert = 10{,}000$).
THMC effectively combines the favorable scaling with dimension exhibited by HMC with the advantages of tempering techniques for multimodal sampling.
Indeed, it can be viewed as a combination of HMC and the tempered transitions method proposed by \citet{neal1996sampling} (see the Supplementary Section~\ref{sec:tempered_transitions_link} for further discussion).

We have developed an automatic tuning algorithm for our method by leveraging a stability property under a time-scale transformation, as outlined in Equations~\ref{eqn:tbar} and \ref{eqn:HEM_bar}.
ATHMC requires minimal customization; the only aspect that typically requires attention is the specification of the search scope for isolated modes.

Unlike some other methods for multimodal sampling, our THMC does not require prior knowledge of mode locations.
Instead, it leverages the gradient of the log target density to guide the search trajectory toward isolated modes. 
However, if such mode-specific knowledge is available, our method could incorporate strategies such as those proposed by \citet{tawn2020weight} and \citet{andrieu2011nonlinear} to further enhance sampling efficiency.

\smallskip

\noindent\textbf{Acknowledgements:} This paper was partially supported by the General Research Fund of the University of Kansas Office of Research.
The author thanks Dr. Yves Atchad\'e for his comments on an earlier draft of this manuscript. 

\appendix

\section{Proof of Proposition~\ref{prop:THMC_reversibleMC}}
Let $\Psi_\kappa$ for $\kappa \in \{\tfrac{1}{2}, \dots, K-\tfrac{1}{2}\}$ denote the map defined by a leapfrog step with step size $\epsilon_\kappa = e^{2a\eta_\kappa} \bar\epsilon$, described by lines \ref{line:THMC_leapfrog1}--\ref{line:THMC_leapfrog3} in Algorithm~\ref{alg:THMC_sim}.
Then the trajectory constructed by Algorithm~\ref{alg:THMC_sim} can be expressed as
\[
  \Psi_\alpha := \Psi_{K-\frac{1}{2}} \circ \Psi_{K-\frac{3}{2}} \circ \cdots \circ \Psi_{\frac{3}{2}} \circ \Psi_{\frac{1}{2}}.
\]
We first verify that $\Psi_\alpha$ is time-reversible.
It is straightforward to check that each $\Psi_\kappa$ is time reversible: $\Psi_\kappa \circ \mathcal T \circ \Psi_\kappa \circ \mathcal T (x,p)) = (x,p)$, $\forall (x,p) \in \mathsf X \times \mathsf P$, where $\mathcal T(x,p) = (x, -p)$.
Since $\mathcal T$ is an involution, this condition can be expressed as $\Psi_\kappa \circ \mathcal T \circ \Psi_\kappa = \mathcal T$.
Moreover, since the temperature schedule is symmetric, that is, $\eta_\kappa = \eta_{K-\kappa}$, we have $\Psi_\kappa = \Psi_{K-\kappa}$, $\forall \kappa$.
We observe that
\[
  \begin{split}
    &\Psi^{K-\frac{1}{2}} \circ \Psi^{K-\frac{3}{2}} \circ \cdots \circ \Psi^{\frac{3}{2}} \circ \Psi^{\frac{1}{2}} \circ \mathcal T \circ \Psi^{K-\frac{1}{2}} \circ \Psi^{K-\frac{3}{2}} \circ \cdots \circ \Psi^{\frac{3}{2}} \circ \Psi^{\frac{1}{2}} \\
    &= \Psi^{K-\frac{1}{2}} \circ \Psi^{K-\frac{3}{2}} \circ \cdots \circ \Psi^{\frac{3}{2}} \circ (\Psi^{\frac{1}{2}} \circ \mathcal T \circ \Psi^{\frac{1}{2}}) \circ \Psi^{K-\frac{3}{2}} \circ \cdots \circ \Psi^{\frac{3}{2}} \circ \Psi^{\frac{1}{2}} \\
    &= \Psi^{K-\frac{1}{2}} \circ \Psi^{K-\frac{3}{2}} \circ \cdots \circ \Psi^{\frac{3}{2}} \circ \mathcal T \circ \Psi^{K-\frac{3}{2}} \circ \cdots \circ \Psi^{\frac{3}{2}} \circ \Psi^{\frac{1}{2}} \\
    &= \cdots = \mathcal T.
  \end{split}
\]
Hence, $\Psi_\alpha$ is time-reversible.

Next, we verify that $\Psi_\alpha$ is symplectic.
It is again straightforward to check that each $\Psi_\kappa$ is symplectic:
\[
  (D\Psi_\kappa)^\top J^{-1} (D\Psi_\kappa) = J^{-1}, \qquad \text{where} J^{-1} = \begin{pmatrix} 0 & -I_d \\ I_d & 0 \end{pmatrix}.
\]
We then have
\[
  \begin{split}
    (D\Psi_\alpha)^\top J^{-1} (D\Psi_\alpha)
    &= \left\{ (D\Psi_{K-\frac{1}{2}}) \cdot \cdots \cdot (D\Psi_{\frac{1}{2}}) \right\}^\top J^{-1} \left\{ (D\Psi_{K-\frac{1}{2}}) \cdot \cdots \cdot (D\Psi_{\frac{1}{2}}) \right\}\\
    &= (D\Psi_{\frac{1}{2}})^\top \cdot \cdots \cdot (D\Psi_{K-\frac{1}{2}})^\top J^{-1} (D\Psi_{K-\frac{1}{2}}) \cdot \cdots \cdot (D\Psi_{\frac{1}{2}})\\
    &= (D\Psi_{\frac{1}{2}})^\top \cdot \cdots \cdot (D\Psi_{K-\frac{3}{2}})^\top J^{-1} (D\Psi_{K-\frac{3}{2}}) \cdot \cdots \cdot (D\Psi_{\frac{1}{2}})\\
    &= \cdots = J^{-1}.
  \end{split}
\]
From the symplecticness of $\Psi_\alpha$, it follows that $\Psi_\alpha$ conserves the volume element: $dx(t_K) dp(t_K) = dx(0) dp(0)$ \citep[Section~38B]{arnold1989mathematical}.

Finally, we recall that the Metropolis-Hastings acceptance ratio is given by
\[
  \exp\left\{ - H(x(t_K), p(t_K)) + H(x(0), p(0)) \right\} = \frac{ \pi(x(t_K)) \cdot \phi(p(t_K); 0, M) }{ \pi(x(0)) \cdot \phi(p(0); 0, M) },
\]
where $\phi(p; 0, M)$ is the multivariate normal density with mean 0 and covariance $M$, evaluated at $p$.
Therefore, by Proposition~2 of \citet{park2020markov} or by Proposition~2 of \citet{neklyudov2020involutive}, Markov chains constructed by tempered HMC (Algorithm~\ref{alg:THMC}) is reversible and has the target $\pi$ as an invariant density.

\bibliographystyle{apalike}
\bibliography{ref}

\end{document}

\maketitle

%\abstract{This document provides additional information for \emph{Sampling from high-dimensional, multimodal distributions using automatically-tuned, tempered Hamiltonian Monte Carlo}.}

\tableofcontents

\newcounter{example}[section]
\renewcommand{\theexample}{\thesection.\arabic{example}}

%%%%%%%%%%%%%%%%%%%%%%%%%%%%

\section{Connection between tempered HMC (Algorithm~\ref{alg:THMC}) and the velocity scaling method by \citet[Section~5.5.7]{neal2011mcmc}}\label{sec:supp_neal2011}
In this section, we provide details on the connection between our tempered HMC algorithm (Algorithm~\ref{alg:THMC}) and \citet[Section~5.5.7]{neal2011mcmc}'s velocity scaling method.
Our method numerically simulates the dynamics described in \eqref{eqn:HEM_alpha_t} as described in Algorithm~\ref{alg:THMC_sim}.
Specifically, it iteratively carries out leapfrog steps with step sizes $\epsilon_{k-\frac{1}{2}} = e^{2a\eta_{k-\frac{1}{2}}} \bar \epsilon$ where $\bar\epsilon$ is a reference step size.
The optimal value for $a$ is $\frac{2}{\gamma+2}$, where $\gamma$ is the polynomial degree of the local growth of $U(x)$.

In Section~\ref{sec:equivalence}, we demonstrated the relationship bewteen the continuous-time Hamiltonian dynamics for $H_\alpha(x,p,t)$ and a continuous-time version of the velocity scaling method by considering a time scale change $d\breve t = \alpha^{-1/2} dt = e^{-\eta} dt$ and transformed momentum $\breve p = e^\eta p$.
We directly verify below that our numerical simulation scheme using the leapfrog step size $\epsilon_{k-\frac{1}{2}} = e^{\eta_{k-\frac{1}{2}}} \bar \epsilon$ (corresponding to $a = \tfrac{1}{2}$) is equivalent to the velocity scaling method simulating \eqref{eqn:ODE_p_scaling} using a constant leapfrog step size $\bar \epsilon$.
This implies that \citet{neal2011mcmc}'s velocity scaling method is optimal when $\gamma = 2$, but sup-optimal otherwise.

Numerical simulation of the Hamiltonian dynamics for $H_\alpha(x,p,t)$ is described by lines \ref{line:THMC_leapfrog1}--\ref{line:THMC_leapfrog3} of Algorithm~\ref{alg:THMC_sim}:
\begin{equation}
  \begin{split}
    &p(t_{k-1} + \tfrac{1}{2} \epsilon_{k-\frac{1}{2}}) = p(t_{k-1}) - \tfrac{1}{2} \epsilon_{k-\frac{1}{2}} \cdot \alpha_{k-\frac{1}{2}}^{-1} \frac{\partial U}{\partial x}(x(t_{k-1}))\\
    &x(t_{k-1} + \epsilon_{k-\frac{1}{2}}) = x(t_{k-1}) + \epsilon_{k-\frac{1}{2}} \cdot M^{-1} p(t_{k-1}+\tfrac{1}{2}\epsilon_{k-\frac{1}{2}})\\
    &p(t_{k-1} + \epsilon_{k-\frac{1}{2}}) = p(t_{k-1} + \tfrac{1}{2} \epsilon_{k-\frac{1}{2}}) - \tfrac{1}{2} \epsilon_{k-\frac{1}{2}} \cdot \alpha_{k-\frac{1}{2}}^{-1} \frac{\partial U}{\partial x}(x(t_{k-1} + \epsilon_{k-\frac{1}{2}})).
  \end{split}
  \label{eqn:THMC_leapfrog}
\end{equation}
Recall that we write $t_k = t_{k-1} + \epsilon_{k-\frac{1}{2}}$.
For $k=1,\dots, K$, we let
\[
  \breve p(t_{k-1}+\tfrac{1}{2}\epsilon_{k-\frac{1}{2}}) = e^{\eta_{k-\frac{1}{2}}} p(t_{k-1} + \tfrac{1}{2} \epsilon_{k-\frac{1}{2}}).
\]
Additionally, for $k=0,1,\dots, K$, we define
\[
  \begin{split}
    \breve p(t_k-) &= e^{\eta_{k-\frac{1}{2}}} p(t_k),\\
    \breve p(t_k+) &= e^{\eta_{k+\frac{1}{2}}} p(t_k).
  \end{split}
\]
Here $\breve p(t_k-)$ represents the momentum right after the $k$-th leapfrog step, and $\breve p(t_k+)$ the momentum right before carring out the $k+1$-st leapfrog step.
In between the $k$-th and $k+1$-st leapfrog steps, the momentum, or velocity, is scaled by a factor of $e^{\eta_{k+\frac{1}{2}} - \eta_{k-\frac{1}{2}}} = \xi^{\pm 1}$:
\[
  \breve p(t_k+) = e^{\eta_{k+\frac{1}{2}} - \eta_{k-\frac{1}{2}}} \breve p(t_k-).
\]
This momentum scaling corresponds to a piecewise-linear log-temperature schedule given by $\eta_k = \frac{2\eta_*}{K} \min(k, K-k)$, where
\[
  \eta_{k+\frac{1}{2}} - \eta_{k-\frac{1}{2}}
  = \begin{cases}
      \frac{2\eta_*}{K} = \log \xi & \text{ if } k\leq \frac{K-1}{2}\\
      0 & \text{ if } k = \frac{K}{2}\\
      \frac{-2\eta_*}{K} = -\log \xi & \text{ if } k \geq \frac{K+1}{2}\\
    \end{cases}
\]
Since
\[
  \breve p(t_{k-1} + \tfrac{1}{2} \epsilon_{k-\frac{1}{2}}) = e^{\eta_{k-\frac{1}{2}}} p(t_{k-1} + \tfrac{1}{2} \epsilon_{k-\frac{1}{2}})
\]
and
\[
  \breve p(t_{k-1}+) = e^{\eta_{k-\frac{1}{2}}} p(t_{k-1}),
\]
multiplying $e^{\eta_{k-\frac{1}{2}}}$ on both sides of the first and the third equations of \eqref{eqn:THMC_leapfrog}, we obtain
\[
  \begin{split}
    \breve p(t_{k-1} + \tfrac{1}{2} \epsilon_{k-\frac{1}{2}})
    &= \breve p(t_{k-1}+) - \frac{1}{2} e^{\eta_{k-\frac{1}{2}}} \epsilon_{k-\frac{1}{2}} \cdot \alpha_{k-\frac{1}{2}}^{-1} \frac{\partial U}{\partial x}(x(t_{k-1})) \\
    &= \breve p(t_{k-1}+) - \frac{1}{2} \bar\epsilon \cdot \frac{\partial U}{\partial x}(x(t_{k-1})), \\
    x(t_{k-1} + \tfrac{1}{2} \epsilon_{k-\frac{1}{2}})
    &= x(t_{k-1}) + \epsilon_{k-\frac{1}{2}} \cdot M^{-1} p(t_{k-1} + \tfrac{1}{2} \epsilon_{k-\frac{1}{2}}) \\
    &= x(t_{k-1}) + \bar \epsilon \cdot M^{-1} \breve p(t_{k-1} + \tfrac{1}{2} \epsilon_{k-\frac{1}{2}}),\\
    \breve p(t_k-)
    &= \breve p(t_{k-1}+\tfrac{1}{2}\epsilon_{k-\frac{1}{2}}) - \frac{1}{2} e^{\eta_{k-\frac{1}{2}}} \epsilon_{k-\frac{1}{2}} \cdot \alpha_{k-\frac{1}{2}}^{-1} \frac{\partial U}{\partial x}(x(t_k)) \\
    &= \breve p(t_{k-1}+\tfrac{1}{2}\epsilon_{k-\frac{1}{2}}) - \frac{1}{2} \bar\epsilon \cdot \frac{\partial U}{\partial x}(x(t_k)), \\
  \end{split}
\]
because $\epsilon_{k-\frac{1}{2}} = e^{\eta_{k-\frac{1}{2}}} \bar \epsilon$ and $\alpha_{k-\frac{1}{2}} = e^{2\eta_{k-\frac{1}{2}}}$.
After the $k$-th leapfrog step, the momentum is scaled by
\[
  \breve p(t_k+) = e^{\eta_{k+\frac{1}{2}} - \eta_{k-\frac{1}{2}}} \breve p(t_k-).
\]
The initial momentum is drawn from $p(0) \sim \N(0,M)$, but the first leapfrog step starts with $p(0+) = e^{\eta_{\frac{1}{2}}} \cdot p(0)$.
The last leapfrog step ends with $p(t_K-)$, but the final momentum used for checking acceptance of the proposal is obtained by $p(t_K) = e^{\eta_K - \eta_{K-\frac{1}{2}}} \breve p(t_K-) =  e^{-\eta_{K-\frac{1}{2}}} \breve p(t_K-)$.
These steps match the velocity scaling method described in \citet[Section~5.5.7]{neal2011mcmc}.

\section{Connection between the tempered transitions method by \citet{neal1996sampling} and our tempered Hamiltonian Monte Carlo}\label{sec:tempered_transitions_link}
In this section, we explain the connection between the tempered transitions method by \citet{neal1996sampling} and our tempered Hamiltonian Monte Carlo (THMC) method (Algorithm~\ref{alg:THMC}).
The tempered transitions method by applies a sequence of transition kernels having varied tempered distributions as stationary distributions, increasing the chance that the end state is on a different mode than the mode the initial state was in.
The sequence of states are denoted by 
\[
\hat x_0 \underset{\hat T_1}{\longrightarrow} \hat x_1 \underset{\hat T_2}{\longrightarrow} \cdots \hat x_{n-1} \underset{\hat T_n}{\longrightarrow} \bar x_n
\underset{\check T_{n}}{\longrightarrow} \check x_{n-1} \underset{\check T_{n-1}}{\longrightarrow} \cdots \check x_1 \underset{\check T_1}{\longrightarrow} \check x_0,
\]
where $\hat x_0$ is the initial state and $\check x_0$ is the final state, which becomes the proposed candidate for the next state of the Markov chain constructed by the method.
Denoting the density of the target distribution that we want to sample from by $\pi_0(x)$ and the density of the $k$-th auxiliary distribution by $\pi_k(x_k)$, the transition kernels $\hat T_1, \dots, \hat T_n, \check T_n, \dots, \check T_1$ are constructed such that the following relationship is satisfied:
\begin{equation}
\pi_k(x_k) dx_k \hat T_k(x_k, dx_k') = \pi_k(x_k') dx_k' \check T_k(x_k', dx_k).
\label{eqn:tempered_transitions_detailed_balance}
\end{equation}
The acceptance probability for the proposed candidate $\check x_0$ is given by
\begin{equation}
\min\left[ 1,\,
\frac{\pi_1(\hat x_0)}{\pi_0(\hat x_0)} \cdot \frac{\pi_2(\hat x_1)}{\pi_1(\hat x_1)} \cdot \cdots \cdot \frac{\pi_n(\hat x_{n-1})}{\pi_{n-1}(\hat x_{n-1})} \cdot \frac{\pi_{n-1}(\check x_{n-1})}{\pi_n(\check x_{n-1})} \cdot \cdots \cdot \frac{\pi_0(\check x_0)}{\pi_1(\check x_0)} \right].
\label{eqn:tempered_accprob}
\end{equation}
It can be shown that the density $\pi_0$ is invariant, for any choice of auxiliary densities $\{\pi_k; 1\leq k \leq n\}$.
In tempered transitions, the auxiliary distributions often represent the distribution at various temperature levels.
The choice $\pi_k(x) \propto \pi^{\beta_k}(x)$ is often used, where $\beta_k$ denotes the $k$-th inverse temperature, facilitating global mixing for multimodal distributions.

The kernels $\hat T_k$ and $\check T_k$ can be constructed using the Metropolis-Hastings (M-H) strategy as follows.
Consider a sequence of kernels $Q_k$, $1\leq k \leq 2n$, for the following proposals:
\begin{equation}
  \hat x_0 \underset{Q_1}{\longrightarrow} \hat x_1,~
  \hat x_1 \underset{Q_2}{\longrightarrow} \hat x_2,~
  \cdots,~
  \hat x_{n-1} \underset{Q_n}{\longrightarrow} \bar x_n,~
  \bar x_n \underset{Q_{n+1}}{\longrightarrow} \check x_{n-1},~
  \cdots,~
  \check x_1 \underset{Q_{2n}}{\longrightarrow} \check x_0.
\label{eqn:MH_tempered_transitions}
\end{equation}
The kernels $\hat T_k(\hat x_{k-1}; d\hat x_k)$, $1\leq k \leq n$, are constructed by a proposaling $\hat x_k$ using the kernel $Q_k$ and then accepting it with probability 
\begin{equation}
  \min\left(1, \frac{\pi_k(\hat x_k) d\hat x_k Q_{2n-k+1}(\hat x_k, d\hat x_{k-1})}{\pi_k(\hat x_{k-1})d\hat x_{k-1} Q_k(\hat x_{k-1}, d\hat x_k)}\right).
\label{eqn:tempered_proposal_MHratio1}
\end{equation}
The kernels $\check T_k(\check x_k, d\check x_{k-1})$, $1\leq k\leq n$, are constructed by proposing $\check x_{k-1}$ using $Q_{2n-k+1}$ and accepting it with probability
\begin{equation}
  \min\left(1, \frac{\pi_k(\check x_{k-1}) d\check x_{k-1} Q_{k}(\check x_{k-1}, d\check x_k)}{\pi_k(\check x_k) d\check x_k Q_{2n-k+1}(\check x_k, d\check x_{k-1})}\right).
\label{eqn:tempered_proposal_MHratio2}
\end{equation}
Here we understand $\hat x_n = \check x_n = \bar x_n$.
It is straightforward to check that \eqref{eqn:tempered_proposal_MHratio1} and \eqref{eqn:tempered_proposal_MHratio2} satisfy \eqref{eqn:tempered_transitions_detailed_balance}.

Consider a sequence of Monte Carlo draws, $\hat x_1, \dots, \hat x_{n-1}, \bar x_n, \check x_{n-1}, \dots, \check x_0$, obtained through the kernels $Q_1, \dots, Q_{2n}$ as described in \eqref{eqn:MH_tempered_transitions}, but without the intermediate Metropolis-Hastings acceptance/rejection steps.
The acceptance probability for the final state $\check x_0$ is then given by
\begin{equation}
  \min\left[1,\,
    \frac{\pi_0(\check x_0) d\check x_0 Q_1(\check x_0, d\check x_1) \cdots Q_n(\check x_{n-1}, d\bar x_n) Q_{n+1}(\bar x_n, d\hat x_{n-1}) \cdots Q_{2n}(\hat x_1, d\hat x_0)}
    {\pi_0(\hat x_0) d\hat x_0 Q_1(\hat x_0, d\hat x_1) \cdots Q_n(\hat x_{n-1}, d\bar x_n) Q_{n+1}(\bar x_n, d\check x_{n-1}) \cdots Q_{2n}(\check x_1, d\check x_0)} \right].
  \label{eqn:MHratio_combined}
\end{equation}
This sampling scheme can be compared with the tempered transitions method, where each intermediate draw is accepted or rejected with probability \eqref{eqn:tempered_proposal_MHratio1} or \eqref{eqn:tempered_proposal_MHratio2} and the final draw $\check x_0$ is accepted or rejected with probability \eqref{eqn:tempered_accprob}.
Indeed, the Metropolis-Hastings ratio in \eqref{eqn:MHratio_combined} is equal to the product of those in \eqref{eqn:tempered_proposal_MHratio1} and \eqref{eqn:tempered_proposal_MHratio2} for $1\leq k \leq n$ and the final M-H ratio in \eqref{eqn:tempered_accprob}.

Tempered Hamiltonian Monte Carlo (Algorithm~\ref{alg:THMC}) is equivalent to a method using a sequence of proposals kernels $Q_1, \dots, Q_{2n}$ that are given by deterministic maps
$Q_k(x_{k-1}, p_{k-1}; dx_k dp_k)$ where
\[
(x_k, p_k) = \Psi_{k-\frac{1}{2}}(x_{k-1}, p_{k-1}),
\]
as described by lines~\ref{line:THMC_leapfrog1}--\ref{line:THMC_leapfrog3} of Algorithm~\ref{alg:THMC_sim}.
Here the pair $(x_k, p_k)$ for $k=0,1,\dots, K$, with $K=2n$, are the intermediate states $\hat x_k$ or $\check x_k$ in \eqref{eqn:MH_tempered_transitions}.
We show that the Metropolis-Hastings ratio for this sequence of proposals is given by
\[
  \min\left[ 1, e^{-H(x_{2n}, p_{2n}) + H(x_0, p_0)} \right].
\]
To see this, consider a sequence of maps
\[
  (x_0, p_0) \underset{\Psi_{\frac{1}{2}}}\longrightarrow (x_1, p_1) \underset{\Psi_{\frac{3}{2}}}\longrightarrow \cdots \underset{\Psi_{2n-\frac{1}{2}}}\longrightarrow (x_{2n}, p_{2n}) \underset{\mathcal T}\longrightarrow (x_{2n}, -p_{2n}),
\]
where $\mathcal T$ is the momentum reflection operator.
Since we employ a symmetric temperature schedule satisfying $\eta_\kappa = \eta_{K-\kappa}$, we have $\Psi_\kappa = \Psi_{K-\kappa} = \Psi_{2n-\kappa}$ for $\kappa \in \{\tfrac{1}{2}, \dots, {K-\tfrac{1}{2}}\}$.
Thus we can see that, if we apply the same sequence of maps to $(x_{2n}, -p_{2n})$, we obtain the initial pair $(x_0, p_0)$:
\[
  (x_{2n}, -p_{2n}) \underset{\Psi_{2n-\frac{1}{2}}=\Psi_{\frac{1}{2}}}\longrightarrow (x_{2n-1}, -p_{2n-1}) \underset{\Psi_{2n-\frac{3}{2}}=\Psi_{\frac{3}{2}}}\longrightarrow \cdots \underset{\Psi_{\frac{1}{2}}=\Psi_{2n-\frac{1}{2}}}\longrightarrow (x_0, -p_0) \underset{\mathcal T}\longrightarrow (x_0, p_0).
\]
From this, we also see that $|dx_0 dp_0| = |dx_{2n} dp_{2n}|$.
Hence, the Metropolis-Hastings ratio is given by
\[
  \frac{e^{-H(x_{2n}, -p_{2n})}}{e^{-H(x_0,p_0)}} \left|\frac{dx_{2n} d(-p_{2n})}{dx_0 dp_0}\right|
  = e^{-H(x_{2n}, p_{2n}) + H(x_0, p_0)},
\]
since $H(x,p) = H(x,-p)$ by construction.

%%%%%%%%%%%%%%%%%%%%%%%%%%
%%% AUXILIARY STRATEGY %%%
%%%%%%%%%%%%%%%%%%%%%%%%%%
\section{Auxiliary strategy when the support is disconnected}\label{sec:relatedstrategies}

\begin{figure}
  \centering
  \includegraphics[width=.89\textwidth]{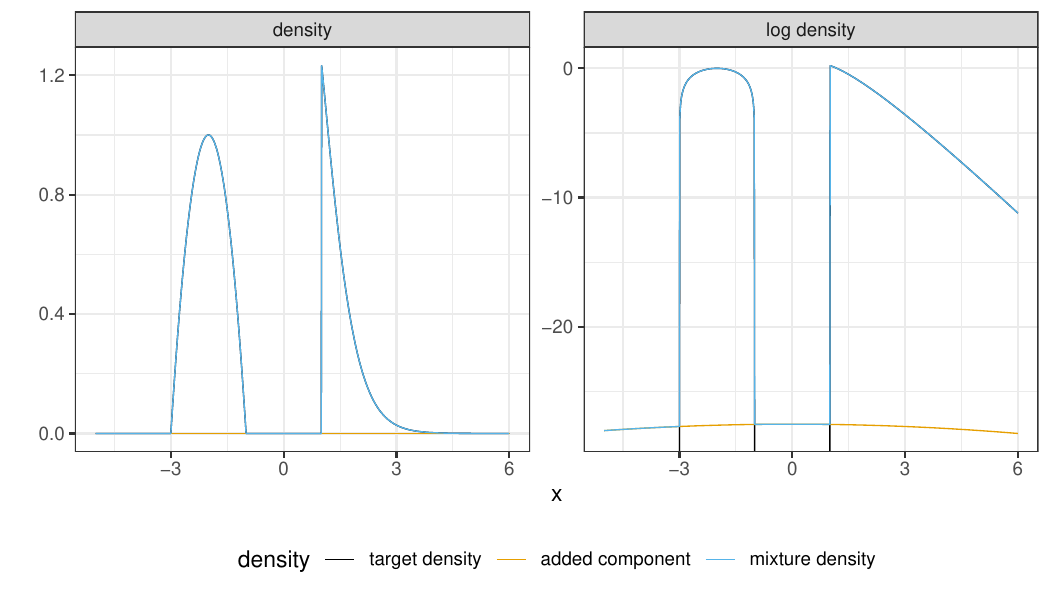}
  \caption{An illustrative diagram showing the effect of adding a small mixture component.
    The target density $\pi(x)$ and the mixture density $\pi^+(x)$ are visually indistinguishable on the natural (non-log) scale, because the weight of the added mixture component $\nu = e^{-25}$ is tiny.
    However, $\log\{\pi^+(x)\}$ is lower bounded by $\log\{\nu g(x)\}$, so finite everywhere.
  }
  \label{fig:mixture_target}
\end{figure}

In this section, we propose an auxiliary strategy for applying tempered HMC when the support of the target distribution is separated.
If the support of the target density $\text{supp}(\pi) := \{x \in \mathsf X ; \pi(x) > 0 \}$ has disconnected components, the Hamiltonian path started from one density component cannot reach other disconnected components, since the potential energy $U(x) = -\log \pi(x)$ is infinite on $\text{supp}(\pi)^c = \{x; \pi(x) = 0\}$.
%In this section, we consider an auxiliary strategy for Algorithm~\ref{alg:THMC} that enables jumps between the disconnected components.
We consider addition of a small mixture component to the unnormalized target density,
\[
  \pi^+(x) = \pi(x) + \nu g(x).
\]
Here $\pi(x)$ is the unnormalized target density function, $\nu$ a small positive constant, and $g(x)$ the probability density of the added mixture component.
We assume that $g(x)$ can be evaluated pointwise.
This mixture component is intended to bridge separated density components.
Figure~\ref{fig:mixture_target} shows $\pi(x)$, $\nu g(x)$, and the sum $\pi^+(x)$ where
\[
\begin{split}
  \text{supp}(\pi) &= (-3, -1) \cup (1,\infty),\\
  \pi^+(x) &= \pi(x) + e^{-25} \phi(x; 0, 5^2).
\end{split}
\]
Since $\log \pi^+(x)$ is lower bounded by $\log (\nu g(x))$, tempered HMC (Algorithm~\ref{alg:THMC}) can enable jumps between the two separated components of $\text{supp}(\pi)$.
Once a sample Markov chain has been constructed by Algorithm~\ref{alg:THMC} targeting $\pi^+(x)$, draws from the original target density $\pi(x)$ can be obtained using rejection sampling.
Provided that the states of the constructed Markov chain can be considered as draws from $\pi^+$, rejection sampling accepts a state $x$ with probability
\[
\frac{\pi(x)}{\pi(x) + \nu g(x)}.
\]
It can be readily checked that the accepted draws can be considered as samples from the original target density:
\[
\frac{\pi(x) + \nu g(x)}{Z+\nu} \cdot \frac{\pi(x)}{\pi(x) + \nu g(x)} \propto \frac{\pi(x)}{Z}.
\]
In order to reduce the number of draws lost by the rejection sampling, the mixture weight $\nu$ can be chosen such that $\nu g(x) \ll \pi(x)$ for typical posterior draws.
Section~\ref{sec:tuning} shows that Algorithm~\ref{alg:THMC} can enable jumps across a potential energy barrier whose height $\Delta$ scales exponentially with the maximum increase in the log-temperature schedule $\eta(t)$.
Thus for a typical posterior draw $x_1$ from $\pi$ and a point $x_2$ in $\text{supp}(\pi)^c$, we can choose $\nu$ such that the difference in the potential energy, which is approximately
\[
\log\pi^+(x_1) - \log\pi^+(x_2) \approx \log\pi(x_1) - \log (\nu g(x_2)),
\]
is on the order of $e^{\gamma a \eta_*}$.
Therefore, the mixture weight $\nu$ can be chosen exponentially small.
For such small values of $\nu$, it is likely that all MCMC draws from $\pi^+(x)$ are accepted, making the rejection sampling practically unnecessary.

The mixture density $g(x)$ should be chosen such that the graph of $\log g(x)$ is more flat than that of $\log \pi(x)$, so that the disconnected components of supp($\pi$) are bridged by $g(x)$.
However, if it is excessively flat, the constructed Hamiltonian paths may unnecessarily reach far beyond the region where most of the probability mass of $\pi(x)$ is placed.
A reasonable choice in practice may be to let $g(x)$ be the density of a normal distribution that covers most of the region where the support of $\pi(x)$ is expected to be located.

%
%%%%%%%%%%%%%%%%%%%%%%
\section{Hamiltonian Monte Carlo with trajectory bouncing}\label{sec:supp_bounce}
We describe a modified Hamiltonian Monte Carlo algorithm for target distributions with bounded support.
Specifically, we assume that the support can be expressed as $A_1\times A_2 \times \cdots \times A_d$ where $A_j$, $j\in 1\col d$, are intervals with finite upper bound, lower bound, or both.
We denote by $(x_j, p_j)$ the position-momentum pair for the $j$-th coordinate.

The modified HMC algorithm is identical to standard HMC, except in the way the trajectory is constructed.
Suppose that, after the half-step leapfrog update on the position, some components of $x$ lie outside the corresponding interval constaints.
Then the corresponding momentum components are negated, by multiplying them by $-1$.
One full leapfrog step with momentum bouncing is summarized in Algorithm~\ref{alg:HMC_reflect}.
This modified scheme is time reversible, like the original leapfrog method, so the resulting Markov chain is reversible with respect to the target distribution restricted to the specified bounds.
This momentum bouncing technique can also be incorporated into tempered HMC.

\begin{algorithm}[H]
  \SetKwInOut{Input}{Input}
  \Input{
    Position-momentum pair at time $t_{k-1}$, $(x(t_{k-1}), p(t_{k-1}))$;~
    Interval constraints, $A_j$, $j \in 1\col d$.
  }
  Let $p(t_{k-1} + \tfrac{1}{2} \epsilon) = p(t_{k-1}) - \tfrac{1}{2} \epsilon \cdot \frac{\partial U}{\partial x}(x(t_{k-1}))$\\
  Let $x(t_{k-1} + \tfrac{1}{2} \epsilon) = x(t_{k-1}) + \tfrac{1}{2} \epsilon M^{-1} p(t_{k-1} + \tfrac{1}{2}\epsilon)$\\
  \For {$j$ in 1\col d} {
    \If {$x_j(t_{k-1} + \tfrac{1}{2} \epsilon) \notin A_j$} {
      Let $p(t_{k-1} + \tfrac{1}{2} \epsilon) \gets -p(t_{k-1} + \tfrac{1}{2} \epsilon)$
    }
  }
  Let $x(t_{k-1} + \epsilon) = x(t_{k-1} + \tfrac{1}{2}\epsilon) + \frac{1}{2} \epsilon M^{-1} p(t_{k-1}+\tfrac{1}{2}\epsilon)$\\
  Let $p(t_{k-1} + \epsilon) = p(t_{k-1}+\tfrac{1}{2} \epsilon) - \tfrac{1}{2} \epsilon \cdot \frac{\partial U}{\partial x}(x(t_{k-1} + \epsilon))$\\\label{line:THMC_leapfrog3}
  \caption{A leapfrog step with bouncing at the boundary.}
  \label{alg:HMC_reflect}
\end{algorithm}

%
%%%%%%%%%%%%%%%%%%%%%%%%%%%
\section{Additional figures for Section~\ref{sec:comparison}}\label{sec:supp_compfig}
\FloatBarrier

\begin{figure}[tp]
\begin{knitrout}
\definecolor{shadecolor}{rgb}{0.969, 0.969, 0.969}\color{fgcolor}

{\centering \includegraphics[width=\maxwidth]{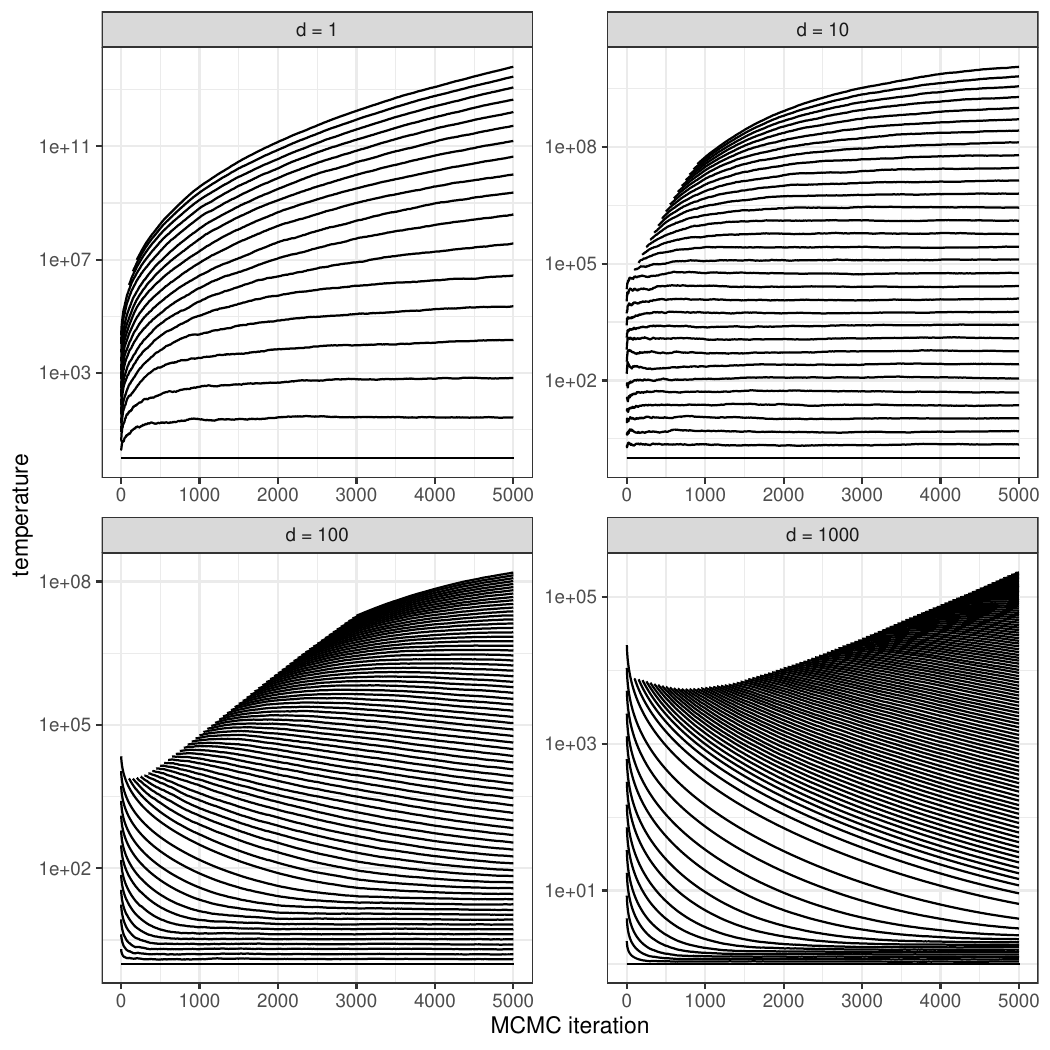} 

}

\end{knitrout}
\caption{Trace plots of the adaptively tuned temperature sequences in parallel tempering for the mixture of Gaussian models with isotropic covariances, considered in Section~\ref{sec:comparison}.}
\label{fig:pt_iso_tempseq}
\end{figure}

In this section, we provide additional supplementing Section~\ref{sec:comparison}, where we compared ATHMC with parallel tempering (PT) and tempered sequential Monte Carlo (TSMC).
Figure~\ref{fig:pt_iso_tempseq} displays the temperature levels adaptively tuned for the PT algorithm used to sample from a mixture of Gaussian models with isotropic covariances, $\Sigma_1 = \Sigma_2 = I$, and equal mixture weights, $w=0.5$, as considered in Section~\ref{sec:comparison}.

The tuning process began with 15 temperature levels.
The number of chains were adaptively tuned until the chain at the highest temperature level had visited both $(-\infty, -10000)$ and $(10000, \infty)$ in at least half of the $d$ coordinates during the past 100 MCMC iterations.
If this condition was not met, an additional chain was added above the current highest temperature chain every 50 iterations.
The process stopped once this search condition had been satisfied 100 times.

Figure~\ref{fig:pt_iso_tempseq} shows that the resulting temperature levels have approximately constant gaps on the log scale, although the spacing was not perfectly uniform.
As the dimension $d$ of the target distribution increased, the number of chains required after tuning also increased.
This is because the Metropolis-Hastings acceptance rate for swaps between adjacent chains tends to decrease with higher dimensionality, requiring finer temperature spacing.
Consequently, tuning the temperature sequence took longer for higher-dimensional targes.

Trace plots for $d=1, 10, 100$ show that the number of chains stabilized within the first 5000 iterations, whereas for $d=1000$, the tuning process extended beyond 5000 iterations, with noticeable adjustments still occurring at that point.
These results indicate that tuning the temperature sequence can be computationally intensive for high dimensional target distributions.

\begin{figure}[tp]
\begin{knitrout}
\definecolor{shadecolor}{rgb}{0.969, 0.969, 0.969}\color{fgcolor}

{\centering \includegraphics[width=\maxwidth]{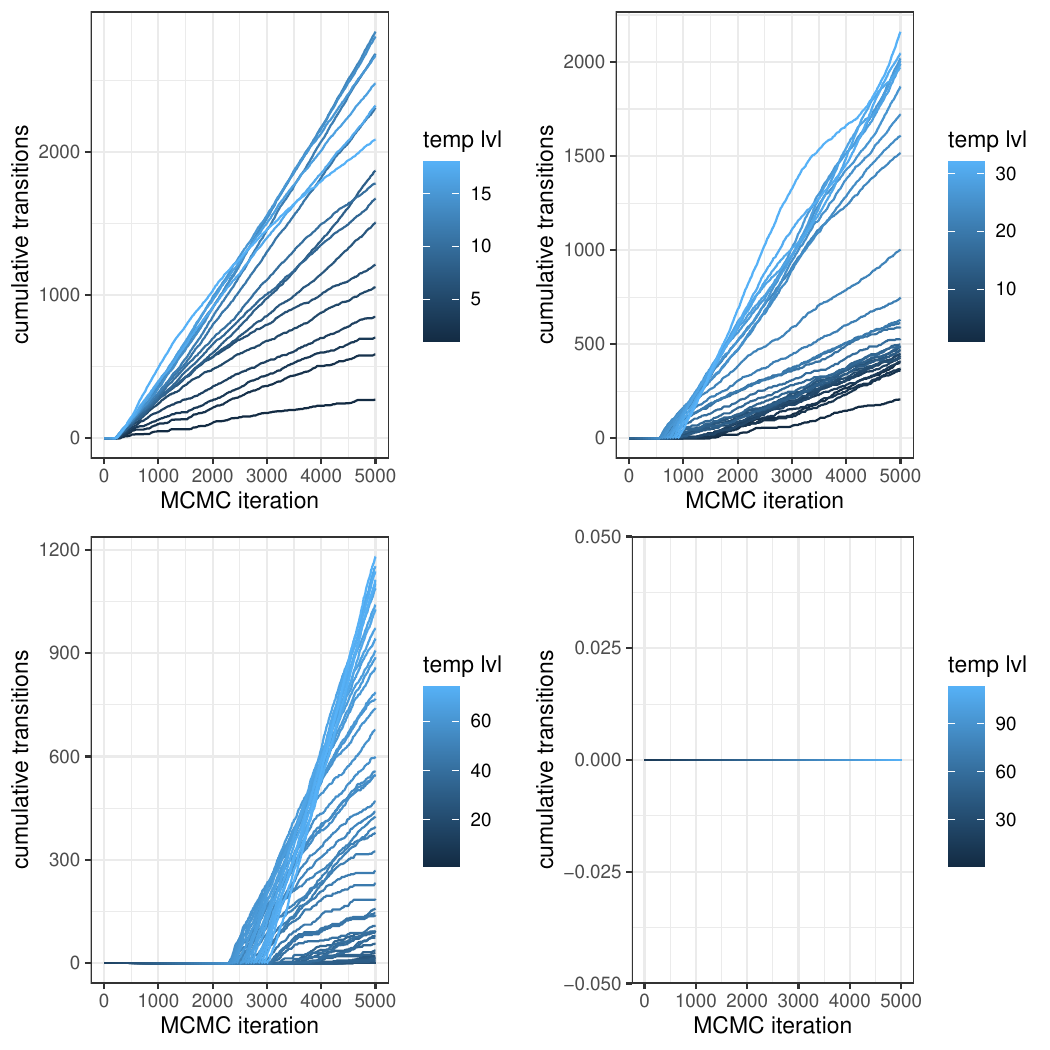} 

}

\end{knitrout}
\caption{Trace plots of the cumulative number of transitions for each parallel chain for the mixture of Gaussian models with isotropic covariances, considered in Section~\ref{sec:comparison}.}
\label{fig:pt_iso_cumtr}
\end{figure}

Figure~\ref{fig:pt_iso_cumtr} shows the cumulative number of transitions between the two dominant modes for each chain.
Chains at the higher temperature levels tend to exhibit the most frequent transitions, while those at lower temperature levels undergo transitions less frequentlly.
As the dimension increases, transitions between modes begin to occur later in the sampling process.
This delay indicates that successful sampling from both modes---evidenced by inter-mode transitions at the lowest temperature level---only occurs after the temperature sequence has been properly tuned.
For $d=1000$, tuning was not completed within the first 5000 iterations, and no inter-mode transitions occurred in any of the parallel chains during that period.

\begin{figure}[tp]
  \includegraphics[width=\textwidth]{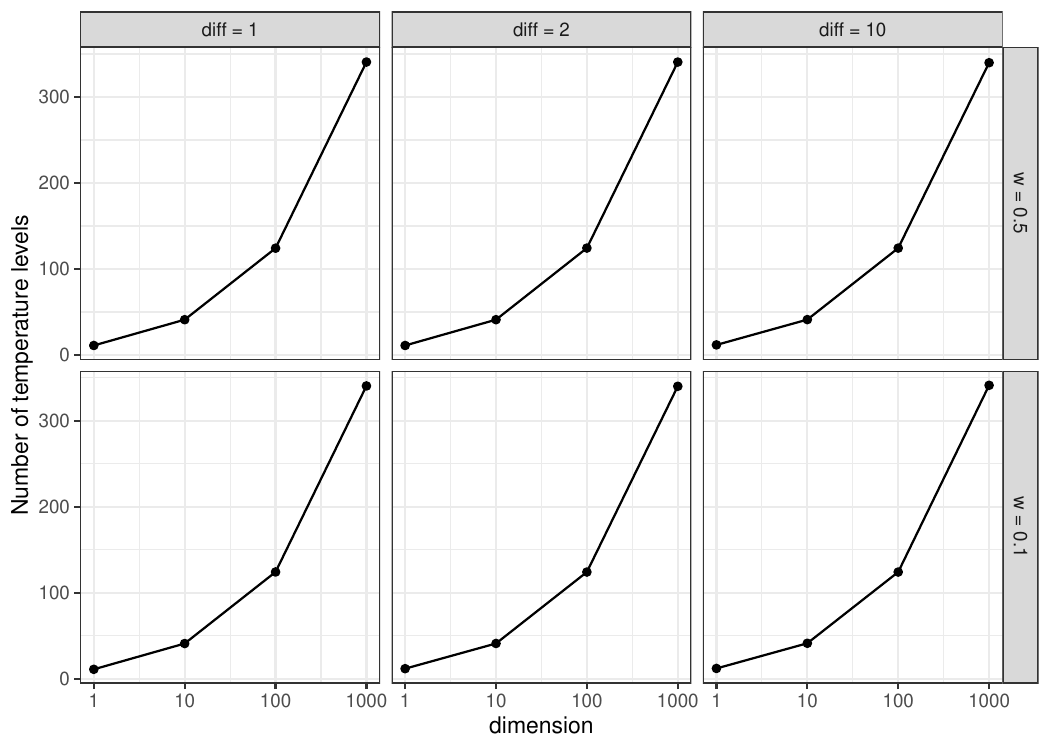}
  \caption{Number of temperature levels automatically selected by tempered SMC across different experimental settings.}
  \label{fig:supp_ntemps_tsmc_iso}
\end{figure}

Figure~\ref{fig:supp_ntemps_tsmc_iso} shows the number of temperature levels automatically selected by the adaptive TSMC scheme we employed for the mixture of Gaussian target distribution.
The number of temperature levels increased with the dimension, but its dependence on the mixture weight $w$ and the mode scale difference $c$ was minimal.

\begin{figure}[tp]
  \includegraphics[width=\textwidth]{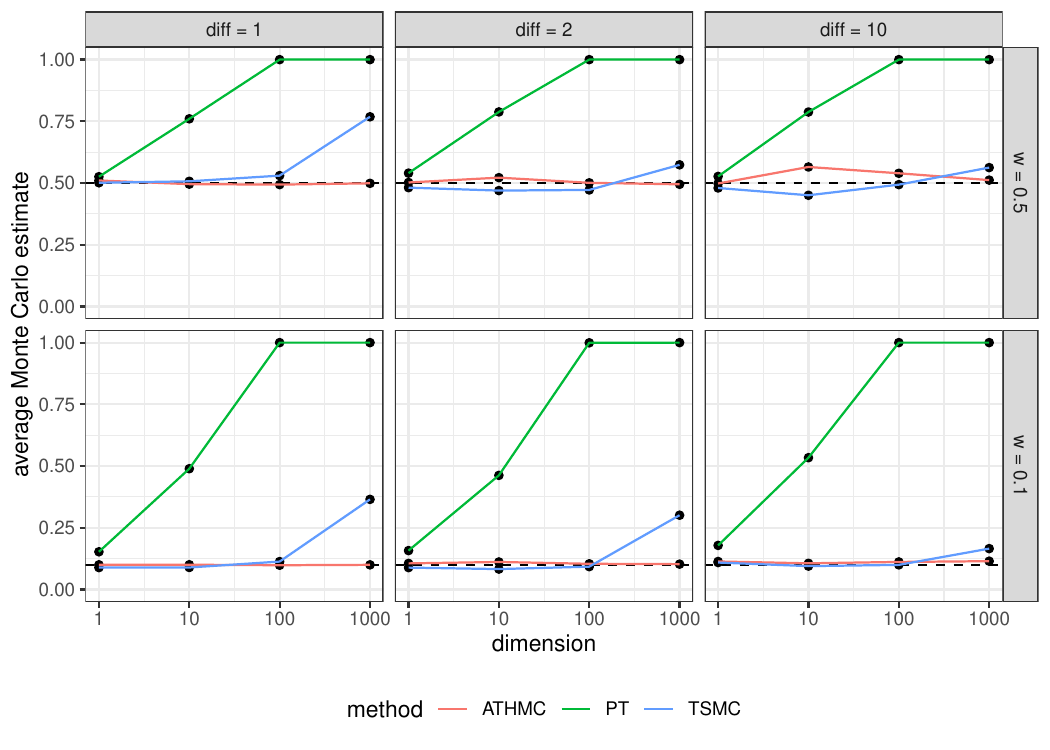}
  \caption{Average of 20 Monte Carlo estimates of $w$ from ATHMC, parallel tempering (PT), and tempered sequential Monte Carlo (TSMC).
  The true values of the mixture weight, $w$, are indicated by horizontal dashed lines. }
\label{fig:supp_comparison_iso_bias}
\end{figure}

Figure~\ref{fig:supp_comparison_iso_bias} displays the average Monte Carlo estimates of $w$ across the 20 replications.
All ATHMC estimates were close to the true values of $w$, whereas estimates from the other methods showed increasing bias as the dimension increased.

\begin{figure}[tp]
\begin{knitrout}
\definecolor{shadecolor}{rgb}{0.969, 0.969, 0.969}\color{fgcolor}

{\centering \includegraphics[width=\maxwidth]{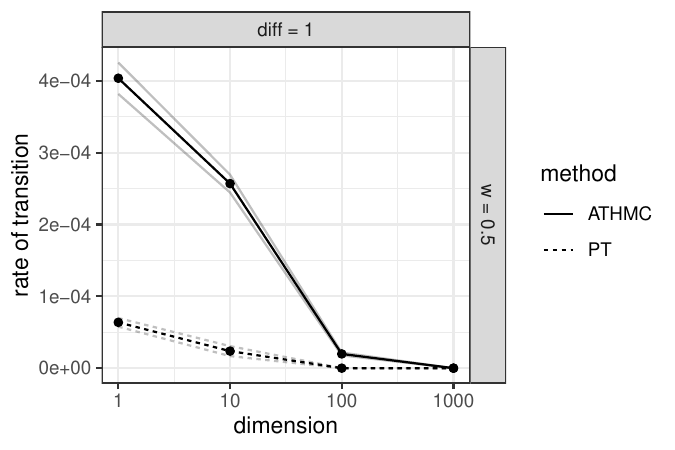} 

}

\end{knitrout}
\caption{Number of inter-mode transitions divided by the total number of leapfrog steps, for mixture of two Gaussian densities with anisotropic covariances.
  ATHMC (solid line) and parallel tempering with adaptive tuning (dashed line) are compared where the dimension $d$ varied from $1$ to $1000$.
  The average transition rate over 20 replications is shown, with $\pm 1$ standard deviation indicated by upper and lower bounding lines in light gray.
}
\label{fig:supp_athmc_pt_transition_rate_aniso}
\end{figure}

\begin{figure}[tp]
\begin{knitrout}
\definecolor{shadecolor}{rgb}{0.969, 0.969, 0.969}\color{fgcolor}

{\centering \includegraphics[width=\maxwidth]{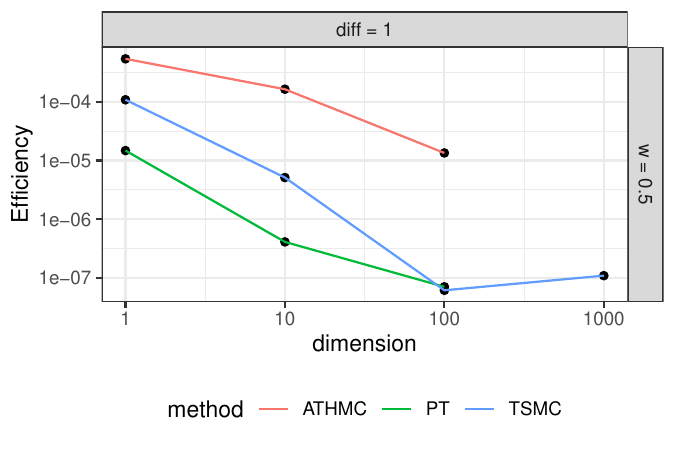} 

}

\end{knitrout}
\caption{Monte Carlo efficiency, evaluated as the effective number of draws divided by the total number of leapfrog steps, for ATHMC, PT, and TSMC for the anisotropic case.
  For ATHMC and PT, the efficiency for $d=1000$ was not computed, as there occurred no inter-mode transitions. }
\label{fig:supp_comparison_eff_aniso}
\end{figure}

We applied ATHMC, PT, and TSMC to a mixture of Gaussian target distributions where the covariance matrices $\Sigma_1$ and $\Sigma_2$ are anisotropic.
To reduce the computational cost of matrix multiplications, which scale as $O(d^3)$ with the dimension $d$, we aligned the principal components of both covariance matrices with the coordinate axes.
The inverse of the coordinate-wise variances for each matrix was drawn independently from a chi-squared distribution with 10 degrees of freedom.
To ensure that both modes had the same overall scale, the coordinate-wise variances were rescaled so that both covariance matrices had determinant 1.
The dimension varied over $d=1, 10, 100$ and $1000$, while the distance between the two modes was fixed at 10000, and the mixture weight set to $w=0.5$.

Figure~\ref{fig:supp_athmc_pt_transition_rate_aniso} shows the transition rates for ATHMC and PT under this anisotropic setting.
Each method was run for 5000 MCMC iterations.
For $d=1$ and $10$, the transition rates for ATHMC were substantially higher than those for PT.
For $d=100$, while PT exhibited no inter-mode transitions, ATHMC achieved meaningfully frequent transitions.
At $d=1000$, neither methods produced any transitions between modes.

Similarly to the isotropic case, we evaluated Monte Carlo efficiency for ATHMC, PT, and TSMC by dividing the effective number of draws by the number of leapfrog steps.
The effective number of draws was computed as $w(1-w)$ divided by the MSE, which was estimated using 20 replications.
We used Markov chains of length 5000 for ATHMC and PT, and an ensemble of 5000 particles for TSMC.
However, efficiency was not computed for ATHMC and PT in dimension $d=1000$ because there were no transitions between modes.
Figure~\ref{fig:supp_comparison_eff_aniso} shows the efficiency evaluated for the three methods.
The efficiency for ATHMC was substantially higher than that of the other two methods up to $d=100$.

\FloatBarrier

%

%%%%%%%%%%%%%%%%%%%%%%%%%%%%
\section{Additional figures for Section~\ref{sec:applications}}\label{sec:supp_applfig}
\subsection{Additional figures for Section~\ref{sec:Bayesian_mixture}}\label{sec:supp_bayesmixfig}
\FloatBarrier
\begin{figure}[tp]
\begin{knitrout}
\definecolor{shadecolor}{rgb}{0.969, 0.969, 0.969}\color{fgcolor}

{\centering \includegraphics[width=\maxwidth]{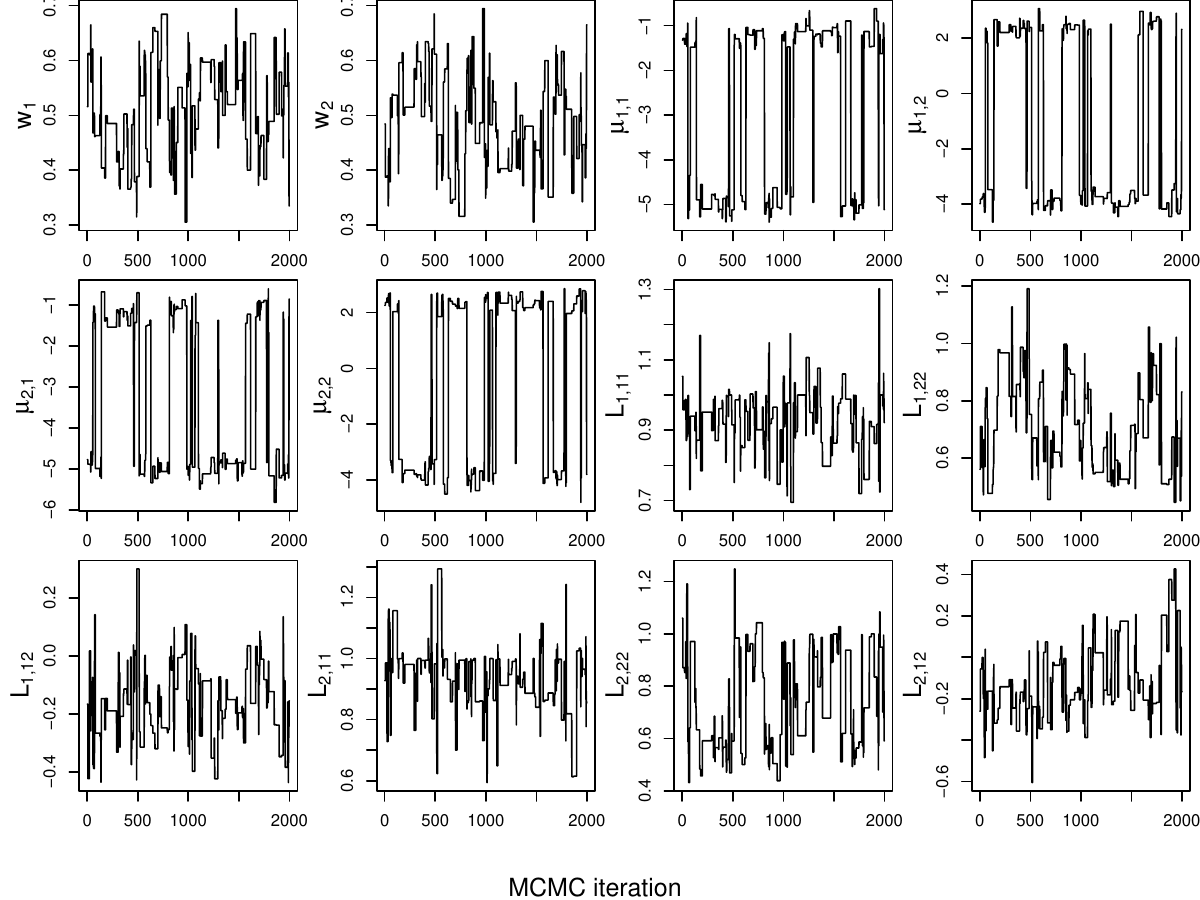} 

}

\end{knitrout}
\caption{Trace plots for the model parameters in the Bayesian normal mixture model considered in Section~\ref{sec:Bayesian_mixture}.}
\label{fig:bayesmix_traceplots}
\end{figure}

\begin{figure}[tp]
\begin{knitrout}
\definecolor{shadecolor}{rgb}{0.969, 0.969, 0.969}\color{fgcolor}

{\centering \includegraphics[width=\maxwidth]{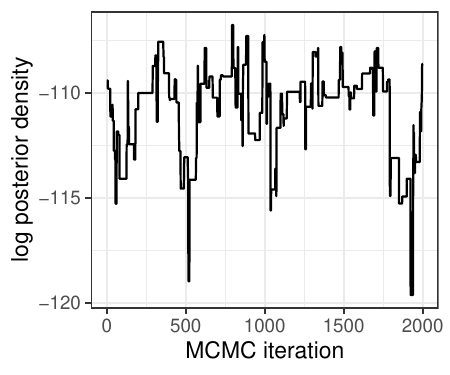} 

}

\end{knitrout}
\caption{Trace plot of the log-posterior density from the Markov chain generated using ATHMC for the Bayesian mixture model.}
\label{fig:bayesmix_logposterior}
\end{figure}

\FloatBarrier

\subsection{Additional figures for Section~\ref{sec:ising}}\label{sec:supp_ising}
\FloatBarrier

\begin{figure}[tp]
\begin{knitrout}
\definecolor{shadecolor}{rgb}{0.969, 0.969, 0.969}\color{fgcolor}

{\centering \includegraphics[width=\maxwidth]{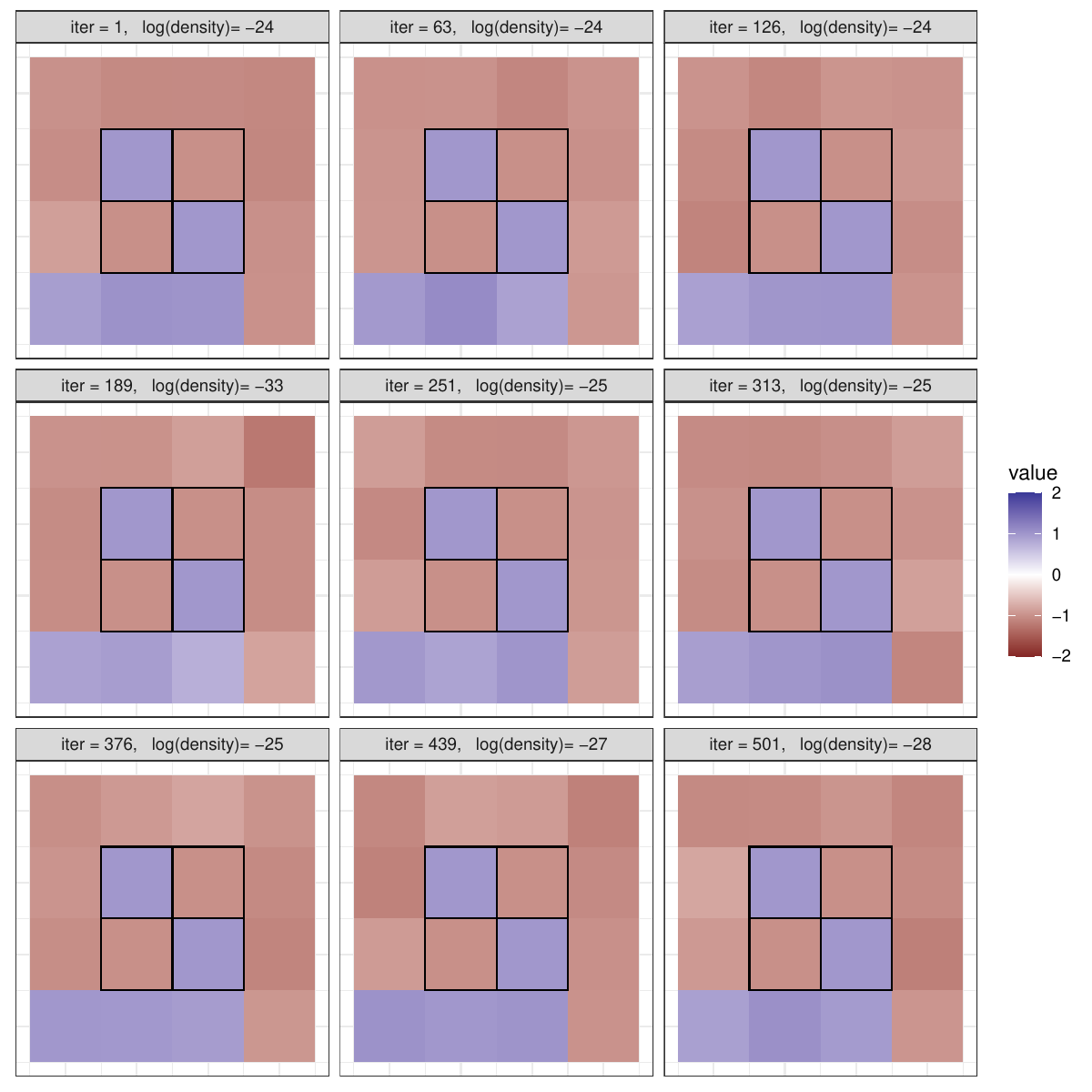} 

}

\end{knitrout}
\caption{Spin configurations sampled using standard HMC for the Ising model.}
\label{fig:ising_hmc}
\end{figure}

In Section~\ref{sec:ising}, we considered an ising model on a spin lattice comprising 16 sites.
When sampled using ATHMC, 48 changes in the signs of the spins were observed over 500 iterations.
In contrast, when HMC was used without tempering, no sign changes occurred.
Figure~\ref{fig:ising_hmc} shows nine spin configurations sampled using standard HMC over 500 iterations, all of which exhibit the same pattern.

\FloatBarrier

\subsection{Additional figures for Section~\ref{sec:sensor_localization}}\label{sec:supp_sensorfig}
\FloatBarrier

\begin{figure}
\begin{knitrout}
\definecolor{shadecolor}{rgb}{0.969, 0.969, 0.969}\color{fgcolor}

{\centering \includegraphics[width=\maxwidth]{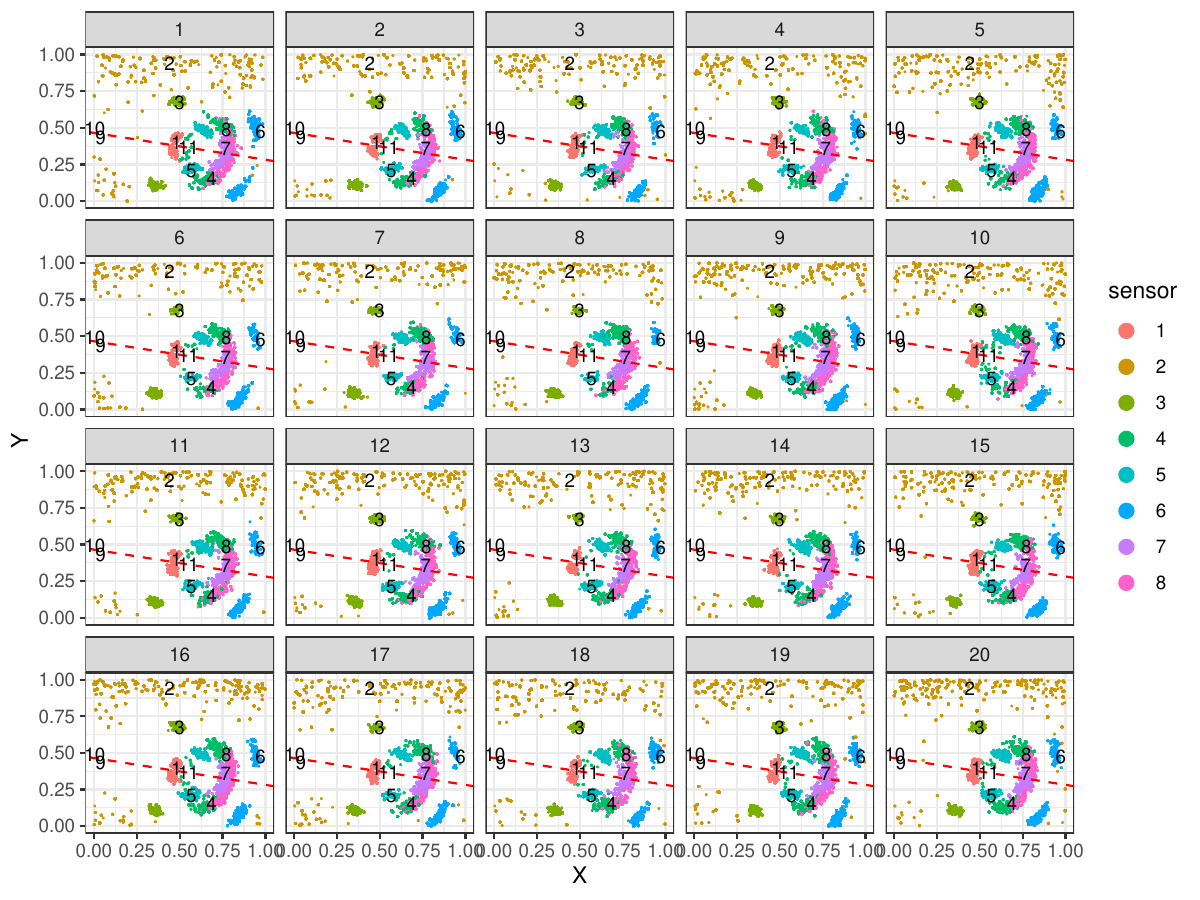} 

}

\end{knitrout}
\caption{
  The marginal sample points for the eight unknown sensor locations in 20 Markov chains constructed by ATHMC for the sensor self-localization example with the posterior density given by \eqref{eqn:sensor_posterior}.
  The true location of each sensor is marked by its number ID.
  A line approximately connecting the three sensors of known locations (9, 10, 11) is marked by a red dashed line.
  The true posterior density should be bimodal, where the marginal sensor locations for the two modes should be approximately symmetric with respect to the red dashed line.
  %The parameters $R=0.3$ and $\sigma_e=0.02$ were assumed to be known.
}
\label{fig:sensor_marginal_ATHMC_only}
\end{figure}

\begin{figure}
\begin{knitrout}
\definecolor{shadecolor}{rgb}{0.969, 0.969, 0.969}\color{fgcolor}

{\centering \includegraphics[width=\maxwidth]{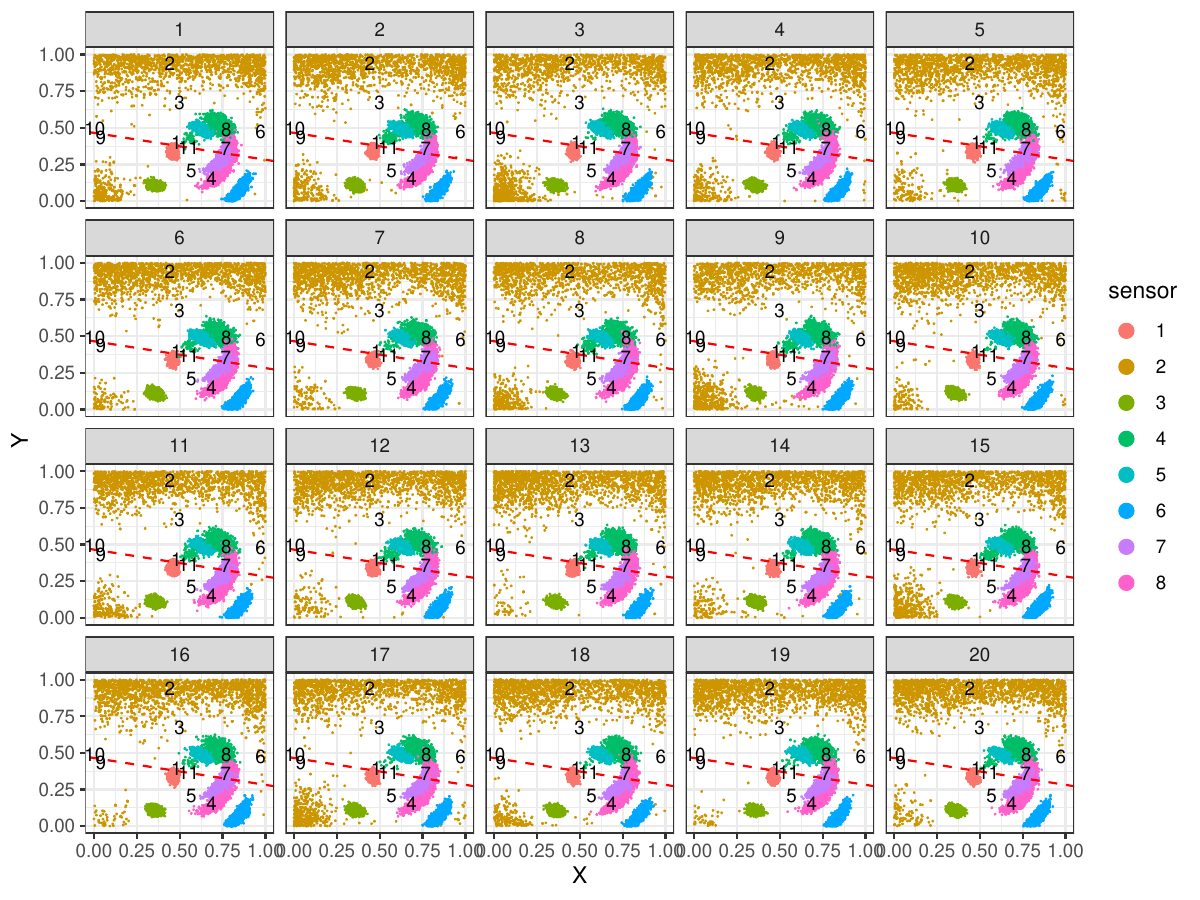} 

}

\end{knitrout}
\caption{
  The marginal sample points for the eight unknown sensor locations in 20 Markov chains constructed by standard HMC for the sensor self-localization example with the posterior density given by \eqref{eqn:sensor_posterior}.
  %The parameters $R=0.3$ and $\sigma_e=0.02$ were assumed to be known.
  The true posterior density should be bimodal, but all 20 Markov chains remains confined to a single mode (compare with Figure~\ref{fig:sensor_marginal_ATHMC_only}.)
}
\label{fig:sensor_marginal_HMC_only}
\end{figure}

In this supplementary section, we provide some additional details and figures for Section~\ref{sec:applications}.
In Section~\ref{sec:applications}, we considered a sensor network self-localization problem where the locations of eight sensors with unknown positions are estimated by noisy, pairwise distances between the sensors.

Figure~\ref{fig:sensor_marginal_ATHMC_only} shows the estimated positions for the eight sensors obtained in 20 Markov chains generated by tempered HMC with adaptive tuning.
The posterior distribution is strongly multimodal, and all 20 chains correctly visit both modes.
To see this, note that there are two isolated point clouds for both sensor \#3 (dark green) and for sensor \#6 (blue), roughly symmetric about the dashed line passing through the three sensors of known locations.
Figure~\ref{fig:sensor_marginal_HMC_only} shows the estimated positions for the same eight sensors obtained by 20 Markov chains independently constructed by running standard Hamiltonian Monte Carlo.
All 20 chains explored only a single mode, as can be noted by the fact that there is a single point cloud for each sensor \#3 or \#6.
These results show that our ATHMC facilitates transitions between isolated modes, whereas standard HMC does not.
The fact that the point clouds look relatively denser in Figure~\ref{fig:sensor_marginal_HMC_only} for chains constructed by standard HMC is because the acceptance probability was on average higher for chains constructed by standard HMC than those by ATHMC.
The ATHMC method focuses on making global transitions between isolated modes, and thus the average acceptance probability is relatively low.
However, local explorations within each mode can be readily carried out by complementing ATHMC by occasionally employing standard HMC kernels.

%
%$$
Next, we assumed $R$ and $\sigma_e$ to be unknown and estimated them jointly with the unknown sensor locations.
We used ATHMC within the Gibbs sampler.
For each iteration of the Gibbs sampler, the sensor locations $\mathbf x_{1:8}$ were updated by running one iteration of ATHMC targeting $\pi(\mathbf x_{1:8}|R, \sigma_e)$, and then $R$ and $\sigma_e$ were updated using standard HMC, targeting $\pi(R|\mathbf x_{1:8}, \sigma_e)$ and $\pi(\sigma_e | \mathbf x_{1:8}, R)$, respectively.
%The density $\pi(\mathbf x_{1:8}|R, \sigma_e)$ targeted by ATHMC changed over the iterations of the Gibbs sampler.

\begin{figure}
\begin{knitrout}
\definecolor{shadecolor}{rgb}{0.969, 0.969, 0.969}\color{fgcolor}

{\centering \includegraphics[width=\maxwidth]{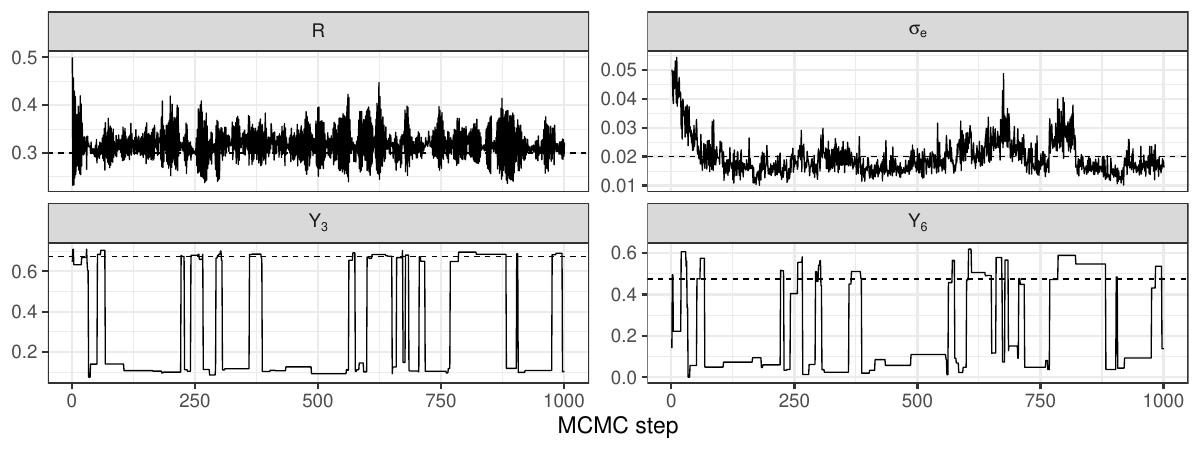} 

}

\end{knitrout}
\caption{
  The traceplots for $R$, $\sigma_e$, and the $y$-coordinates of the third and the sixth sensors for a chain constructed by ATHMC within Gibbs.
  The values of $R$ and $\sigma_e$ were considered unknown.
  The true value for each variable is indicated by a dashed horizontal line.
}
\label{fig:sensor_traceplots_gibbs}
\end{figure}

We constructed twelve independent chains targeting the joint posterior distribution for $\mathbf x$, $R$, and $\sigma_e$ using ATHMC within the Gibbs sampler.
The prior distributions for $R$ and $\sigma_e$ were given by the exponential distributions with rates $1/0.5$ and $1/0.05$, respectively.
Figure~\ref{fig:sensor_traceplots_gibbs} shows the traceplots of $Y_3$, $Y_6$, $R$, and $\sigma_e$ for one of the twelve constructed chains.
The sample draws for $Y_3$ and $Y_6$ show that both posterior modes were visited frequently.
The sample draws for $R$ and $\sigma_e$ were close to the values we used to generate the data.

\FloatBarrier
%

%%%%%%%%%%%%%%%%%%%%%%%%%%
%%% MASS-ENHANCED HMC %%%
%%%%%%%%%%%%%%%%%%%%%%%%%%
\section{HMC with fixed, increased temperature (Algorithm~\ref{alg:enhancedHMC})}\label{sec:supp_enhancedHMC}
In Section~\ref{sec:motivation} of the main text, we briefly discussed the idea of using constant, but increased temperature $\alpha>1$ for simulating paths in HMC.
In the current section, we delve into this approach, summarized in Algorithm~\ref{alg:enhancedHMC}.
This method simulates the Hamiltonian dynamics for the modified Hamiltonian
\[
  H_\alpha(x,p) = \frac{1}{2} p^\top M^{-1} p + \alpha^{-1} U(x).
\]
Each pair $(x(n\epsilon), p(n\epsilon))$ is accepted if $\Lambda < \exp(-H\{x(n\epsilon), p(n\epsilon)\} + H\{x(0), p(0)\})$, where $\Lambda$ is a Uniform$(0,1)$ draw.
Note that the acceptance probability depends on the original Hamiltonian $H(x,p)$, not the modified Hamiltonian $H_\alpha(x,p)$.
Here, the single draw $\Lambda$ is used for check acceptance for \emph{all} pairs.
This is the key difference of the sequential-proposal strategy proposed by \citet{park2020markov} compared to the standard Metropolis-Hastings strategy.
\citet{park2020markov} gives a proof of the fact that the original target distribution with unnormalized density $\pi(x)$ is invariant for the resulting Markov chain.

The use of the sequential-proposal strategy is critical for Algorithm~\ref{alg:enhancedHMC}, because, unlike standard HMC, the acceptance probability varies greatly along the path.
The original Hamiltonian $H(x,p) = K(p) + U(x) = \tfrac{1}{2} p^\top M^{-1} p + U(x)$ varies along the path approximately as follows:
\begin{equation}
\begin{split}
  H(x(n\epsilon), p(n\epsilon))
  &= K(p(n\epsilon)) + U(x(n\epsilon)) \\
  &= K(p(n\epsilon)) + \alpha^{-1} U(x(n\epsilon)) + (1-\alpha^{-1}) U(x(n\epsilon)) \\
  &\approx K(p(0)) + \alpha^{-1} U(x(0)) + (1-\alpha^{-1}) U(x(n\epsilon)) \\
  &= K(p(0)) + U(x(0)) + (1-\alpha^{-1}) \{ U(x(n\epsilon)) - U(x(0)) \}.
\end{split}
\label{eqn:ham_increment}
\end{equation}
Here we use the fact that the modified Hamiltonian $H_\alpha(x,p) = K(p) + \alpha^{-1} U(x)$ is approximately conserved along the path.
Equation~\ref{eqn:ham_increment} implies that if $\alpha \gg 1$, the amount of change in Hamiltonian $H$ is close to the change in the potential energy.
Since $\alpha > 1$, the trajectory may reach high potential energy regions that are rarely visited under the original target distribution.
When the trajectory stays in these regions, $U(x(n\epsilon)) - U(x(0))$ is large positive, and \eqref{eqn:ham_increment} indicates that the acceptance probability for $x(n\epsilon)$ is exponentially small.
The sequential-proposal strategy continues path simulation until a low potential energy region is reached where acceptable states are found.

This method (Algorithm~\ref{alg:enhancedHMC}) can be successful for low-dimensional multimodal target distributions (say $d\leq 5$) or for some special high-dimensional distributions such as a mixture of Gaussian distributions where all mixture components have the same covariance matrix $\Sigma$.
However, in general, this approach can be ineffective in high dimensions, since the simulated trajectory with increased temperature visits regions of low potential energy only rarely.
For this reason, we developed our tempered Hamiltonian Monte Carlo algorithm (Algorithm~\ref{alg:THMC}) that gradually increases and then decreases the mass with which the Hamiltonian dynamics is simulated, so that at the end of the path the particle may settle down at a low potential energy region.

\subsection{Examples: mixtures of multivariate normal distributions} \label{sec:multimodalMVN}
\begin{figure}
  \centering
  \captionsetup{width=0.89\linewidth}
  \includegraphics[width=0.89\linewidth]{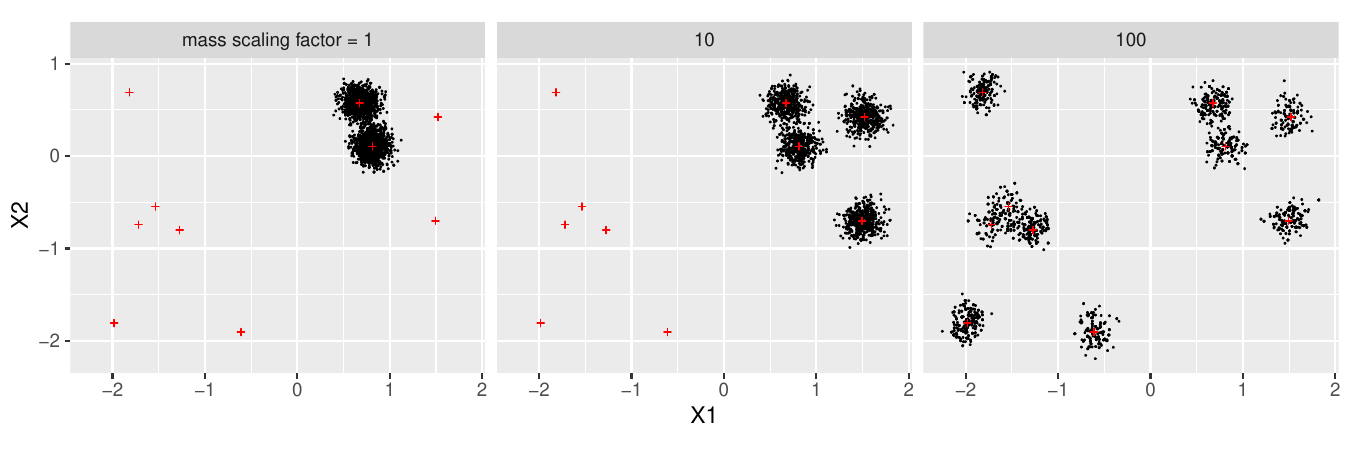}
  \caption{Sample draws of Markov chains constructed by HMC with constant, increased temperature (Algorithm~\ref{alg:enhancedHMC} with various temperature levels ($\alpha=1$, 10, and 100) for the target distribution considered in Example~\ref{ex:enhanced_bimodal_2d_alpha}.
    The centers of the ten Gaussian components are marked by red '+' signs.
  }
  \label{fig:mvnorm_2d_sample_pts}
\end{figure}
Here we demonstrate how Algorithm~\ref{alg:enhancedHMC} performs for mixtures of normal distributions.
Examples~\ref{ex:enhanced_bimodal_2d_alpha} and \ref{ex:enhanced_bimodal_dim} show that the method can construct globally mixing Markov chains for mixtures of high-dimensional normal distributions in the special case where all mixture components have the same covariance that is equal to the mass matrix $M$.
Example~\ref{ex:enhanced_unimodal_misspecified_M} show that in general, however, the method may fail if the mixture covariances are different.

We first note that if every mixture component has the same covariance given by $\Sigma$, then by choosing the mass matrix $M$ equal to $\Sigma^{-1}$, one can effectively turn the target distribution into a mixture of normal components with a shared covariance $I$.
To see this, consider a transformation of variables
\begin{equation}
  x' := \Sigma^{-1/2} x, \qquad p' := \Sigma^{1/2} p
  \label{eqn:linear_transformation_xp}
\end{equation}
where $\Sigma^{-1/2}$ is the inverse of the symmetric square-root matrix of $\Sigma$.
This linear transformation make the target distribution a mixture of normal components with covariance $I$.
These transformed variables $(x', p')$ satisfy the Hamiltonian equation of motion with unit mass, as follows:
\[
\begin{split}
  \frac{dx'}{dt} &= \Sigma^{-1/2} \frac{dx}{dt} = \Sigma^{-1/2} M^{-1} p = \Sigma^{1/2} p,\\
  \frac{dp'}{dt} &= \Sigma^{1/2} \frac{dp}{dt} = -\Sigma^{1/2} \frac{\partial U}{\partial x} = - \frac{\partial U}{\partial x'}.
\end{split}
\]
Thus for the case where all mixture components have the same covariance, we can without loss of generality assume that the common covariance is equal to $I$.

\paragraph{Example~\theexample} \refstepcounter{example} \label{ex:enhanced_bimodal_2d_alpha}
Figure~\ref{fig:mvnorm_2d_sample_pts} shows the sample draws for a two-dimensional mixture distribution of ten Gaussian components.
The target density is given by
\[
  \pi(x) = \frac{1}{10}\sum_{j=1}^{10} \phi(x; \mu_j, \sigma^2 I), \quad x \in \mathsf R^2,
\]
where the centers of the density components $\mu_j$, $j\in 1\col 10$, are distributed randomly across the square $[-2,2]^2$ as shown in Figure~\ref{fig:mvnorm_2d_sample_pts}.
The standard deviation of each component $\sigma$ was 0.1.
We constructed Markov chains of length two thousand using Algorithm~\ref{alg:enhancedHMC} with three different values of the mass scale factors, $\alpha = 1$, 10, and 100.
The leapfrog step size varied linearly with the square root of the mass scale factor ($\epsilon = 0.1\cdot \sqrt{\alpha}$).
The first acceptable proposal was taken for the next state of the Markov chain (i.e., $N=1$), and each iteration was terminated if no acceptable proposal was found among the first $N_{\max}=10$ proposals.
The covariance matrix for the momentum distribution was equal to the identity matrix ($M=I$).

The numerical results shown in Figure~\ref{fig:mvnorm_2d_sample_pts} show that using temperature $\alpha>1$ enables jumps between separated density components.
When we run sequential-proposal HMC \emph{without} tempering (i.e., $\alpha=1$), the Markov chain could reach only one density component nearest to the initial component.
When $\alpha=10$, the sample Markov chain reached more density components but not remotely separated ones.
When $\alpha=100$, the chain reached all components.
\hfill$\Box$

\begin{figure}
  \centering
  \captionsetup{width=0.89\linewidth}
  \begin{subfigure}[b]{0.89\linewidth}
    \centering
    \includegraphics[width=\linewidth]{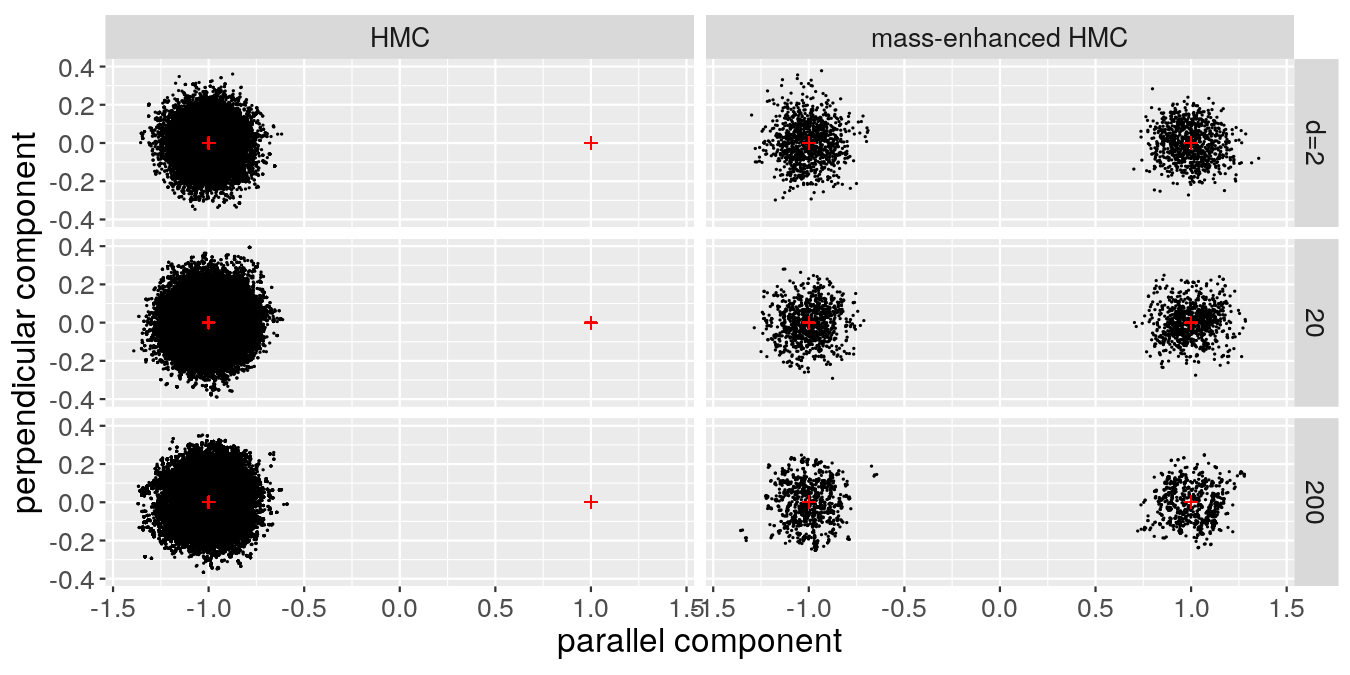}
    \caption{Sample draws of the constructed Markov chains.
      The $x$-axis shows the projection onto the direction from one density mode to the other (i.e., $\mu_{2,d}-\mu_{1,d}$), and the $y$-axis shows the projection onto a randomly selected direction perpendicular to $\mu_{2,d}-\mu_{1,d}$.
      The experimental settings are the same as those in Figure~\ref{fig:mvnorm_d_closest_comp}.
    }
    \label{fig:mvnorm_d_parallel_perp}
  \end{subfigure}

  \vspace{3ex}
  \begin{subfigure}[b]{0.89\linewidth}
    \includegraphics[width=\linewidth]{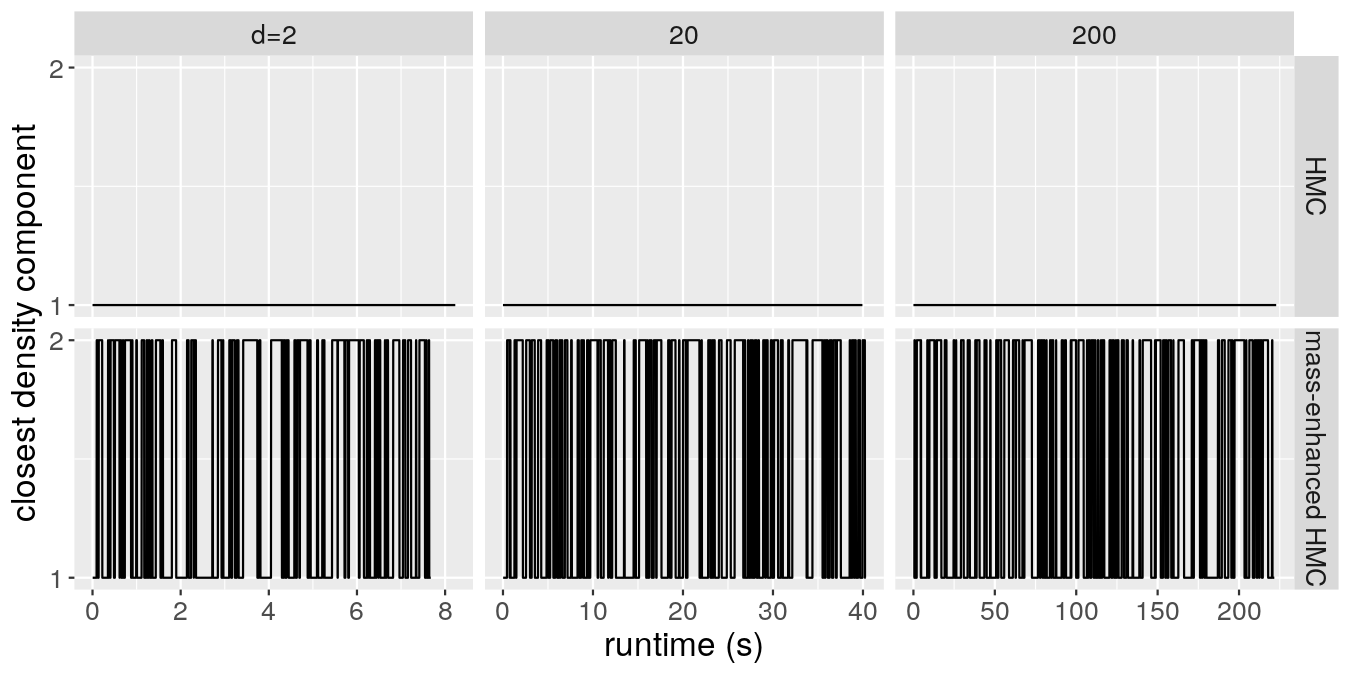}
    \caption{Trace plots of the closest density component (1 or 2) for standard HMC (top) and HMC with $\alpha=100$ (Algorithm~\ref{alg:enhancedHMC}).
      Algorithm~\ref{alg:enhancedHMC} ran for 2000 iterations, and standard HMC was run for approximately the same amount of time.
      The leapfrog step size of 0.1 was used for both methods.
    }
    \label{fig:mvnorm_d_closest_comp}
  \end{subfigure}
  \caption{Markov chains constructed by HMC and mass-enhanced HMC (Algorithm~\ref{alg:enhancedHMC}) where the target density is given by \eqref{eqn:d_bimodal_MVN} ($d=2$, 20, 200).}
  \label{fig:mvnorm_d}
\end{figure}
\paragraph{Example~\theexample} \refstepcounter{example} \label{ex:enhanced_bimodal_dim}
Next we demonstrate how Algorithm~\ref{alg:enhancedHMC} scales with increasing space dimension $d$.
Since it is sufficient to demonstrate that the algorithm facilitates jumps from one density component to another, we set the target distribution to be a mixture of two Gaussian components where the distance between the center of the components and the standard deviation of both components are fixed:
\begin{equation}
  \pi_d(x_d) = \frac{1}{2} \sum_{j=1}^2 \phi(x_d; \mu_{j,d}, \sigma^2 I_d), \quad x_d \in \mathsf R^d,
  \label{eqn:d_bimodal_MVN}
\end{equation}
where $\Vert \mu_{1,d} - \mu_{2,d} \Vert = 2$, $\sigma \equiv 0.1$ for $d = 2$, 20, and 200.
The directions of $\mu_{1,d}-\mu_{2,d}$ were randomly generated from spherically uniform distributions.
We ran both a standard HMC and HMC with $\alpha=100$ (Algorithm~\ref{alg:enhancedHMC}) for these target distributions.
For HMC with increased temperature, the number of acceptable states found in each iteration was $N=1$, and the maximum number of leapfrog steps in each iteration was $N_{\max}=1000$.
HMC with $\alpha=100$ ran for $2000$ iterations, and standard HMC ran for roughly the same amount of time.
The proposals in standard HMC were obtained by making five leapfrog steps.
The leapfrog step size of 0.1 was used for both methods.

Figure~\ref{fig:mvnorm_d_closest_comp} shows the closest density component (1 or 2) for each state of the Markov chains constructed by both methods.
Whereas standard HMC failed to transition between the two modes, HMC with $\alpha=100$ made frequent inter-mode transitions.
Figure~\ref{fig:mvnorm_d_parallel_perp} shows obtained samples projected to a two dimensional affine plane containing both mode centers. 
Considering that the second density component has most of its mass on an exponentially small part in high dimensions, HMC with $\alpha=100$ exhibits a remarkable efficiency in sampling from the multimodal target distribution even in $d=200$.
As pointed out previously, this is due to the fact that the mass-enhanced HMC method searches the target space by exploiting the geometry of the log target density function via the Hamiltonian dynamics.
See Section~\ref{sec:enhancedHMC_cases} for further discussion.
\hfill $\Box$ \bigskip

\paragraph{Example~\theexample} \refstepcounter{example} \label{ex:enhanced_unimodal_misspecified_M}
Previous examples showed that when the mixture components have the same covariance, HMC with $\alpha>1$ can construct Markov chains that hop frequently between the components.
Example~\ref{ex:enhanced_bimodal_dim} in particular showed that the method works well for high-dimensional target distributions.
In general, however, when the mixture components have different covariances, HMC with increased temperature fails to facilitate jumps between the components.

\begin{figure}
  \centering
  \includegraphics[width=\linewidth]{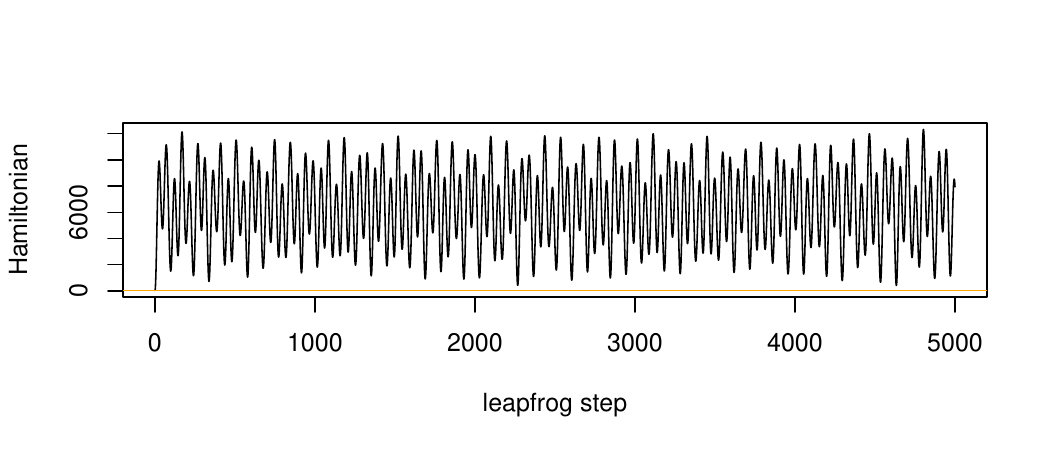}
  \caption{The Hamiltonian evaluated along an example path for the ten-dimensional Gaussian target distribution $\N(0, \Sigma)$ considered in Example~\ref{ex:enhanced_unimodal_misspecified_M} where the mass matrix $M$ is not equal to $\Sigma$. }
  \label{fig:trajectory_enhancedHMC_unimodal_misspecified_M}
\end{figure}
If the covariances of the mixture components are not the same, then obviously, the mass matrix $M$ cannot be equal to the inverse of every mixture covariance.
Therefore it suffices to show that if $M$ is not equal to the inverse of the covariance matrix $\Sigma$, Algorithm~\ref{alg:enhancedHMC} fails to transition between isolated modes in high dimensions.
We demonstrate that the Hamiltonian $H(x,p)$ tend to increase above 0 in this case.
Consider a (non-mixutre) Gaussian target distribution with covariance $\Sigma$.
Without loss of generality, we assume that $\Sigma$ is a diagonal matrix.
We consider a ten dimensional space and assume that the square roots of the diagonal entries of $\Sigma$ are randomly drawn from the uniform distribution between 0.5 and 4.
The mass matrix $M$ is equal to the identity matrix, and the temperature $\alpha$ is equal to 2000.
We used leapfrog step size $\epsilon = 0.1 \sqrt{\alpha} = \sqrt {20}$.
Figure~\ref{fig:trajectory_enhancedHMC_unimodal_misspecified_M} shows the original Hamiltonian evaluated along the trajectory.
The Hamiltonian is consistently higher than the initial value, indicated by the yellow horizontal line.
This indicates that, when the mass matrix $M$ is not equal to the inverse of the covariance matrix, the simulated paths with enhanced mass may not reach an acceptable state for a very long time.
A theoretical analysis in Section~\ref{sec:enhancedHMC_cases_misspecified_M} shows that this is the case in general when $M$ is not a scalar multiple of $\Sigma^{-1}$.
\hfill $\Box$ \bigskip

\begin{figure}
  \captionsetup{width=.89\linewidth}
  \newcommand{\pathAlphaFigScale}{0.85}
  \centering
  \begin{subfigure}[b]{\pathAlphaFigScale\linewidth}
    \includegraphics[width=\linewidth]{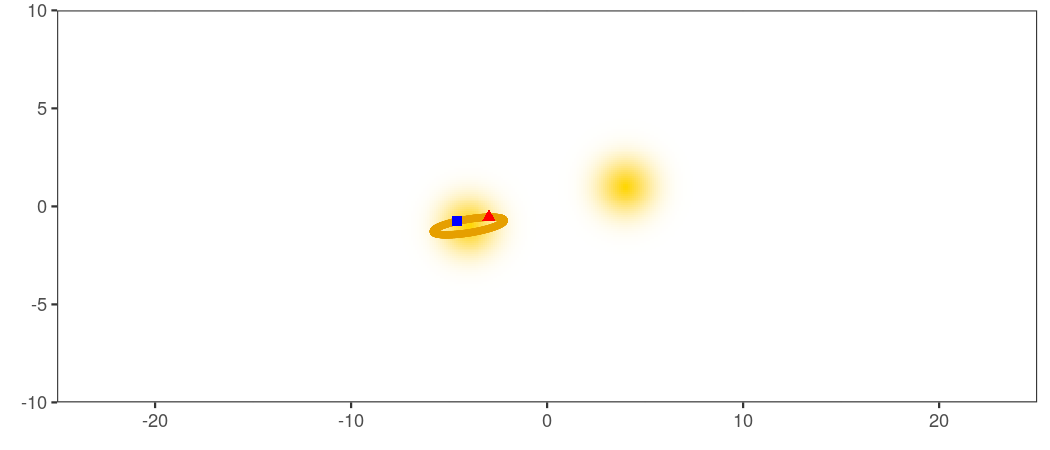}
    \caption{$\alpha = 1$}\label{fig:path_Mscale1}
  \end{subfigure}
  \hfill\\
  \begin{subfigure}[b]{\pathAlphaFigScale\linewidth}
    \includegraphics[width=\linewidth]{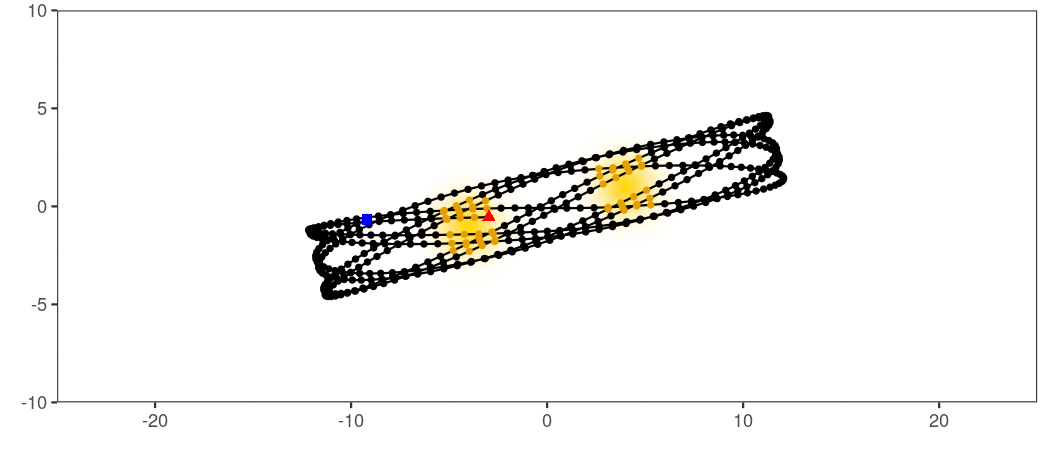}
    \caption{$\alpha = 30$}\label{fig:path_Mscale30}
  \end{subfigure}
  \hfill\\
  \begin{subfigure}[b]{\pathAlphaFigScale\linewidth}
    \includegraphics[width=\linewidth]{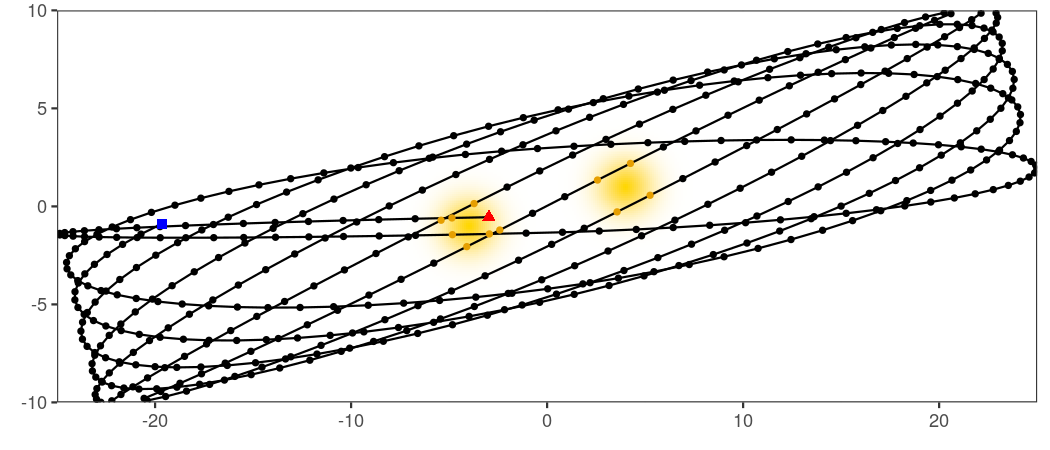}
    \caption{$\alpha = 200$}\label{fig:path_Mscale200}
  \end{subfigure}
  \caption{Numerically simulated Hamiltonian paths with various temperature levels $\alpha$ for a mixture of two normal target density considered in Example~\ref{ex:enhanced_bimodal_trajectory}.
    The two modes of the target distribution are represented by yellow color gradient.
    Acceptable points when the uniform $(0,1)$ random number $\Lambda$ is equal to 0.5 is shown as orange dots.
    The initial position is marked by a red triangle, and the 500-th position by a blue square.}
  \label{fig:path_Mscale}
\end{figure}

In Algorithm~\ref{alg:enhancedHMC}, the temperature $\alpha$ implicitly determines the search scope for separated density components. %, because it affects where the simulated Hamiltonian paths can reach.
Since $U(x(t)) + \breve K(v(t))$ is approximately conserved on the simulated path, every point on the path satisfies
\begin{equation}
  U(x(t)) \lesssim U(x(0)) + \alpha K(p(0)),
  \label{eqn:U_upperbound}
\end{equation}
where $K(p(0)) \sim \frac{1}{2}\chi_d^2$.
%Moreover, along the Hamiltonian path from $x(0)$ to $x'$, the potential energy $U$ cannot be greater than the right hand side of \eqref{eqn:U_upperbound}.
Therefore, with increasing $\alpha$, the area reachable by the simulated paths are expanded, and the depth of the potential energy barrier that can be crossed is increased.

\paragraph{Example~\theexample} \refstepcounter{example} \label{ex:enhanced_bimodal_trajectory}
Figure~\ref{fig:path_Mscale} shows the simulated Hamiltonian paths with constant temperature $\alpha = 1, 30$, and $200$ where the target distribution $\pi$ is a mixture of two Gaussian density components:
\begin{equation}
  \pi(x) = \frac{1}{2} \phi\left\{x; (-4,-1), I_{2\times 2}\right\} + \frac{1}{2} \phi\left\{x; (4,1), I_{2\times 2}\right\}.
  \label{eqn:bimodal_example_pi}
\end{equation}
Here $\phi(x; \mathbf \mu, I_{2\times 2})$ denotes the multivariate Gaussian density with mean $\mathbf \mu$ and the covariance matrix equal to the two dimensional identity matrix.
All three paths start with the same initial momentum $p(0)$.
The points on the numerically simulated paths are marked by orange dots if they are acceptable according to the criteria $H(x,v) < H\{x(0), p(0)\} - \log \Lambda$, where the Uniform(0,1) random number $\Lambda$ is assumed to take value 0.5.

When $\alpha=1$ (Figure~\ref{fig:path_Mscale1}), the Hamiltonian path does not leave the density component where the current state of the Markov chain is located.
However, the simulated paths can move across the region of low target density between the two modes when $\alpha=30$ (Figure~\ref{fig:path_Mscale30}).
If $\alpha$ is even greater ($\alpha=200$), the simulated Hamiltonian path reaches a wider area (Figure~\ref{fig:path_Mscale200}).
Since the simulated path traverses a wider area when $\alpha=200$, the proportion of acceptable states along the path may be smaller than when $\alpha=30$.
Thus tuning $\alpha=30$ may lead to more efficient sampling than $\alpha = 200$ for this target density given by \eqref{eqn:bimodal_example_pi}.
However, if there was another density component at a remote location (say at $(-20, -5)$), the simulation with $\alpha=200$ would be able to reach it whereas that with $\alpha=30$ would not.
Therefore, the mass enhancement ratio $\alpha$ may be chosen based on the desired search scope for isolated modes.
The search area is dependent on the geometry of the surface of the log target density function.
In Figure~\ref{fig:path_Mscale30} and \ref{fig:path_Mscale200}, the simulated Hamiltonian paths stay along the direction connecting the two density modes, where the potential energy is relatively low.
Algorithm~\ref{alg:enhancedHMC} is assisted by the local geometry of the potential energy landscape so that it can efficiently search for isolated modes regardless of the space dimension.

\begin{figure}
  \captionsetup{width=0.89\linewidth}
  \centering
  \includegraphics[width=0.89\linewidth]{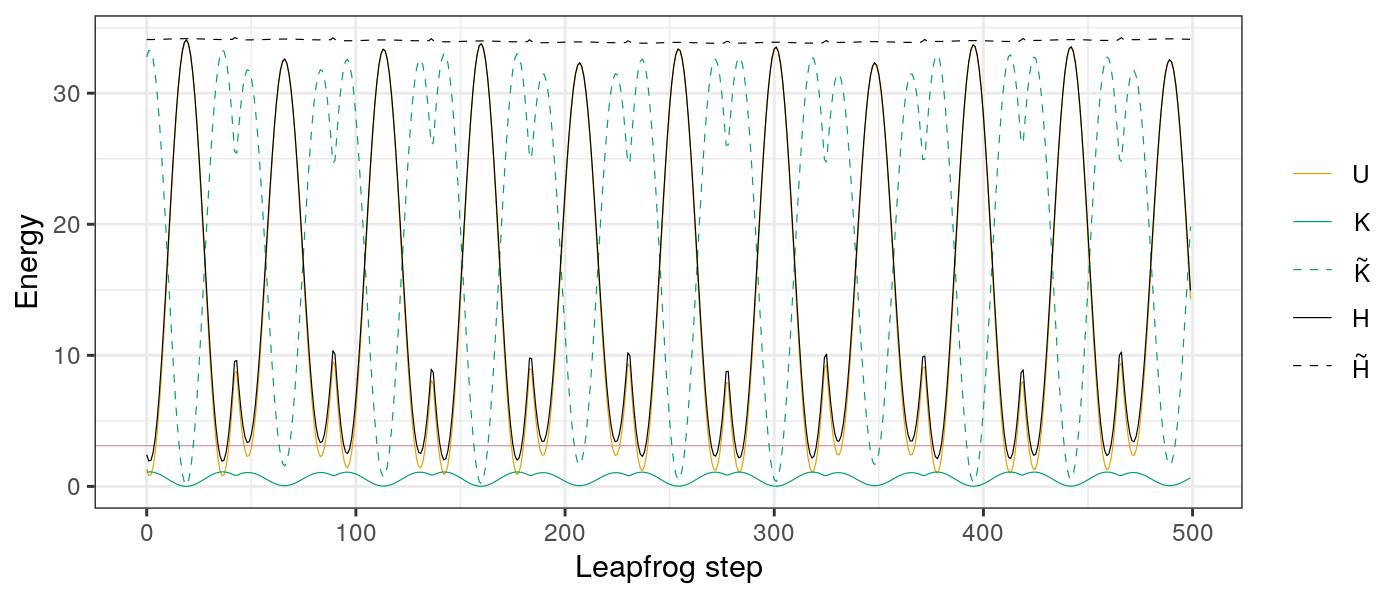}
  \caption{
    The traceplots for $U(x(t))$, $K(v(t))$, $\breve K(v(t))$, $H\{x(t), v(t)\}$, and $\breve H\{x(t), v(t)\}$ for the simulated path shown in Figure~\ref{fig:path_Mscale30}.
    The horizontal line at around 3 marks the acceptability criterion $H\{x(0), v(0)\} - \log(0.5)$, assuming that $\Lambda = 0.5$.
  }
  \label{fig:HMCms_Eplot_Mscale30}
\end{figure}

\hfill $\Box$ \bigskip

The parameter $N$ in Algorithm~\ref{alg:enhancedHMC} should be chosen large enough that the simulated path may leave the local mode it starts from.
This makes it more likely that the next state of the constructed Markov chain, which is the $N$-th acceptable proposal along the path, is found in a different density component.
In Figure~\ref{fig:path_Mscale30}, the simulated path starts from the red triangle and leaves the acceptable region around the first density component in four steps.
Thus in this case, if $N$ is suitably large, say $N\geq 4$, the next state of the constructed Markov chain may be found near the other mode located at $(4,1)$.

The parameter $N_{\max}$ sets an upper bound on the simulation time per iteration.
Therefore, it needs to be large enough to allow the simulated path to reach a remote density component.
However, in some cases such as when the log density surface is flat, the simulated path may head to a wrong direction for a prolonged amount of time.
Moreover, in rare cases where the current state of the Markov chain is close to the maximum of the target density and the drawn uniform random number $\Lambda$ is close to one, acceptable points may only be found in a very small area of $\mathsf X$ where the target density is higher than the starting point.
The parameter $N_{\max}$ needs to be set appropriately so as to avoid spending an excessive amount of time on finding an acceptable candidate in these unfavorable cases.

\subsection{Theoretical explanations of when mass-enhanced HMC does and does not work}\label{sec:enhancedHMC_cases}
\subsubsection{The case where mass-enhanced HMC works}\label{sec:enhancedHMC_cases_isotropic}
The fact that mass-enhanced HMC can construct globally mixing Markov chains for high-dimensional target distributions of the form
\[
\pi(x) = \frac{1}{2} \sum_{j=1}^2 \phi(x; \mu_j, I), \quad x \in \mathsf R^d
\]
as in Example~\ref{ex:enhanced_bimodal_dim} can be mathematically explained as follows.
The target density is proportional to
\[
\pi(x) \propto e^{-\Vert x-\mu_1 \Vert^2/2} + e^{-\Vert x-\mu_2 \Vert^2/2} =: c_1(x) + c_2(x),
\]
where without loss of generality we assume that
\[
\mu_1 = (\mu_1^1, 0, \dots, 0), \quad \mu_2 = (\mu_2^1, 0, \dots, 0).
\]
For simplicity we suppose $M=I$.
We have
\[
\frac{\partial U}{\partial x} = -\frac{\partial}{\partial x} \log \pi(x) = \frac{c_1(x) \cdot (x-\mu_1) + c_2(x) \cdot (x-\mu_2)}{c_1(x) + c_2(x)},
\]
or
\[
\frac{\partial U}{\partial x_i} = \begin{cases} \frac{c_1(x) (x_1 - \mu_1^1) + c_2(x) (x_1 - \mu_2^1)}{c_1(x) + c_2(x)} & \text{ for }i=1\\ x_i & \text{ for }i\geq 2.\end{cases}
\]
Thus for $i\geq 2$ the Hamiltonian equations of motion for $(x_i, v_i)$ can be solved independently:
\[
\frac{dx_i}{dt} = v_i, \quad \frac{dv_i}{dt} = -{\breve M}^{-1} \frac{\partial U}{\partial x_i} = -\alpha^{-1} x_i.
\]
The solution is given by
\[
x_i(t) = A_i \sin(\omega t + \varphi_i), \quad v_i(t) = A_i \omega \cos(\omega t+ \varphi_i)
\]
where $\omega = \sqrt{\alpha^{-1}}$.
The amplitude $A_i$ satisfies
\[
A_i^2 = x_i(0)^2 + \omega^{-2} v_i(0)^2 = x_i(0)^2 + \alpha v_i(0)^2,
\]
and the initial phase $\varphi_i \in [0,2\pi)$ satisfies
\[
x_i(0) = A_i \sin(\varphi_i), \quad v_i(0) = A_i \omega \cos(\varphi_i)
\]
Since $\omega = \sqrt{\alpha^{-1}} \gg 1$ and both $x_i(0)$ and $v_i(0)$ are random draws from $\N(0,1)$, the value of $\sin\varphi$ is close to zero and $\cos\varphi$ is close to one.

For $i=1$, we can consider a division of the space into three regions, $R_1= \{x; c_1(x) \gg c_2(x)\}$, $R_2 = \{x; c_1(x) \ll c_2(x)\}$, and $R_3 = (R_1 \cup R_2)^c$.
Since both $c_1(x)$ and $c_2(x)$ are exponential quadratic functions of $x$, most points in the space belongs to either the first or the second region.
Suppose without loss of generality that $x(0)$ belongs to $R_1$.
In this region, $(x_1(t), v_1(t))$ approximately has a similar form as that for $i\leq2$, namely
\[
x_1(t) - \mu_1^1 = A_1 \sin(\omega t + \varphi_1), \quad v_1(t) = A_1 \omega \cos(\omega t + \varphi_1).
\]
As the path enters the region $R_3$, the amplitude and the phase change in a way that is hard to predict, but after the path passes through $R_3$ and enters $R_2$ the solution becomes again approximately sinusoidal:
\[
x_1(t) - \mu_2^1 = A_1' \sin(\omega t + \varphi_1'), \quad v_1(t) = A_1' \omega \cos(\omega t + \varphi_1').
\]
The Hamiltonian in $R_2$ can be expressed as
\[
\begin{split}
H(t) &\approx \frac{1}{2} \Vert x-\mu_2 \Vert^2 + \frac{1}{2} \Vert v \Vert^2 \\
&= \frac{1}{2} (x_1-\mu_2^1)^2 + \frac{1}{2} v_1(t)^2 + \sum_{i\geq 2} \left[ \frac{1}{2} x_i(t)^2 + \frac{1}{2} v_i(t)^2 \right] \\
&= \left\{\frac{1}{2} A_1'^2 \sin^2(\omega t + \varphi_1') + \frac{1}{2} A_1'^2 \omega^2 \cos^2(\omega t + \varphi_1')\right\}
+ \left\{ \sum_{i\geq 2} \left[ \frac{1}{2} A_i^2 \sin^2(\omega t + \varphi_i) + \frac{1}{2} A_i^2 \omega^2 \cos^2(\omega t + \varphi_i) \right] \right\}\\
&:= H_1(t) + H_{i\geq 2}(t)
\end{split}
\]
Since the angular frequency is the same and equal to $\omega$ for all $i\geq 2$, the second term can be expressed as
\begin{equation}
H_{i\geq 2}(t) = B^0 + B^1 \cos(2\omega t + \theta)
\label{eqn:combined_phase}
\end{equation}
for some constants $B^0, B^1$, and $\theta$.
Therefore, $\Delta H_{i\geq 2}(t) := H_{i\geq 2}(t) - H_{i\geq 2}(0)$ periodically becomes zero or negative value with frequency $\omega/\pi$, regardless of the dimension $d$.
The quantity $H_1(t)$ is also periodic with frequency $\omega/\pi$.
Therefore if $\varphi_1'$ is such that both $\Delta H_1(t)$ and $\Delta H_{i\geq2}(t)$ can be simultaneously small at some point, that point can be accepted with a reasonably large probability.
As $\alpha$ increases, $H_{i\geq 2}(0)$ becomes closer to the minimum of its cycle, and the combined phase $\theta$ in \eqref{eqn:combined_phase} approaches $\pi$.
Thus the range of $\omega t$ for which $\Delta H_{i\geq 2}$ is below zero becomes narrower, and the chance that a point is accepted during the time a path stays in $R_2$ decreases.
In this case, many re-entering into $R_2$ may be necessary before the phase $\varphi_1'$ takes a fortunate value that allows a point on the path to be accepted.
In conclusion, as the distance between the two modes $\Vert \mu_1 - \mu_2 \Vert$ increases, a larger value of $\alpha$ will need to be used in order to enable the simulated path to reach a different mode, and the probability of accepted jump from one mode to another decreases.
However, the jump probability is not sensitive to the space dimension $d$; this conclusion is consistent with the results shown in Figure~\ref{fig:mvnorm_d}.

\subsubsection{The case where mass-enhanced HMC does not work}\label{sec:enhancedHMC_cases_misspecified_M}
Now we consider the case where mass-enhanced HMC does not work well; Example~\ref{ex:enhanced_unimodal_misspecified_M} illustrates this case.
Consider again a unimodal Gaussian distribution $\N(0, \Sigma)$ where $\Sigma$ is not equal to a scalar multiple of the mass inverse matrix $M^{-1}$.
Using the linear transformation \eqref{eqn:linear_transformation_xp}, we can simplify the case such that $M=I$ and $\Sigma$ is anisotropic (i.e., not a scalar multiple of the identity matrix).
Without loss of generality, suppose that $\Sigma$ is diagonal with entries $\sigma_{ii}^2$, $i=1,\dots, d$ and that these $d$ variances are all different.
The solution to the Hamiltonian equations of motion
\[
\frac{d x_i}{dt} = v_i, \quad \frac{d v_i}{dt} = - {\breve M}^{-1} \frac{\partial U}{\partial x_i} = - \alpha^{-1} \sigma_{ii}^{-2} x_i
\]
for the $i$-th component is given by
\[
x_i(t) = A_i \sin(\omega_i t+ \varphi_i), \quad v_i(t) = A_i \omega_i \cos(\omega_i t+ \varphi_i)
\]
where $\omega_i = \sqrt{\alpha^{-1}} \sigma_{ii}^{-1}$.
If $\alpha \gg 1$, the $d$ components are almost in sync, because
\[
\frac{\sin(\varphi_i)}{\cos(\varphi_i)} = \frac{x_i(0)}{v_i(0)} \omega_i = \frac{x_i(0)}{v_i(0)} \sqrt{\alpha^{-1}} \sigma_{ii}^{-1} \approx 0, \quad \forall i.
\]
However, as time progresses, the $d$ components become asynchronized due to the fact that $\omega_i$ are all different.
It takes an exponentially long time in $d$ for all the components to be in sync again.
Therefore for $\alpha \gg 1$ and large $d$, the increase in Hamiltonian $\Delta H(t)$ is consistently large for a very long time.
This phenomenon was demonstrated by Example~\ref{ex:enhanced_unimodal_misspecified_M}.

\clearpage
\bibliographystyle{abbrvnat}
\bibliography{ref}